%
%
\def\unredoffs{} \def\redoffs{\voffset=-.31truein\hoffset=-.59truein}
\def\speclscape{\special{ps: landscape}}
\newbox\leftpage \newdimen\fullhsize \newdimen\hstitle \newdimen\hsbody
\tolerance=1000\hfuzz=2pt
\catcode`\@=11 
\def\bigans{b }
\def\answ{b }
\ifx\answ\bigans\message{(This will come out unreduced.}
\magnification=1200\unredoffs\baselineskip=16pt plus 2pt minus 1pt
\hsbody=\hsize \hstitle=\hsize 
\else\message{(This will be reduced.} \let\l@r=L
\magnification=1000\baselineskip=16pt plus 2pt minus 1pt \vsize=7truein
\redoffs \hstitle=8truein\hsbody=4.75truein\fullhsize=10truein\hsize=\hsbody
\output={\ifnum\pageno=0 
  \shipout\vbox{\speclscape{\hsize\fullhsize\makeheadline}
    \hbox to \fullhsize{\hfill\pagebody\hfill}}\advancepageno
  \else
  \almostshipout{\leftline{\vbox{\pagebody\makefootline}}}\advancepageno 
  \fi}
\def\almostshipout#1{\if L\l@r \count1=1 \message{[\the\count0.\the\count1]}
      \global\setbox\leftpage=#1 \global\let\l@r=R
 \else \count1=2
  \shipout\vbox{\speclscape{\hsize\fullhsize\makeheadline}
      \hbox to\fullhsize{\box\leftpage\hfil#1}}  \global\let\l@r=L\fi}
\fi
%
\newcount\yearltd\yearltd=\year\advance\yearltd by -1900

\def\Title#1#2{\nopagenumbers\abstractfont\hsize=\hstitle\rightline{#1}%
\vskip 1in\centerline{\titlefont #2}\abstractfont\vskip .5in\pageno=0}
\def\Date#1{\vfill\leftline{#1}\tenpoint\supereject\global\hsize=\hsbody%
\footline={\hss\tenrm\folio\hss}}
%

\def\draftmode{\message{ DRAFTMODE }\def\draftdate{{\rm preliminary draft:
\number\month/\number\day/\number\yearltd\ \ \hourmin}}%
\headline={\hfil\draftdate}\writelabels\baselineskip=20pt plus 2pt minus 2pt
 {\count255=\time\divide\count255 by 60 \xdef\hourmin{\number\count255}
  \multiply\count255 by-60\advance\count255 by\time
  \xdef\hourmin{\hourmin:\ifnum\count255<10 0\fi\the\count255}}}
\def\nolabels{\def\wrlabeL##1{}\def\eqlabeL##1{}\def\reflabeL##1{}}
\def\writelabels{\def\wrlabeL##1{\leavevmode\vadjust{\rlap{\smash%
{\line{{\escapechar=` \hfill\rlap{\sevenrm\hskip.03in\string##1}}}}}}}%
\def\eqlabeL##1{{\escapechar-1\rlap{\sevenrm\hskip.05in\string##1}}}%
\def\reflabeL##1{\noexpand\llap{\noexpand\sevenrm\string\string\string##1}}}
\nolabels
%
\global\newcount\secno \global\secno=0
\global\newcount\meqno \global\meqno=1
\def\newsec#1{\global\advance\secno by1\message{(\the\secno. #1)}
\global\subsecno=0\eqnres@t\noindent{\bf\the\secno. #1}
\writetoca{{\secsym} {#1}}\par\nobreak\medskip\nobreak}
\def\eqnres@t{\xdef\secsym{\the\secno.}\global\meqno=1\bigbreak\bigskip}
\def\sequentialequations{\def\eqnres@t{\bigbreak}}\xdef\secsym{}
\global\newcount\subsecno \global\subsecno=0
\def\subsec#1{\global\advance\subsecno by1\message{(\secsym\the\subsecno. #1)}
\ifnum\lastpenalty>9000\else\bigbreak\fi
\noindent{\it\secsym\the\subsecno. #1}\writetoca{\string\quad 
{\secsym\the\subsecno.} {#1}}\par\nobreak\medskip\nobreak}
\def\appendix#1#2{\global\meqno=1\global\subsecno=0\xdef\secsym{\hbox{#1.}}
\bigbreak\bigskip\noindent{\bf Appendix #1. #2}\message{(#1. #2)}
\writetoca{Appendix {#1.} {#2}}\par\nobreak\medskip\nobreak}
%
%
\def\eqnn#1{\xdef #1{(\secsym\the\meqno)}\writedef{#1\leftbracket#1}%
\global\advance\meqno by1\wrlabeL#1}
\def\eqna#1{\xdef #1##1{\hbox{$(\secsym\the\meqno##1)$}}
\writedef{#1\numbersign1\leftbracket#1{\numbersign1}}%
\global\advance\meqno by1\wrlabeL{#1$\{\}$}}
\def\eqn#1#2{\xdef #1{(\secsym\the\meqno)}\writedef{#1\leftbracket#1}%
\global\advance\meqno by1$$#2\eqno#1\eqlabeL#1$$}
%
\newskip\footskip\footskip14pt plus 1pt minus 1pt 
\def\footnotefont{\ninepoint}\def\f@t#1{\footnotefont #1\@foot}
\def\f@@t{\baselineskip\footskip\bgroup\footnotefont\aftergroup\@foot\let\next}
\setbox\strutbox=\hbox{\vrule height9.5pt depth4.5pt width0pt}
\global\newcount\ftno \global\ftno=0
\def\foot{\global\advance\ftno by1\footnote{$^{\the\ftno}$}}
%
\newwrite\ftfile   
\def\footend{\def\foot{\global\advance\ftno by1\chardef\wfile=\ftfile
$^{\the\ftno}$\ifnum\ftno=1\immediate\openout\ftfile=foots.tmp\fi%
\immediate\write\ftfile{\noexpand\smallskip%
\noexpand\item{f\the\ftno:\ }\pctsign}\findarg}%
\def\footatend{\vfill\eject\immediate\closeout\ftfile{\parindent=20pt
\centerline{\bf Footnotes}\nobreak\bigskip\input foots.tmp }}}
\def\footatend{}
%
%
\global\newcount\refno \global\refno=1
\newwrite\rfile
\def\ref{[\the\refno]\nref}
\def\nref#1{\xdef#1{[\the\refno]}\writedef{#1\leftbracket#1}%
\ifnum\refno=1\immediate\openout\rfile=refs.tmp\fi
\global\advance\refno by1\chardef\wfile=\rfile\immediate
\write\rfile{\noexpand\item{#1\ }\reflabeL{#1\hskip.31in}\pctsign}\findarg}
\def\findarg#1#{\begingroup\obeylines\newlinechar=`\^^M\pass@rg}
{\obeylines\gdef\pass@rg#1{\writ@line\relax #1^^M\hbox{}^^M}%
\gdef\writ@line#1^^M{\expandafter\toks0\expandafter{\striprel@x #1}%
\edef\next{\the\toks0}\ifx\next\em@rk\let\next=\endgroup\else\ifx\next\empty%
\else\immediate\write\wfile{\the\toks0}\fi\let\next=\writ@line\fi\next\relax}}
\def\striprel@x#1{} \def\em@rk{\hbox{}} 
\def\lref{\begingroup\obeylines\lr@f}
\def\lr@f#1#2{\gdef#1{\ref#1{#2}}\endgroup\unskip}
\def\semi{;\hfil\break}
\def\addref#1{\immediate\write\rfile{\noexpand\item{}#1}} 
\def\startrefs#1{\immediate\openout\rfile=refs.tmp\refno=#1}
\def\xref{\expandafter\xr@f}\def\xr@f[#1]{#1}
\def\refs#1{\count255=1[\r@fs #1{\hbox{}}]}
\def\r@fs#1{\ifx\und@fined#1\message{reflabel \string#1 is undefined.}%
\nref#1{need to supply reference \string#1.}\fi%
\vphantom{\hphantom{#1}}\edef\next{#1}\ifx\next\em@rk\def\next{}%
\else\ifx\next#1\ifodd\count255\relax\xref#1\count255=0\fi%
\else#1\count255=1\fi\let\next=\r@fs\fi\next}
%

%
\newwrite\ffile\global\newcount\figno \global\figno=1
\def\fig{fig.~\the\figno\nfig}
\def\nfig#1{\xdef#1{fig.~\the\figno}%
\writedef{#1\leftbracket fig.\noexpand~\the\figno}%
\ifnum\figno=1\immediate\openout\ffile=figs.tmp\fi\chardef\wfile=\ffile%
\immediate\write\ffile{\noexpand\medskip\noexpand\item{Fig.\ \the\figno. }
\reflabeL{#1\hskip.55in}\pctsign}\global\advance\figno by1\findarg}
\def\xfig{\expandafter\xf@g}\def\xf@g fig.\penalty\@M\ {}
\def\figs#1{figs.~\f@gs #1{\hbox{}}}
\def\f@gs#1{\edef\next{#1}\ifx\next\em@rk\def\next{}\else
\ifx\next#1\xfig #1\else#1\fi\let\next=\f@gs\fi\next}
\newwrite\lfile
{\escapechar-1\xdef\pctsign{\string\%}\xdef\leftbracket{\string\{}
\xdef\rightbracket{\string\}}\xdef\numbersign{\string\#}}

\def\writestop{\def\writestoppt{\immediate\write\lfile{\string\pageno%
\the\pageno\string\startrefs\leftbracket\the\refno\rightbracket%
\string\def\string\secsym\leftbracket\secsym\rightbracket%
\string\secno\the\secno\string\meqno\the\meqno}\immediate\closeout\lfile}}
\def\writestoppt{}\def\writedef#1{}
\def\seclab#1{\xdef #1{\the\secno}\writedef{#1\leftbracket#1}\wrlabeL{#1=#1}}
\def\subseclab#1{\xdef #1{\secsym\the\subsecno}%
\writedef{#1\leftbracket#1}\wrlabeL{#1=#1}}
\newwrite\tfile \def\writetoca#1{}
\def\leaderfill{\leaders\hbox to 1em{\hss.\hss}\hfill}
\def\writetoc{\immediate\openout\tfile=toc.tmp 
   \def\writetoca##1{{\edef\next{\write\tfile{\noindent ##1 
   \string\leaderfill {\noexpand\number\pageno} \par}}\next}}}

%
\catcode`\@=12 
%
\edef\tfontsize{\ifx\answ\bigans scaled\magstep3\else scaled\magstep4\fi}
\font\titlerm=cmr10 \tfontsize \font\titlerms=cmr7 \tfontsize
\font\titlermss=cmr5 \tfontsize \font\titlei=cmmi10 \tfontsize
\font\titleis=cmmi7 \tfontsize \font\titleiss=cmmi5 \tfontsize
\font\titlesy=cmsy10 \tfontsize \font\titlesys=cmsy7 \tfontsize
\font\titlesyss=cmsy5 \tfontsize \font\titleit=cmti10 \tfontsize
\skewchar\titlei='177 \skewchar\titleis='177 \skewchar\titleiss='177
\skewchar\titlesy='60 \skewchar\titlesys='60 \skewchar\titlesyss='60
\def\titlefont{\def\rm{\fam0\titlerm}
\textfont0=\titlerm \scriptfont0=\titlerms \scriptscriptfont0=\titlermss
\textfont1=\titlei \scriptfont1=\titleis \scriptscriptfont1=\titleiss
\textfont2=\titlesy \scriptfont2=\titlesys \scriptscriptfont2=\titlesyss
\textfont\itfam=\titleit \def\it{\fam\itfam\titleit}\rm}
 \ifx\answ\bigans\else scaled\magstep1\fi
\ifx\answ\bigans\def\abstractfont{\tenpoint}\else
\font\abssl=cmsl10 scaled \magstep1
\font\absrm=cmr10 scaled\magstep1 \font\absrms=cmr7 scaled\magstep1
\font\absrmss=cmr5 scaled\magstep1 \font\absi=cmmi10 scaled\magstep1
\font\absis=cmmi7 scaled\magstep1 \font\absiss=cmmi5 scaled\magstep1
\font\abssy=cmsy10 scaled\magstep1 \font\abssys=cmsy7 scaled\magstep1
\font\abssyss=cmsy5 scaled\magstep1 \font\absbf=cmbx10 scaled\magstep1
\skewchar\absi='177 \skewchar\absis='177 \skewchar\absiss='177
\skewchar\abssy='60 \skewchar\abssys='60 \skewchar\abssyss='60
\def\abstractfont{\def\rm{\fam0\absrm}
\textfont0=\absrm \scriptfont0=\absrms \scriptscriptfont0=\absrmss
\textfont1=\absi \scriptfont1=\absis \scriptscriptfont1=\absiss
\textfont2=\abssy \scriptfont2=\abssys \scriptscriptfont2=\abssyss
\textfont\itfam=\bigit \def\it{\fam\itfam\bigit}\def\footnotefont{\tenpoint}%
\textfont\slfam=\abssl \def\sl{\fam\slfam\abssl}%
\textfont\bffam=\absbf \def\bf{\fam\bffam\absbf}\rm}\fi
\def\tenpoint{\def\rm{\fam0\tenrm}
\textfont0=\tenrm \scriptfont0=\sevenrm \scriptscriptfont0=\fiverm
\textfont1=\teni  \scriptfont1=\seveni  \scriptscriptfont1=\fivei
\textfont2=\tensy \scriptfont2=\sevensy \scriptscriptfont2=\fivesy
\textfont\itfam=\tenit \def\it{\fam\itfam\tenit}\def\footnotefont{\ninepoint}%
\textfont\bffam=\tenbf \def\bf{\fam\bffam\tenbf}\def\sl{\fam\slfam\tensl}\rm}
\font\ninerm=cmr9 \font\sixrm=cmr6 \font\ninei=cmmi9 \font\sixi=cmmi6 
\font\ninesy=cmsy9 \font\sixsy=cmsy6 \font\ninebf=cmbx9 
\font\nineit=cmti9 \font\ninesl=cmsl9 \skewchar\ninei='177
\skewchar\sixi='177 \skewchar\ninesy='60 \skewchar\sixsy='60 
\def\ninepoint{\def\rm{\fam0\ninerm}
\textfont0=\ninerm \scriptfont0=\sixrm \scriptscriptfont0=\fiverm
\textfont1=\ninei \scriptfont1=\sixi \scriptscriptfont1=\fivei
\textfont2=\ninesy \scriptfont2=\sixsy \scriptscriptfont2=\fivesy
\textfont\itfam=\ninei \def\it{\fam\itfam\nineit}\def\sl{\fam\slfam\ninesl}%
\textfont\bffam=\ninebf \def\bf{\fam\bffam\ninebf}\rm} 
%
%

\hyphenation{anom-aly anom-alies coun-ter-term coun-ter-terms}
\def\inv{^{\raise.15ex\hbox{${\scriptscriptstyle -}$}\kern-.05em 1}}

\def\Dsl{\,\raise.15ex\hbox{/}\mkern-13.5mu D} 
\def\dsl{\raise.15ex\hbox{/}\kern-.57em\partial}
\def\del{\partial}

\def\tr{{\rm tr}} \def\Tr{{\rm Tr}}
\font\bigit=cmti10 scaled \magstep1
\def\lspace{\ifx\answ\bigans{}\else\qquad\fi}
\def\lbspace{\ifx\answ\bigans{}\else\hskip-.2in\fi} 
\def\boxeqn#1{\vcenter{\vbox{\hrule\hbox{\vrule\kern3pt\vbox{\kern3pt
        \hbox{${\displaystyle #1}$}\kern3pt}\kern3pt\vrule}\hrule}}}
\def\mbox#1#2{\vcenter{\hrule \hbox{\vrule height#2in
                \kern#1in \vrule} \hrule}}  
%
 \def\CO{{\cal O}} 
    
\def\CL{{\cal L}} \def\CH{{\cal H}}  
  \def\CD{{\cal D}} 
\def\e#1{{\rm e}^{^{\textstyle#1}}}

\def\vev#1{\langle #1 \rangle}

\def\darr#1{\raise1.5ex\hbox{$\leftrightarrow$}\mkern-16.5mu #1}

\def\half{{\textstyle{1\over2}}} 
\def\roughly#1{\raise.3ex\hbox{$#1$\kern-.75em\lower1ex\hbox{$\sim$}}}
%
\lref\cartersanda{A. B. Carter and A. I. Sanda, \pr{D23}{1981}{1567}}
\lref\messiah{A. Messiah, {\sl Quantum Mechanics}, v.~II, North-Holland Pub.\
Co., Amsterdam, 1962, p.~994}
\lref\bjdrell{J. D. Bjorken and S. D. Drell, {\sl Relativistic Quantum
Fields}, McGraw-Hill Book Co., 1965}
\lref\jpsia{J.J. Aubert et al., \prl{33}{1974}{1404}}
\lref\jpsib{J.-E. Augustin et al., \prl{33}{1974}{1406}}
\lref\psip{G.S. Abrams et al., \prl{33}{1453}{1974}}
\lref\siegrist{J. Siegrist et al., \prl{36}{1976}{700}}
\lref\dasp{W. Braunschweig et al., \pl{57B}{1975}{407}}
\lref\crysball{R. Partridge et al., \prl{45}{1980}{1150}}
\lref\dmsnmrki{G. Goldhaber et al., \prl{37}{1976}{301}}
\lref\upsifermi{S.W. Herb et al., \prl{39}{1977}{252}}
\lref\upsipluto{Ch.~Berger et al., \pl{76B}{1978}{243}}
\lref\upsidasp{C.W. Darden et al., \pl{76B}{1978}{246}}
\lref\upsipdasp{C.W. Darden et al., \pl{78B}{1978}{364}}
\lref\upsipdesyh{J.K. Bienlein et al., \pl{78B}{1978}{360}}
\lref\upsicesra{D. Andrews et al., \prl{44}{1980}{1108}\semi
T. B\"ohringer et al.,  \prl{44}{1980}{1111}}
\lref\upsicesrb{D. Andrews et al., \prl{45}{1980}{219}\semi
G. Finocchiaro et al.,  \prl{45}{1980}{222}}
\lref\bmesoncleo{S. Behrends et al., \prl{50}{1983}{881}}
\lref\witten{E. Witten, \np{B122}{1977}{109}}
\lref\altarelli{G. Altarelli and L. Maiani, \pl{52B}{1974}{351}}
\lref\gaillard{M. K. Gaillard and B. W. Lee, \prl{33}{1974}{108}}
\lref\eichtenhill{E. Eichten and B. Hill, \pl{B234}{1990}{511}}
\lref\grinstein{B. Grinstein, \np{B339}{1990}{253}}
\lref\georgi{H. Georgi, \pl{B240}{1990}{447}}
\lref\dgg{M. Dugan, M. Golden and B. Grinstein, \pl{B282}{142}{1992}}
\lref\grinsteinannrevs{B. Grinstein, 
{\it Ann.\ Rev.\ Nucl.\ Part.\ Sc.,\/}\hfil\break {\bf 42} (1992) 101}
\lref\iwspectrum{N. Isgur and M. B. Wise, \prl{66}{1991}{1132}}
\lref\falkspin{A. F. Falk, \np{B378}{1992}{79}}
\lref\fggw{A. F.  Falk, et al, \np{B343}{1990}{1}}
\lref\isgurwisea{N. Isgur and M.B. Wise, \pl{B232}{1989}{113};
\pl{B237}{1990}{527}}
\lref\nussinov{S. Nussinov and W. Wetzel, \pr{D36}{1987}{130}}
\lref\volosh{M.B. Voloshin and M.A. Shifman, \sjnp{47}{1988}{511}}
\lref\eichtenfeinberg{E. Eichten 
and F. L. Feinberg, \prl{43}{1979}{1205};\semi {\it idem,\/}
  \pr{D23}{1981}{2724}}
\lref\lepagethacker{G. Lepage and B.A. Thacker, \np{B{\rm
(Proc.~Suppl.)4}}{1988}{199}}
\lref\carone{C. Carone, \pl{253B}{1991}{408}}
\lref\decoupling{T. Appelquist and J. Carrazone, \pr{D11}{1975}{2856}}
\lref\falkgrinsteina{A. F. Falk and B. Grinstein, \pl{B249}{1990}{314}}
\lref\falkgrinsteinb{A. Falk and B. Grinstein, \pl{B247}{1990}{406}}
\lref\eichtenhillb{E. Eichten and B. Hill, \pl{B243}{1990}{427}}
\lref\volotwo{M.B. Voloshin and M.A. Shifman, \sjnp{45}{1987}{292}}
\lref\politzerwiseone{H.D. Politzer and M.B. Wise, \pl{B206}{1988}{681}}
\lref\ggw{H. Georgi, B. Grinstein and M. B. Wise, \pl{252B}{1990}{456}}
\lref\chog{P. Cho and B. Grinstein, \pl{B285}{1992}{153}}
\lref\luke{M.E. Luke, \pl{B252}{1990}{447}}
\lref\chris{Boucaud, P., et al, \pl{B220}{1989}{219}\semi
Allton, C. R., et al, \np{B349}{1991}{598}}
\lref\baryonrefs{N. Isgur and M.B. Wise, 
\np{B348}{1991}{276}\semi H. Georgi, \np{B348}{1991}{293}\semi 
T. Mannel, W. Roberts and Z. Ryzak, \np{B355}{1991}{38};
\pl{B255}{1991}{593}}
\lref\anatoli{G.P. Korchemskii and  A.V. Radyushkin ,
\pl{B279}{1992}{359}}
%
%
\lref\georgibook{{A good introduction can be found in  H. Georgi, {\sl
Weak Interactions and Modern Particle Physics\/}(The Benjamin/Cummings
Publishing Company, California,1984)}}
\lref\chiral{{G. Burdman, J. F. Donoghue, \pl{B280}{1992}{287}\semi 
M. B. Wise,  \pr{D45}{1992}{2188}}}
\lref\yanetal{{T.-M. Yan, H.-Y. Cheng, C.-Y. Cheung, G.-L. Lin,  
Y. C. Lin and H.-L. Yu, \pr{D46}{1992}{1148}} }
\lref\accmor{{The ACCMOR Collaboration (S. Barlag {\it et al\/}), 
\pl{B278}{1992}{480}}} 
\lref\cleobfs{{The CLEO Collaboration (F. Butler {\it et al\/}),
\prl{69}{1992}{2041}} }
\lref\jimcho{{J. Amundson, C. Boyd, E. Jenkins, M. Luke, A.
Manohar, J. Rosner, M. Savage and M. Wise, \pl{B296}{1992}{415}\semi 
P. Cho and H. Georgi, \pl{B296}{1992}{408}}}
\lref\yanetalbaryons{{H.-Y. Cheng, C.-Y. Cheung, G.-L. Lin,  
Y. C. Lin, T.-M. Yan and H.-L. Yu, \pr{D46}{1992}{5060}}}
\lref\morecleo{The CLEO collaboration (S. Henderson, {\it et al,\/})
\pr{D45}{1992}{2212}}
\lref\cleofds{The  CLEO Collaboration (D. Acosta,  {\it et al,\/}),
{\it First Measurement Of $\Gamma(D_s^+\to\mu^+\nu) /
\Gamma(D_s^+\to\phi\pi^+)$\/,} CLNS-93-1238, Aug 1993.}
\lref\fiveus{{B. Grinstein, E. Jenkins, A. Manohar, M. Savage and
M. Wise, \np{B380}{1992}{369}}}
\lref\goity{{J. Goity, \pr{D46}{1992}{3929}}}
\lref\ratioofbs{B. Grinstein, \prl{71}{1993}{3067}}
\lref\soni{See, for example, C. Bernard, J. Labrenz and  A. Soni,
\physrev{D49}{1994}{2536}\semi  
C.  Dominguez, \pl{B318}{1993}{629}} 
\lref\PDG{{Particle Data Group, \pr{D45}{1992}{ } }}
\lref\cleomass{{The Cleo Collaboration (D. Bortoletto {\it et al\/})
\prl{69}{1992}{2046}}}
\lref\cusb{{The CUSB-II Collaboration (J. Lee-Franzini
{\it et al\/}), \prl{65}{1990}{2947} }}
\lref\cleotwomass{{The Cleo II Collaboration (D.S. Akerib
{\it et al\/}), \prl {67}{1991}{1692}}}
\lref\rosnerwise{{J. Rosner and M. Wise, \pr{D47}{1992}{343}}}
\lref\fgl{A. Falk, B. Grinstein and M. Luke, \np{B357}{1991}{185}}
\lref\ransath{L. Randall and E. Sather, \pl{B303}{1993}{345}}
\lref\jenk{E. Jenkins, \np{B412}{1994}{181}}
\lref\ranwise{L. Randall and M. Wise, \pl{B303}{1993}{135}}
\lref\chowwise{C-K. Chow, M.B. Wise, \physrev{D48}{1993}{5202}
(hep-ph/9305229)} 
\lref\savjen{E. Jenkins and M.J. Savage, \pl{B281}{1992}{331}}
\lref\falkben{ A. Falk and B. Grinstein, \np{B416}{1994}{771}}
\lref\flf{A. Falk, \np{B378}{1992}{79}\semi
A. Falk and M. Luke, \pl{B292}{1992}{119}}
\lref\falk{A. Falk, \pl{B305}{1993}{268}}
%
\lref\gronauwyler{M. Gronau and D. Wyler, \pl{B253}{1991}{172}}
%
\lref\dstwoobsrved{Y. Kubota, et al., the CLEO Collaboration, 
CLNS-94-1266, Jan 1994,hep-ph/9403325}
\lref\cleobsgamma{M.S. Alam, et al., the CLEO Collaboration,
{\it First Measurement Of The Rate For The Inclusive Radiative Penguin
Decay  $b\to s\gamma$,\/}  CLNS-94-1314, Dec 1994.
}
\lref\boydbg{C. G.  Boyd and B. Grinstein, {\sl Chiral and Heavy Quark
Symmetry Violation in B Decays}, UCSD/PTH 93-46, SMU-HEP/94-03,
SSCL--Preprint--532, Feb.~1994 (hep-ph/9402340)}
\lref\dunietz{I. Dunietz, {\sl Extracting CKM parameters from B
decays,\/ }in Proceedings of the Workshop on B Physics at Hadron
Accelerators, Snowmass Colo., 1993}
\lref\pdgXCIV{Review of Particle Properties, \physrev{D50}{1}{1994}}
\lref\goldenhill{M. Golden and B. Hill, \pl{B254}{1991}{225}}
\relax\Title{\vbox{\hbox{UCSD/PTH 95-05}}}{
\centerline{An Introduction To Heavy Mesons}}
\centerline{Benjamin
Grinstein\footnote{}{\hskip-2.5ex$^{\S}$\hskip1ex{}Lectures presented
at the VI Mexican School Of Particles And Fields, Villahermosa, 
Tabasco, 3--7 October, 1994}\footnote{$^{\dagger}$}{bgrinstein@ucsd.edu}}
\bigskip\centerline{Department of Physics}
\centerline{University of California, San Diego}
\centerline{La Jolla, California 92093-0319}
\vskip .3in
Introductory lectures on heavy quarks and heavy quark effective field
theory. Applications to inclusive semileptonic decays and to
interactions with light mesons are covered in detail.
\Date{August 1995} 
%
%
%
%
\headline={\ifnum\pageno=1\firstheadline\else
\ifodd\pageno\rightheadline \else\leftheadline\fi\fi}
\def\firstheadline{\hfil}
\def\rightheadline{\hfil}
\def\leftheadline{\hfil}
	\footline={\ifnum\pageno=1\firstfootline\else\otherfootline\fi}
\def\firstfootline{\rm\hss\folio\hss}
\def\otherfootline{\hfil}
\parindent=1.5pc
\hsize=6.0truein
\vsize=8.5truein
\nopagenumbers
\centerline{\ninebf AN INTRODUCTION TO HEAVY MESONS}
\vglue 0.8cm
\centerline{\ninerm BENJAMIN GRINSTEIN}
\baselineskip=13pt
\centerline{\nineit Department Of Physics, University of California, San Diego}
\baselineskip=12pt
\centerline{\nineit La Jolla, CA 92093-0319, USA}
\vglue 0.8cm
\centerline{\ninerm ABSTRACT}
\vglue 0.3cm
{\rightskip=3pc
 \leftskip=3pc
 \ninerm\baselineskip=12pt\noindent
Introductory lectures on heavy quarks and heavy quark effective field
theory. Applications to inclusive semileptonic decays and to
interactions with light mesons are covered in detail.
\vglue 0.6cm}
%
\baselineskip=14pt plus 2pt minus 1pt
%
%
\def\normconv{(2.1)}
\def\cclag{(2.2)}
\def\deltalageff{(2.3)}
\def\normalgreenfs{(2.4)}
\def\normalC{(2.5)}
\def\heffcsdutree{(2.6)}
\def\heffculesstree{(2.7)}
\def\heffcsdubeyond{(2.8)}
\def\charmcoeffs#1{\hbox {$(2.9#1)$}}
\def\twotermhamil{(2.10)}
\def\intuitive{3.1}
\def\efflagsec{3.2}
\def\eIIi{(3.1)}
\def\eIIii{(3.2)}
\def\eIIiii{(3.3)}
\def\eIIiv{(3.4)}
\def\pisv{(3.5)}
\def\greentwo{(3.6)}
\def\greenthreefull{(3.7)}
\def\greenthree{(3.8)}
\def\gammaisv{(3.9)}
\def\boost{(3.10)}
\def\eIIv{(3.11)}
\def\symmtrssec{3.3}
\def\hqeteffl{(3.12)}
\def\largemass{(3.13)}
\def\projonQ{(3.14)}
\def\leffagain{(3.15)}
\def\vectprop{(3.16)}
\def\vectorleff{(3.17)}
\def\vecconstraint{(3.18)}
\def\totangmom{(3.19)}
\def\massratios{(3.20)}
\def\dblts{(3.21)}
\def\wignerdblts{(3.22)}
\def\cvrepofsts{3.5}
\def\melements{(3.23)}
\def\uvrep{(3.24)}
\def\comba{(3.25)}
\def\combb{(3.26)}
\def\reposmesons{(3.27)}
\def\ohyes#1{\hbox {$(3.28#1)$}}
\def\fmesondefd{(3.29)}
\def\mesonnorm{(3.30)}
\def\fvectdefd{(3.31)}
\def\ftildedefd{(3.32)}
\def\IVii{(3.33)}
\def\ftoftilde{(3.34)}
\def\fBoverfD{(3.35)}
\def\eIVi{(3.36)}
\def\eIVii{(3.37)}
\def\eIViii{(3.38)}
\def\eIViv{(3.39)}
\def\eIVv{(3.40)}
\def\semilepdecsec{3.7}
\def\fulff#1{\hbox {$(3.41#1)$}}
\def\eVii#1{\hbox {$(3.42#1)$}}
\def\eViii{(3.43)}
\def\eVv{(3.44)}
\def\eVvi#1{\hbox {$(3.45#1)$}}
\def\eVvii{(3.46)}
\def\eVviii{(3.47)}
\def\eVix{(3.48)}
\def\baryonvev{(3.49)}
\def\lambdaffszeroth{(3.50)}
\def\oneloopsec{3.8}
\def\eIIiloop{(3.51)}
\def\eIIvagain{(3.52)}
\def\eIIvi{(3.53)}
\def\fulltoefffig{fig.~1}
\def\fourptfnctfig{fig.~2}
\def\threeptfnctfig{fig.~3}
\def\eIIvii{(3.54)}
\def\eIIviii{(3.55)}
\def\eIIviiii{(3.56)}
\def\xtrnlcs{3.9}
\def\fullcurrent{(3.57)}
\def\ex{(3.58)}
\def\exi{(3.59)}
\def\exii{(3.60)}
\def\exiii{(3.61)}
\def\fulltoeffcurr{(3.62)}
\def\exiv{(3.63)}
\def\exv{(3.64)}
\def\exvii{(3.65)}
\def\exviii{(3.66)}
\def\exiv{(3.67)}
\def\exx{(3.68)}
\def\exxi{(3.69)}
\def\exxii{(3.70)}
\def\exxiiidupl{(3.71)}
\def\exxiii{(3.72)}
\def\exxvi{(3.73)}
\def\exxviv{(3.74)}
\def\exxvv{(3.75)}
\def\factorone{(3.76)}
\def\factortwo{(3.77)}
\def\exxvviii{(3.78)}
\def\replacevec{(3.79)}
\def\replaceaxi{(3.80)}
\def\rdefined{(3.81)}
\def\eVxiv{(3.82)}
\def\eVxv{(3.83)}
\def\eVxvi{(3.84)}
\def\oneoverm{4}
\def\eVIi{(4.1)}
\def\eVIii{(4.2)}
\def\eVIiii{(4.3)}
\def\eVIiv{(4.4)}
\def\eVIv{(4.5)}
\def\eVIvi{(4.6)}
\def\eVIvii{(4.7)}
\def\eVIviii{(4.8)}
\def\eVIix{(4.9)}
\def\eVIx{(4.10)}
\def\eVIxi{(4.11)}
\def\eVIxii{(4.12)}
\def\colormagmom{(4.13)}
\def\coefsinlag{(4.14)}
\def\ccoefsa{(4.15)}
\def\ccoefsb{(4.16)}
\def\subhlcur{(4.17)}
\def\fullsubhlcur{(4.18)}
\def\opexamples{(4.19)}
\def\subhhcur{(4.20)}
\def\fullsubhhcur{(4.21)}
\def\replacement{(4.22)}
\def\exampleope{(4.23)}
\def\exampleopegives{(4.24)}
\def\deltavec{(4.25)}
\def\aovera{(4.26)}
\def\bcoefs{(4.27)}
\def\bsum{(4.28)}
\def\lukesthm{4.4}
\def\magneticvev{(4.29)}
\def\magneticzero{(4.30)}
\def\baryonvevb{(4.31)}
\def\BandA{(4.32)}
\def\baryonvevc{(4.33)}
\def\Aandzeta{(4.34)}
\def\eVIviii{(4.35)}
\def\baryonffs{(4.36)}
\def\baryonffnorm{(4.37)}
\def\inclusiverate{(5.1)}
\def\phspdefd{(5.2)}
\def\ellmu{(5.3)}
\def\hmunudefd{(5.4)}
\def\tmunudefd{(5.5)}
\def\handtffs#1{\hbox {$(5.6#1)$}}
\def\hfromt{(5.7)}
\def\fiveone{fig.~4}
\def\fivetwo{fig.~5}
\def\inthtof{(5.8)}
\def\fivethree{fig.~6}
\def\fivefour{fig.~7}
\def\avggnrlzd{(5.9)}
\def\fivefive{fig.~8}
\def\opetree{(5.10)}
\def\hqetope{(5.11)}
\def\fivesix{fig.~9}
\def\opepoles{(5.12)}
\def\csgolike{(5.13)}
\def\oneopdone{(5.14)}
\def\twoopdone{(5.15)}
\def\eqsuperf{(6.1)}
\def\calmeq{(6.2)}
\def\eqgterm{(6.3)}
\def\eqgexpand{(6.4)}
\def\polediagfig{fig.~10}
\def\eqeffcurr{(6.5)}
\def\eqlcoup{(6.6)}
\def\fDdefd{(6.7)}
\def\fDSdefd{(6.8)}
\def\chiralqQcurr{(6.9)}
\def\fdsubsfig{fig.~11}
\def\eqratiodecays{(6.10)}
\def\fsubdsubs{(6.11)}
\def\eqratioratio{(6.12)}
\def\xisoverxiud{(6.13)}
\def\eqspinsymviol{(6.14)}
\def\eqlthree{(6.15)}
\def\eqhyperfsplit{(6.16)}
\def\eqmassrel{(6.17)}
\def\eqspinindepmass{(6.18)}
\def\eqranwise#1{\hbox {$(6.19#1)$}}
\def\eqranwisef{(6.20)}
\def\eqkpirel{(6.21)}
\def\eqSsuperf{(6.22)}
%
\def\IFBIG#1#2{\ifx\answ\bigans{#1}\else{#2}\fi}
\input epsf%
\def\INSERTFIG#1#2#3{\vbox{\vbox{\hfil\epsfbox{#1}\hfill}%
{\narrower\noindent%
\multiply\baselineskip by 3%
\divide\baselineskip by 4%
{\ninerm Figure \xfig#2 }{\ninesl #3 \medskip}}
}}%
\def\PASTEFIG#1#2#3{\vbox{\vtop to #1 truein{ } %
{\narrower\noindent%
\multiply\baselineskip by 3%
\divide\baselineskip by 4%
{\ninerm Figure \xfig#2 }{\ninesl #3 \medskip}}
}}%

\def\ihalf{{\scriptstyle{i\over2}}} 
\def\G{\Gamma}
\def\g{\gamma}
\def\e{\epsilon}

\def\im{{\rm Im}}

\def\eg{{\it e.g.,\/\ }}
\def\Dsl{\raise.15ex\hbox{/}\kern-.74em D}
\def\vecpp{{\vec p}\;{}'}

\def\OMIT#1{}
\def\spur{\raise.15ex\hbox{/}\kern-.57em } 
\def\leff{{\cal L}_{\rm eff}}

\def\a{\alpha}
\def\vsl{v \hskip-5pt /}
\def\lsl{\spur {\kern0.1em l}}
\def\tjg{_{\Gamma}}
\def\projp{\left({1+\vsl\over2}\right)}
\def\projpsm{\left(\!{1+\vsl\over2}\!\right)}
\def\projm{\left({1-\vsl\over2}\right)}
\def\projmsm{\left(\!{1-\vsl\over2}\!\right)}
\def\projpm{\left({1\pm\vsl\over2}\right)}
\def\pprojp{\left({1+\vsl'\over2}\right)}

\def\qt#1{\tilde Q_v^{(#1)}}
\def\ccdot{\hbox{\kern-.1em$\cdot$\kern-.1em}}
\def\vvv{v \ccdot v'}

\def\tenmibfonts{
   \global\font\tenmib=cmmib10%
   \global\font\tenbsy=cmbsy10%
   \skewchar\tenmib='177%
   \skewchar\tenbsy='60%
   \gdef\tenmibfonts{\relax}}

\def\mib{
   \tenmibfonts%
   \textfont0=\tenbf\scriptfont0=\sevenbf%
   \scriptscriptfont0=\fivebf%
   \textfont1=\tenmib\scriptfont1=\seveni%
   \scriptscriptfont1=\fivei%
   \textfont2=\tenbsy\scriptfont2=\sevensy%
   \scriptscriptfont2=\fivesy}%



\def\np#1#2#3{\NP{\bf #1} (#2) #3}
\def\pl#1#2#3{\PL{\bf #1} (#2) #3}
\def\prl#1#2#3{\PRL{\bf #1} (#2) #3}
\def\pr#1#2#3{\PR{\bf #1} (#2) #3}
\def\physrev#1#2#3{\PR{\bf #1} (#2) #3}
\def\sjnp#1#2#3{\SJNP{\bf #1} (#2) #3}

\def\NP{{\it Nucl.\ Phys.\ }}
\def\PL{{\it Phys.\ Lett.\ }}
\def\PR{{\it Phys.\ Rev.\ }}

\def\PRL{{\it Phys.\ Rev.\ Lett.\ }}
\def\SJNP{{\it Sov.~J.~Nucl.~Phys.\ }}

\def\frac#1#2{{#1\over#2}}
\def\gs{\roughly{>}}
\def\Dc{\Delta_D}
\def\larr#1{\raise1.5ex\hbox{$\leftarrow$}\mkern-16.5mu #1}
\def\vslash{v\hskip-0.5em /}

\def\ccdot{\hbox{\kern-.1em$\cdot$\kern-.1em}}
\def\vvv{v \ccdot v'}
\def\vv{v \ccdot v'}

\def\g{\gamma}
\def\CO{{\cal O}}

\def\ol{\overline}
\def\Aslash{A\hskip-0.5em /}
\def\imi{{\rm i}}
\def\Tr{{\rm Tr}}
\def\del{\partial}
\def\emnlk{\epsilon_{\mu\nu\lambda\kappa}}

%

\catcode`\@=11 

\global\newcount\exerno \global\exerno=0
\def\exercise#1{\begingroup\global\advance\exerno by1%
\medbreak
\baselineskip=10pt plus 1pt minus 1pt
\parskip=0pt
\hrule\smallskip\noindent{\sl Exercise \the\secno.\the\exerno \/}
{\ninepoint #1}\smallskip\hrule\medbreak\endgroup}

%
\def\newsec#1{\global\advance\secno by1\message{(\the\secno. #1)}
\global\exerno=0%
\global\subsecno=0\eqnres@t\noindent{\bf\the\secno. #1}
\writetoca{{\secsym} {#1}}\par\nobreak\medskip\nobreak}
\def\appendix#1#2{\global\exerno=0%
\global\meqno=1\global\subsecno=0\xdef\secsym{\hbox{#1.}}
\bigbreak\bigskip\noindent{\bf Appendix #1. #2}\message{(#1. #2)}
\writetoca{Appendix {#1.} {#2}}\par\nobreak\medskip\nobreak}

\catcode`\@=12 
%
%


\newsec{Introduction}

The theory of heavy hadrons has received much attention over the last
few years. There are two good reasons for this. On the one hand
experiments probing the properties of heavy hadrons have come of
age. Not a year goes by without some remarkable new measurement, and
1994 is no exception\cleobsgamma. Moreover, SLAC is moving furiously
towards completion of a $B$ factory, which will hopefully operate well
before the end of the decade.  On the other hand, there have been many
theoretical advances that allow clearer interpretation of the
experimental results. These lectures are an introduction to these
theoretical developments.

The lectures are not intended to be encyclopedic in any one subject.
I have decided to try to convey the principal ideas as clearly as I
can, and to give some sample applications here and there. Occasionally
I may describe ``the state of the art'' in a given field, without
necessarily entering into details. I hope to have included enough
references that the reader may follow up on any of these subjects if
so inclined.

Exercises are scattered throughout the lectures. In the age of
electronic typography it was easy enough to display the exercises in
smaller print and separate them with clearly visible horizontal lines.
I hope that the material given in the lectures is sufficient for
obtaining the solution to the problems.

Chapter~2 contains a brief description of what Effective Field
Theories are, at least in the very specific context of weak
interactions. The presentation is somewhat telegraphic. I expect the
student to know something about effective lagrangians. The intention
is to show you how I think about the subject so that the presentation
of the effective lagrangian of heavy quarks, in  Chapter~3, goes down
more easily. 

Chapter~3 describes the Heavy Quark Effective Field Theory (HQET) and
its symmetries, to leading order in the large mass. Some examples and
applications are given. The corrections of order $1/M$ are described
in Chapter~4. 

The last two Chapters concentrate on applications of the HQET that
have received a lot of attention over the last year. This is where
these lectures deviate substantially from my Mexican lectures.
Chapter~5  gives the proof that the  rate for  inclusive semileptonic
$B$ decay is given by the parton decay rate, while Chapter~6
introduces an effective lagrangian of heavy  mesons and pion,
interacting in a chirally invariant way, and respecting heavy quark
symmetries.

\newsec{Preliminaries}
\subsec{Conventions and Notation}
The metric is $(+---)$. Gamma matrices satisfy
$\{\g_\mu,\g_\nu\}= 2g_{\mu\nu}$, $\g^0$ is hermitian, $\g^i$
antihermitian, $\g_5\equiv i\g^0\g^1\g^2\g^3$ and  $\sigma_{\mu\nu}
\equiv\ihalf[\g^\mu,\g^\nu]$. In the Dirac convention $\g^0={\rm
diag}(1,1,-1,-1)$. $\e^{0123}=+1$.

Left and right handed fields are denoted by subscripts,
$$\psi_L=\half(1-\g_5)\psi\qquad\psi_R=\half(1+\g_5)\psi
$$

States have the relativistic normalization,
\eqn\normconv{
\vev{\vec p|\vecpp}=2E\delta(\vec p-\vecpp)
}
unless otherwise noted.

The charged current interaction in the standard model is given by the
lagrangian 
\eqn\cclag{
\CL_{\rm int}={g_2\over\sqrt2}W^{+\mu} \pmatrix{\bar u&\bar c&\bar t\cr}
\g_\mu(1-\g_5)V\pmatrix{d\cr  s\cr b\cr}+ {\rm h.c.}~,}
where $V$ is the $3\times3$ unitary CKM matrix.

\subsec{Effective Lagrangians}
The study of decays of heavy mesons involves two very different types
of physics. On the one hand there is the underlying interactions that
are responsible for the decay. These may be the ordinary weak
interactions of the standard model or some new, presumably weaker,
interactions, \eg extended technicolor or exchange of scalar quarks in
supersymmetric theories. On the other hand there are strong
interactions which modulate the rates of the decays. 
Ultimately we would like to uncover the former, which requires some
level of understanding of the latter.

Since the two physical aspects are conceptually different, it seems
natural to separate them in our computations. This is where effective
lagrangians come in handy. An effective lagrangian for meson
decays will only involve the dynamical degrees of freedom that are
relevant. For example, light quarks ($u,d,s$) and gluons for $K$ meson
decays. The interaction responsible for the decay, a $W$
exchange, is represented as a $\Delta S=1$ four-quark operator. The
fact that the $W$-boson is no longer in the theory does not impair our
ability to predict the decay rate, up to an accuracy or order
$m_K^2/M_W^2$. 

Is this a major step backwards? After all, you may argue, this is going back to
the Fermi theory of weak interactions. Effective lagrangians prove useful
because, 
\item{(i)}The computation of long distance physics is decoupled from the
short distance physics. Thus, one can make a catalog of matrix elements
of operators between meson states. This could then be used to compute
the effects of any fundamental theory, after reducing its lagrangian to an
effective one.
\item{(ii)}They provide a method for the computation of amplitudes when
disparate scales are present. In $K$-meson decays the ratio $m_K/M_W$
is a small number. Logarithms of this ratio invalidate even the
perturbative computation of the decay. Effective lagrangians provide a
method for resummation of these large logs.
\item{(iii)}One can characterize, {\it \`a la} Fermi, all possible
interactions in terms of operators. In other words, one can make {\it
model independent} analysis of the possible effects of new physics.
\item{(iv)}It's the right way to {\it think} about the low energy
effects of very heavy particles.

\subsec{Formulating Effective Lagrangians}
The precise meaning of effective lagrangians is best formulated in
terms of relations between Green functions. Take again the example of
weak interactions at low energies, that is, when all the momenta
involved are much smaller than the $W$--boson mass. Everyone knows
that we can account for the effects of the $W$--boson by adding to the
Lagrangian terms of the form
\eqn\deltalageff{\Delta \leff = {1 \over M^2_W} \kappa \CO~,}
where $\CO$ is a 4-fermion operator and $\kappa$ contains mixing
angles and factors of the weak coupling constant. This is simply the
statement that a Green function $G$ of the original theory (the
standard model including QCD) can be approximated by a Green function
$\widetilde G_{\!\CO}$ of the effective theory (a gauge theory of QCD
and electromagnetism) with an insertion of the effective Lagrangian:
\eqn\normalgreenfs{
G = {1 \over M^2_W} \kappa \widetilde G_{\!\CO} + \ldots\quad.}
The ellipses stand for terms suppressed by additional powers of
$(M_W)^{- 2}$.  This equation replaces the task of computing the more
complicated left side, which depends on $M_W$, by the computation in
the effective theory which is independent of $M_W$, and indeed,
completely free of the $W$--boson dynamical degrees of freedom. On the
right hand side, the factor of $1/M^2_W$ gives the dependence on the
W--boson mass.

The full theory has logarithmic dependence on $M_W$ which has not been
made explicit. Eq.~\normalgreenfs\ is not quite correct. The correct
version is\witten
\eqn\normalC{ G = { 1 \over M^2_W} \kappa C ({M_W / \mu}, g_s) 
\tilde G_{\!\CO} + \ldots\quad.}
The function $C$ is, in this case, also known as the `short distance
QCD effect' first calculated for the process $s\to u \bar u d$ by
Altarelli and Parisi\altarelli, and Gaillard and Lee\gaillard.

Summing up, an effective theory (of either the `normal' or the HQ type) is
a method for extracting explicitly the leading large mass dependence of
amplitudes. Moreover, the rules of computation of the effective theory are
completely independent of the large mass.

\subsec{Computing Effective Lagrangians}
At tree level the computation of effective lagrangians is
straightforward since the short distance QCD effects can be neglected.
This is best explained through an example. Consider transitions
with $\Delta C=-\Delta U=\Delta S=-\Delta D=1$. In the full theory
these take place by the exchange of a $W$-boson coupling to the
charged currents $(\bar c_{L}\g^\mu s_{L})$ and $(\bar d_{L}\g^\mu
u_{L})$. Take the Green function for, say, $\bar c\to\bar s\bar u d$
and expand in powers of  momentum over the $W$-boson mass. This amount
to expanding the $W$-boson propagator; in 'tHooft--Feynman gauge,
$$
-i{g_{\mu\nu}\over p^2-M_W^2}=
i{g_{\mu\nu}\over M_W^2}+\cdots
$$
The effective hamiltonian for the decay is then
\eqn\heffcsdutree{\CH_{\rm eff}=
{4G_F\over\sqrt2}V^*_{ud}V_{cs} (\bar c_{L}\g^\mu s_{L})(\bar
d_{L}\g^\mu u_{L})+\cdots 
}
where we have introduced Fermi's constant $G_F= \sqrt2g_2^2/8M_W^2$.
The ellipsis stand for operators of higher dimension which come from
the expansion of the propagator, replacing $p\to i\partial$, as in
$$(\bar c_{L}\g^\mu s_{L})\partial^2(\bar d_{L}\g^\mu u_{L}) $$
\exercise{Show that the effective hamiltonian for $\Delta B=1$, $\Delta
C=\Delta U=0$, is given at tree level by
\eqn\heffculesstree{
\CH_{\rm eff}=
{4G_F\over\sqrt2}\sum_{q=u,c}\sum_{q'=d,s}V^*_{qb}V_{qq'}
(\bar b_{L\alpha}\g^\mu q_{L\alpha})(\bar q_{L\beta}\g_\mu
q'_{L\beta})
}
where $\alpha$ and $\beta$ are color indices.}

Beyond tree level the simple effective hamiltonian of
Eq.~\heffcsdutree\ is replaced by a sum over operators of 
dimension six and with the same quantum numbers. Neglecting the masses
of the light quarks there is only one more operator:
\eqn\heffcsdubeyond{\CH_{\rm eff}=
{4G_F\over\sqrt2}V^*_{ud}V_{cs}\sum_i c_i\CO_i+\cdots
}
where
$$\eqalign{
\CO_1 &=  (\bar c_{L}\g^\mu s_{L})(\bar d_{L}\g^\mu u_{L})\cr
\CO_2 &=  (\bar c_{L}\g^\mu u_{L})(\bar d_{L}\g^\mu s_{L})\cr}
$$
and the ellipsis stand again for higher dimension operators suppressed
by additional powers of $G_F$. The coefficients $c_i$ encode all the
information about the dependence on the W-mass (beyond the trivial
factor of $G_F$). We already know that at tree level $c_1=1$ and
$c_2=0$. They may be computed as an expansion in $\a_s$. It is best,
however to reorganize the expansion to account for the large ratio of
scales $m_c/M_W$ by assuming $\a_s\ll1$ and $\a_s\log(m_c/M_W)\sim1$,
thus
\eqna\charmcoeffs
$$\eqalignno{
c_1 & = 1 - {\a_s\over2\pi}\log(m_c/M_W) + \cdots \cr
&=
{1\over2}\left[\left({\alpha(m_c)\over\alpha(M_W)}\right)^{-2/b_0} +
\left({\alpha(m_c)\over\alpha(M_W)}\right)^{4/b_0} \right]
+\CO(\a_s^2\log({m_c\over M_W})) &\charmcoeffs a\cr
c_2 & = 0 +{3\a_s\over2\pi}\log(m_c/M_W) + \cdots \cr
&=
{1\over2}\left[\left({\alpha(m_c)\over\alpha(M_W)}\right)^{-2/b_0} -
\left({\alpha(m_c)\over\alpha(M_W)}\right)^{4/b_0} \right]
+\CO(\a_s^2\log({m_c\over M_W}))~. &\charmcoeffs b\cr}
$$
Here $b_0=25/3$ stands for the coefficient of the first term in the
perturbative expansion of the beta function in QCD, $\beta(g)=-b_0
{g^3\over16\pi^2}+\ldots$, in the case of four flavors of quarks. This
is the so-called ``leading-log'' approximation to the coefficients. We
have displayed order of the ``sub-leading-log'' or
``next-to-leading-log'' corrections.

\exercise{Write down the complete list of operators that contributes
to the effective hamiltonian for $\Delta B=1$, $\Delta C=\Delta U=0$.
Neglect the masses of $u$, $d$, $s$ and $c$-quarks, but not of the
$b$-quark. ({\sl Hint:\/} What is the symmetry group of QCD in the
massless limit? How do the operators in Eq.~\heffculesstree\
transform under this symmetry?) }

The computation of the coefficients is itself quite simple but it
detracts from our main focus. A rather similar computation is
presented below for the Heavy Quark Effective Theory; see
Section~\xtrnlcs. For a beautiful exposition of the method, see \witten.
  
\exercise{When you understand the procedure described in
Section~\xtrnlcs\ return here and check Eqs.~\charmcoeffs\null. What
modifications are necesary to account for a fifth quark with mass
$m_b$ intermediate between charm and the weak scale? (Assume $m_c\ll
m_b\ll M_W$).}

\exercise{Show that the effective hamiltonian for $\Delta B=-\Delta D=1$,
$\Delta C=\Delta U=0$ of the previous exercises can be written as the
sum of exactly two terms,
\eqn\twotermhamil{
\CH_w={4G_F\over\sqrt2}(\xi_c O_c + \xi_t O_t)
}
where $\xi_q\equiv V^*_{qb}V_{qd}$, and $O_q$ are linear combinations
of composite operators that may depend on $M_W$ but not on CKM angles.
Use phenomenological information to show that in the effective
hamiltonian for  $\Delta B=-\Delta S=1$, $\Delta C=\Delta U=0$,
depends only on the single mixing angle $V^*_{cb}V_{cs}$.}

\newsec{Heavy Quark Effective Field Theory}
\subsec{Intuitive Introduction}
\subseclab\intuitive
The central idea of the HQET is so simple, it can be described without
reference to a single equation. And it should prove useful to refer
back to the simple intuitive notion, to be presented below, wherever
the formalism and corresponding equations become abstruse.

The HQET is useful when dealing with hadrons composed of one heavy
quark and any number of light quarks. More precisely, the quantum
numbers of the hadrons are unrestricted as far as isospin and
strangeness, but are $\pm1$ for either $B$- or $C$-number.  In what
follows we shall (imprecisely) refer to these as `heavy hadrons'.

The successes of the constituent quark model is indicative of the fact
that, inside hadrons, strongly bound quarks exchange momentum of
magnitude a few hundred MeV. We can think of the typical amount
$\Lambda$ by which the quarks are off--shell in the nucleon as
$\Lambda \approx m_p/3 \approx 330$MeV. In a heavy hadron the same
intuition can be imported, and again the light quark(s) is(are) very
far off--shell, by an amount of order $\Lambda$.  But, if the mass
$M_Q$ of the heavy quark $Q$ is large, $M_Q \gg \Lambda$, then, in
fact, this quark is almost on--shell. Moreover, interactions with the
light quark(s) typically change the momentum of $Q$ by $\Lambda$, but
change the {\it velocity} of $Q$ by a negligible amount, of the order
of $\Lambda/M_Q \ll 1$. It therefore makes sense to think of $Q$ as
moving with constant velocity, and this velocity is, of course, the
velocity of the heavy hadron.

In the rest frame of the heavy hadron, the heavy quark is practically
at rest.  The heavy quark effectively acts as a static source of
gluons. It is characterized by its flavor and color--$SU(3)$ quantum
numbers, but not by its mass. In fact, since spin--flip interactions
with $Q$ are of the type of magnetic moment transitions, and these
involve an explicit factor of $g_s/M_Q$, where $g_s$ is the strong
interactions coupling constant, the spin quantum number itself
decouples in the large $M_Q$ case.  Therefore, {\it the properties of
heavy hadrons are independent of the spin and mass of the heavy source
of color.}

The HQET is nothing more than a method for giving these observations a
formal basis. It is useful because it gives a procedure for making
explicit calculations. But more importantly, it turns the statement
`$M_Q$ is large' into a {\it systematic} perturbative expansion in
powers of $\Lambda/M_Q$. Each order in this expansion involves QCD to
all orders in the strong coupling, $g_s$. Also, the statement of mass
and spin independence of properties of heavy hadrons appears in the
HQET as approximate internal symmetries of the Lagrangian.

Before closing this Section, we point out that these statements apply
just as well to a very familiar and quite different system: the
atom. The r\^ole of the heavy quark is played by the nucleus, and that
of the light degrees of freedom by the electrons (and the
electromagnetic field)\foot{An obvious distinction between the atomic
and hadronic systems is that in the latter the configuration of the
light degrees of freedom is non--computable, due to the difficulties
afforded by the non--perturbative nature of strong interactions. The
methods that we are describing circumvent the need for a detailed
knowledge of the configuration of light degrees of freedom. The price
paid is that the range of predictions is restricted. To emphasize the
non-computable aspect of the configuration of light degrees of
freedom, Nathan Isgur informally referred to it as ``brown muck'', and
the term has somewhat made it into the literature. }.  
That different isotopes have the same chemical properties simply
reflects the nuclear mass independence of the atomic
wave-function. Atoms with nuclear spin $s$ are $2s+1$ degenerate; this
degeneracy is broken when the finite nuclear mass is accounted for,
and the resulting hyperfine splitting is small because the nuclear
mass is so much larger than the binding energy (playing the r\^ole of
$\Lambda$). It is not surprising that, using $M_Q$ independence, the
properties of $B$ and $D$ mesons are related, and using spin
independence, those of $B$ and $B^\ast$ mesons are related, too.

\subsec{The Effective Lagrangian and its Feynman Rules}
\subseclab\efflagsec
We shall focus our attention on the calculation of Green functions in
QCD, with a heavy quark line, its external momentum almost on--shell.
The external momentum of gluons or light quarks can be far off--shell,
but not much larger than the hadronic scale $\Lambda$. This region of
momentum space is interesting because physical quantities
---$S$--matrix elements--- live there. And, as stated in the
introduction, we expect to see approximate symmetries of Green
functions in that region which are not symmetries away from it. That
is, these are approximate symmetries of a sector of the $S$--matrix,
but not of the lagrangian.

The effective Lagrangian $\leff$ is constructed so that it will
reproduce these Green functions, to leading order in $\Lambda/M_Q$. It is
given, for a heavy quark of velocity $v_{\mu}$ ($v^2=1$), 
by\eichtenhill, 
\eqn\eIIi {\leff^{(v)} = \bar Q_{v} i v\ccdot D  Q_v~,}
where the covariant derivative is
\eqn\eIIii {D_{\mu}= \partial_{\mu} + i g_s A^a_{\mu} T^a~,}
and the heavy quark field $Q_{v}$ is a Dirac spinor that satisfies the
constraint
\eqn\eIIiii{\left({1 + \vsl \over 2}\right) Q_v = Q_v~.}
In addition, it is understood that the usual Lagrangian $\CL_{\rm light}$
for gluons and light quarks is added to $\leff^{(v)}$.

We can see how this arises at tree level, as follows\grinstein.
Consider first the tree level 2-point function for the heavy quark
\eqn\eIIiv{G^{(2)}(p) = {i \over \spur p - M_Q}~.}
We are interested in momentum representing a quark of velocity $v_{\mu}$
slightly off--shell:
\eqn\pisv{p_{\mu} = M_Q v_{\mu} + k_{\mu}~.}
Here, `slightly off--shell' means $k_{\mu}$ is of order $\Lambda$, and
independent of $M_Q$. Substituting in Eq.~\eIIiv, and expanding in
powers of $\Lambda/M_Q$, we obtain, to leading order,
\eqn\greentwo{G^{(2)} (p) = i \left({1 + \vsl \over 2}\right) 
{1 \over v\ccdot k} + \CO \left({\Lambda \over M_Q}\right)~.}
We recognize the projection operator of Eq.~\eIIiii, and the
propagator of the lagran\-gian in~\eIIi.

Similarly, the 3-point function (a heavy quark and a gluon) is given by
\eqn\greenthreefull{
G^{(2,1)a}_{\mu} (p,q) = {i \over \spur p-M_Q} (-ig_s  T^a
\gamma^{\nu})  {i \over \spur p + \spur q-M_Q} \Delta_{\nu\mu}(q),
}
where $\Delta_{\nu \mu} (q)$ is the gluon propagator. Expanding as
above, we have
\eqn\greenthree{G_{\mu}^{a(2,1)} (p,q) = \left({1 + \vsl \over 2}\right) 
{i \over v\ccdot k} (-ig_s T^a v^{\nu}) 
{i \over v\ccdot (k + q)} \Delta_{\mu\nu} (q) + \CO\left({\Lambda \over
M_Q}\right),}
where we have used 
\eqn\gammaisv{\left({1 + \vsl \over 2}\right) \gamma_{\nu} \left({1 + 
\vsl \over2}\right)  = \left({1 + \vsl \over 2}\right) v_{\nu}~.}
Again, this corresponds to the vertex obtained from the effective
Lagrangian in Eq.~\eIIi. 
\exercise{Extend these results
to arbitrary tree-level Green functions (but only those with one heavy
quark and all other (light) particles carrying momentum of
order~$\Lambda$).}

The effective Lagrangian in~\eIIi\ is appropriate for the description of a
heavy quark, and indeed a heavy hadron, of velocity $v_{\mu}$.
It does, however, break Lorentz covariance. This is not a surprise, since
we have expanded the Green functions about one particular velocity: in
boosted frames, the expansion in powers of $\Lambda/M_Q$ becomes invalid,
since the boosted momentum $k_{\mu}$ can become arbitrarily large. Lorentz
covariance is recovered, however, if we boost the velocity
\eqn\boost{v_{\mu} \rightarrow \Lambda_{\mu\nu} v_{\nu}}
along with everything else. It will prove useful to keep this simple
observation in mind\foot{In an alternative method, championed by
Georgi\georgi, the effective Lagrangian $\leff$ consists of a sum over
the different velocity Lagrangians, $\leff^{(v)}$, of
Eq.~\eIIi. Lorentz invariance is recovered at the price of
``integrating in'' the heavy degrees of freedom.  This does not lead
to overcounting of states, because the sectors of different velocity
do not couple to each other, a fact that Georgi refers to as a
``velocity superselection rule''. See also \dgg.}.

Just as in the case of an effective theory for weak interactions, when
one goes beyond tree level one must be careful to make explicit any
anomalous mass dependence. When the $W$ is integrated out the
correction was encrypted in some function $C(M_W)$ in Eq.~\normalC.
The situation is entirely analogous in the HQET. We have introduced an
effective Lagrangian $\leff^{(v)}$ such that Green functions
$\widetilde G_v(k;q)$ calculated from it agree, at tree level, with
corresponding Green functions $G(p;q)$, in the QCD to leading order in
the large mass
\eqn\eIIv{
G (p;q) = \widetilde G_v (k;q) + \CO \left({\Lambda / M_Q}\right)
\qquad\qquad\hbox{(tree level)}~.
}
Here, $\Lambda$ stands for any component of $k_{\mu}$ or of the $q$'s,
or for a light quark mass, and $p=M_Q v+k$. We will come back to the
study of the HQET beyond tree level and will make explicit the
anomalous mass dependence in Section~\oneloopsec.

\subsec{Symmetries}
\subseclab\symmtrssec
{\sl Flavor -- $SU(N)$}
The Lagrangian for $N$ species of heavy quarks, all with velocity $v$, is
\eqn\hqeteffl{
\leff^{(v)} = \sum^N_{j=1} \bar Q_v^{(j)} \; i v \ccdot D\; Q^{(j)}_v~.
}
This Lagrangian has a $U(N)$ 
symmetry{\refs{\isgurwisea,\nussinov,\volosh}}. The
subgroup $U(1)^N$ 
corresponds to flavor conservation of the strong interactions,
and was a good symmetry in the original theory. The novelty in the HQET is 
then
the nonabelian nature of the symmetry group. This leads to relations between
properties of heavy hadrons with different quantum numbers. Please note 
that
these will be relations between hadrons of a given velocity, 
even if of different
momentum (since typically $M_{Q_i}  \neq M_{Q_j}$ for $i \neq j$).
Including the $b$ and $c$ quarks in the HQET, so that $N=2$, we see that the
$B$ and $D$ mesons form a doublet under flavor--$SU(2)$.

This flavor--$SU(2)$ is an approximate symmetry of QCD. It is a good
symmetry to the extent that 
\eqn\largemass{
m_c \gg \Lambda\qquad\hbox{and}\qquad 
m_b \gg \Lambda ~.}
These conditions can be met even if $m_b - m_c \gg \Lambda$.
This is in contrast to isospin symmetry,
which holds because $m_d-m_u \ll \Lambda$. 

In the atomic physics analogy of Section~\intuitive, this symmetry implies
the equality of chemical properties of different isotopes of an element.

{\sl  Spin -- $SU(2)$}
The HQET Lagrangian involves only two components of the spinor $Q_v$.
Recall that 
\eqn\projonQ{
\projm Q_v = 0.}
The two surviving components enter the Lagrangian diagonally, {\it i.e.},
there are no Dirac matrices in
\eqn\leffagain{
\leff^{(v)} = \bar Q_v \; iv\ccdot D\; Q_v.}
Therefore, there is an $SU(2)$ symmetry of this Lagrangian which rotates
the two components of $Q_v$ among 
themselves{\refs{\isgurwisea,{\eichtenfeinberg}{--}\lepagethacker}}.

Please note that this ``spin''--symmetry is actually an {\sl internal}
symmetry. That is, for the symmetry to hold no transformation on the 
coordinates
is needed, when a rotation among components of $Q_v$ is made. On
the other hand, to recover Lorentz covariance, one does the usual
transformation on the light--sector, including a Lorentz transformation of
coordinates and in addition a Lorentz transformation on the velocity 
$v_\mu$. A
spin--$SU(2)$ transformation can be added to this procedure, to mimic the
original action of Lorentz transformations.

\exercise{
To make it plain that this symmetry has nothing to do with ``spin'' in the
usual sense, consider the large mass limit for a vector 
particle\carone. Use the massive vector propagator
\eqn\vectprop{
-i\,{g_{\mu\nu} - p_\mu p_\nu/m^2 \over p^2 - m^2} 
}
to obtain the Lagrangian for the HVET (Heavy Vector Effective Theory)
\eqn\vectorleff{ 
\leff^{(v)} = A^{ \dagger}_{v\mu} \; i v \ccdot D A_{v\mu} ~,
}
with the constraint
\eqn\vecconstraint{
(v_\mu v_\nu - g_{\mu\nu}) A_{v\nu }= A_{v\mu} ~.
}
What is the dimension of this effective vector field? Why?
Show that the effective Lagrangian is invariant under an $SU(3)$
group of transformations, rotating the three components of the vector field
among themselves. Note that the ``spin'' symmetry is not associated with
$SU(2)$ in this case.
}

The symmetry of the theory is larger than the product of the flavor and
spin symmetries. If there are $N_S$, $N_F$, and $N_V$ species of heavy
scalars, fermions, and vectors, respectively, all
transforming the same way under color--$SU(3)$,
the symmetry of the effective theory is $SU(N_S + 2N_F + 3N_V$).
\exercise{What is the symmetry group for a theory with $N_S$, $N_F$,
and $N_V$ species of heavy scalars, fermions, and vectors,
respectively, all transforming the same way under color--$SU(3)$, in
$D$ space-time dimensions? (If you can't handle arbitrary $D$, try
$D=2$ and $D=3$).
}

\subsec{Spectrum}

The internal symmetries of the effective Lagrangian are explicitly realized
as degeneracies in the spectrum and as relations between transition
amplitudes. In this Section we will consider the spectrum of
the theory\iwspectrum.

Keep in mind that momenta, and therefore energies and masses, are 
measured
in the HQET relative to $M_Qv_\mu$. Therefore, when we state that in the
HQET the $B$ and $D$ mesons are degenerate, the implication is that the
physical mesons differ in their masses by $m_b - m_c$.

For now let us specialize to the rest frame $v=(1,{\bf 0})$. The total angular
momentum operator ${\bf J}$, {\it i.e.,} the generator of rotations, can be
written as   \eqn\totangmom{
{\bf J} = {\bf L} + {\bf S}~,}
where ${\bf L}$
is the angular momentum operator of the light degrees of freedom, and ${\bf
S}$, the angular momentum operator for the heavy quark, agrees with the
generator of spin--$SU(2)$. Since ${\bf J}$ and ${\bf S}$ are separately
conserved, ${\bf L}$ is also separately conserved. Therefore, the states of the
theory can be labeled by their ${\bf L}$ and ${\bf S}$ quantum numbers 
$(l,m_l ; 
s,m_s)$. Of course, $s= 1/2$, so $m_s$ is $1/2$ or $-1/2$ only.

The simplest state has $l=0$ and, therefore, $J = 1/2$. We will refer
to it as the $\Lambda_Q$, by analogy with the nonrelativistic
potential constituent quark model of the $\Lambda$--baryon, where the
strange quark combines with a $l=0$, $I=0$, combination of the two
light quarks.

Next is the state with $l= 1/2$. It leads to $J = 0$ and $J = 1$.  We
deduce that there is a meson and a vector meson that are degenerate.
For the $b$-quark, the $B$ and $B^\ast$ fit the bill. They are the
lowest lying $B = -1$ states. The lowest lying $C=1$ states are the
$D$ and $D^\ast$ mesons. These again can very well be assigned to our
$J=0$ and $J=1$ multiplet. The difference $M_{D^\ast} - M_D = 145$~MeV
is reasonably smaller than the splitting between the $D^\ast$ and the
next state, the $D_1$, with $M_{D_1} - M_{D^\ast} = 410$~MeV.

The splittings of $B$ and $B^\ast$ and of $D$ and $D^\ast$ result from
spin-$SU(2)$ symmetry breaking effects. These must be corrections of
order $\Lambda/M_Q$ to the HQET predictions. Therefore, one must have
$M_{B^*} - M_B = \Lambda^2/m_b$ and analogously for the $D$--$D^\ast$
pair.  Therefore
\eqn\massratios{
{M_{B^\ast} - M_B\over M_{D^\ast} - M_D} = {m_c \over m_b}~.}
Approximating $m_c$ and $m_b$ by $M_D$ and $M_B$, respectively, we get 
$\sim
1/3$ on the right side, in remarkable agreement with the left side.
Although these results also follow from potential models of constituent 
quarks,
it is important that they can be derived in this generality, and this simply.

The states with $l= 3/2$ have $J=1$ and $2$. The $D_1$ and $D^\ast_2$, with
$M_{D^\ast_2} - M_{D_1} = 40$ MeV, are remarkably closely spaced (and of
course, have the appropriate quantum numbers to form a spin multiplet).

\OMIT{
It is convenient to represent the states in a multiplet in a way that
makes manifest their properties under the spin symmetry.
For fixed $l$, the spin $J$ of the states related by the spin-$SU{(2)}$
is $J= l\pm {1\over2}$. The representation we are looking for is a symbol
$\chi^{(\pm) A}_{\alpha a}$, where
the index $\alpha$ is the heavy quark spin, $a$ and $A$ correspond to the
$z$-components of ${\bf L}$ and ${\bf J}$ respectively, and the values of $ l$
and $J$ are implicitly understood (only the $+/-$ super--index is needed to
distinguish  between the $J= l + {1\over2}$ and   $J= l - {1\over2}$ states).
This problem is nothing but the composition of ${\bf L}$ and ${\bf S}$ into
${\bf J}$, and the solution is in Clebsch-Gordan coefficients 
 \eqn\clebschone{
\chi^{(\pm)A}_{\alpha a}=C\left( l \,a; s\,\alpha\vert
JA\right)=C^{JA}_{\ell\, a, s \,\alpha}~,}
where $s={1\over2}$ and $J= l\pm{1\over2}$. (The last equality defines
a short version of the symbol.)
}

\exercise{A more complete classification of the spectrum would include
parity. What modifications are needed, if any, to include parity as a
quantum number? You should find that there are two possible $\ell=1/2$
multiplets. One corresponds to the $D$ and $D^*$ (or $B$ and $B^*$)
mesons. It is usually argued that the multiplet with opposite parity
would contain very broad resonances that could not be identified as
stable states. Why?  }

While in the infinite mass limit states $|l,m_l;s,m_s\rangle$
have sharp $L^2$, $L_z$, $S^2$
and $S_z$, these are not good
quantum numbers for physical states. Regardless of how small
spin-symmetry breaking effects may be, they force states into linear
combinations of sharp $J^2$, $J_z$, $L^2$ and $S^2$,
$|J,m_J;l,s\rangle$. $SU(2)$-spin transformations connect states of
$J=l+1/2$ with those of $J=l-1/2$. Now
$$|J,m_J;l,s\rangle = \sum |l,m_l;s,m_s\rangle C^{J m_J}_{l m_l,s
m_s}$$
where $C^{J m_J}_{l m_l,s m_s}=C(l m_l;s m_s | J m_J)$ are
Clebsch-Gordan coefficients. The decomposition is useful because we
know how the states on the right transform under spin-$SU(2)$. The
inverse expression, 
$$|l,m_l;s,m_s\rangle=\sum |J,m_J;l,s\rangle (C^{J m_J}_{l m_l,s
m_s})^*$$
gives the linear combinations of physical states with definite
spin-$SU(2)$ numbers.

For example, for the $B$ and $B^*$ multiplet, the $m_l=1/2$ and
$m_l=-1/2$ states that form spin-$SU(2)$ doublets are, respectively
\eqn\dblts{
\psi_{1/2}=\pmatrix{B^*(+)\cr {B^*(0)+B\over\sqrt2}\cr}\qquad{\rm
and}\qquad
\psi_{-1/2}=\pmatrix{ {B^*(0)-B\over\sqrt2}\cr B^*(-)\cr}~.
}
Rotations mix components among these doublets. We can combine them
into a matrix $\Psi_{\alpha a}\equiv (\psi_a)_\alpha$. If
$\CD^{(l)}(R)$ stands for a $2l+1$ dimensional representation of the
rotation $R$, then the action of spin-$SU(2)$ alone is $\Psi\to
\CD^{(1/2)}(R) \Psi$, while a rotation is $\Psi\to
\CD^{(1/2)}(R) \Psi\CD^{(1/2)}(R)^\dagger$. 

This is easily generalized. For arbitrary $l$ there are $2l+1$
doublets of spin-$SU(2)$, $\phi_a$, $a=-l,\ldots,l$. They can be
assembled into a $2\times (2l+1)$ matrix $\Phi_{\alpha
a}=(\phi_a)_\alpha$, which transforms as $\Phi\to
\CD^{(1/2)}(R) \Phi\CD^{(l)}(R)^\dagger$ under rotations. The linear
combination of physical states in $\Phi_{\alpha a}$ can be written as
a sum of at most two terms:
$$\Phi_{\alpha a}=\chi^{(+)A}_{\alpha a}+\chi^{(-)A}_{\alpha a}~,$$
where $\chi^{(+)A}_{\alpha a}$ ($\chi^{(-)A}_{\alpha a}$) is the state
with $J=l+1/2$ ($J=l-1/2$), $A=m_J=\alpha+a$, weighted by the
corresponding Clebsch-Gordan coefficient, $C^{l\pm1/2 A}_{l a, 1/2 \alpha}$.

If $Q_0$ is the two component heavy quark field for $v=(1,\vec0)$
then, according to the Wigner-Eckart theorem the matrix elements of
$ Q_0^\dagger\G Q_0$, for any Pauli matrix $\G$, between $B$ and $B^*$
states are all given in terms of a single reduced matrix element,
$\chi$, times appropriate symmetry factors constructed out of
Clebsch-Gordan coefficients.  Let the state $|\psi\rangle=\sum_{\alpha
a}\Psi_{\alpha a}|\alpha a\rangle$. Then the matrix element
$\vev{\psi'| Q_0^\dagger\G Q_0|\psi} $ must be linear in $\Psi_{\alpha
a}$, $\Psi^{\prime\dagger\beta b}$ and $\G_\gamma^\delta$, and
proportional to the invariant tensor $\vev{\beta b| Q_0^{\dagger\gamma}
Q_{0\delta}|\alpha a}$. There is only one invariant one can construct
out of these tensors.  This can be summarized as follows,
\eqn\wignerdblts{
\vev{\psi'| Q_0^\dagger\G Q_0|\psi} = \chi \Tr \bar\Psi'\G\Psi~.
}
By this we mean that if the state $\psi$ is, say, a $B$-meson, then
the corresponding matrix $\Psi$ on the right hand side is obtained
from the matrix $\Psi_{a\a}$ of Eq.~\dblts\ by setting $B^*=0$ and
$B=1$, and analogously for the other possible choices of the states
$\psi$ and $\psi'$. In the next Section we generalize this result to
the case were the states $\psi$ and $\psi'$ may have different
velocities.

\subsec{Covariant Representation of States}
\subseclab\cvrepofsts
In the Chapters that follow we will be interested in extracting the
consequences of the spin and flavor symmetries of the HQET to a
variety of processes. These processes may involve transitions between
heavy hadrons of different velocities. It is convenient to develop a
formalism that automatically extracts the information encoded in the
symmetries\fggw. I follow the simple presentation of Ref.~\falkspin.

A prototypical example of an application is the
computation of relations between form factors in semileptonic $\bar B$ to $D$
and $D^*$ decays. There one needs to study the matrix elements
\eqn\melements{
\vev{D(v)|\bar c_v \Gamma b_{v'} |B(v')} \quad\hbox{and}\quad
\vev{D^*(v),\e|\bar c_v \Gamma b_{v'} |B(v')}~.
}
We would like to represent these $l=1/2$ mesons as the product
  \eqn\uvrep{
u_Q\bar v_q~,
}
where $u_Q$ is a spinor representing the heavy quark, $\vsl u_Q=u_Q$, and 
$v_q$
is an antispinor representing the light stuff with $l=1/2$, satisfying $\bar
v_q\vsl=\bar v_q$. The product in \uvrep\ is a superposition of states with
$J=0$ and~1. To identify the pseudoscalar meson $P$ and the vector meson
$V(\e)$ with polarization $\e$, $\e\ccdot v=0$, we must form appropriate
linear combinations of the spin up and down spinors. This is most easily done
in the rest frame $v=(1,{\bf 0})$; the result will be generalized to
arbitrary $v$ by boosting. In the Dirac representation the spin operator
 is ${\bf S}=\gamma^5\gamma^0\hbox{\mib$\gamma$}/2$ so that the spinor 
basis
$u^{(1)}_\alpha=\delta_{1\alpha}$ and $u^{(2)}_\alpha=\delta_{2\alpha}$
corresponds to spin up and spin down, and the antispinor basis 
$v^{(1)}_\alpha=-\delta_{3\alpha}$ and  $v^{(2)}_\alpha=-\delta_{4\alpha}$
corresponds to spin down and spin up. With ${\bf S}(u\bar v)= ({\bf
S}u)\bar v + u({\bf S}\bar v)$ it is easy to check that the combination
\eqn\comba{ 
u_Q^{(1)}\bar v^{(1)}_q + u_Q^{(2)}\bar v^{(2)}_q =\pmatrix{0&I\cr
0&0\cr}=\left({1+\gamma^0\over2}\right)\gamma^5
}
has zero spin, while
\eqn\combb{
\eqalign{
u_Q^{(1)}\bar v^{(2)}_q &={1\over\sqrt2}\pmatrix{0&\sigma_1+i\sigma_2\cr
0&0\cr}=\left({1+\gamma^0\over2}\right)\spur\e^{(+)}\cr
u_Q^{(1)}\bar v^{(1)}_q - u_Q^{(2)}\bar v^{(2)}_q &=\pmatrix{0&\sigma_3\cr
0&0\cr}=\left({1+\gamma^0\over2}\right)\spur\e^{(0)}\cr
u_Q^{(2)}\bar v^{(1)}_q &={1\over\sqrt2}\pmatrix{0&\sigma_1-i\sigma_2\cr
0&0\cr}=\left({1+\gamma^0\over2}\right)\spur\e^{(-)}\cr}
}
with $\e^{(\pm)}=(0,1,\pm i,0)$ and $\e^{(0)}=(0,0,0,1)$, have total spin 1,
with third component $1$, $0$ and $-1$, respectively. Thus, for arbitrary
velocity $v$ one obtains the representation for pseudoscalar and vector 
mesons:
\eqn\reposmesons{
\widetilde M(v) =\projp\gamma^5\qquad\qquad
\widetilde M^*(v,\e)=\projp\spur\e~.
}
By construction, the spin symmetry acts on this representation only on the
first index of the matrices $\widetilde M(v)$ and $\widetilde M^*(v,\e)$.

The power of this machinery can now be displayed. Consider the matrix
elements~\melements. Using the above representation of states and noting 
that
the result should transform under the spin symmetry just as the matrix
$\Gamma$, we have
 \eqna\ohyes 
$$\eqalignno{
\vev{D(v)|\bar c_v \Gamma b_{v'}|\bar B(v')} &=-\xi(v\ccdot v'){\rm Tr}\,
\overline{\widetilde D}(v)\Gamma \widetilde B(v') &\ohyes a\cr
\vev{D^*(v)\varepsilon| \bar c_v \Gamma b_{v'}|\bar B(v')} &=-\xi(v\ccdot
v'){\rm Tr}\, \overline{\widetilde D}\null^*(v,\varepsilon)\Gamma
\widetilde B(v')~, &\ohyes b\cr} $$ 
where $ \overline X=\gamma^0 X^\dagger\gamma^0$.  The common factor
$-\xi(\vvv)$ plays the r\^ole of the reduced matrix element in the
Wigner--Eckart theorem.  We will explore the consequences of
Eqs.~\ohyes{}\ in depth in Section~\semilepdecsec.

\exercise{
An even simpler case is that of the $l=0$ multiplet. In this case the
states must transform as a spinor. How would you represent matrix
elements of these states? What about those of the  $l=1$ or $l=3/2$
multiplets? This formalism can be extended\falkspin\ to deal with
multiplets of arbitrary $l$.
}

\subsec{Meson Decay Constants}

The pseudoscalar decay constant is one of the first physical
quantities studied in the context of HQET's. For a heavy-light
pseudoscalar meson $X$ of mass $M_X$, the decay constant $f_X$, we
will see, scales like ${1/\sqrt{M_X}}$. This was known before the
formal development of HQET's, although the arguments relied on models
of strong interactions. The HQET will give us a systematic way of
obtaining this result. Moreover, it will give us the means of studying
corrections to this prediction.

The decay constant $f_X$ is defined through
\eqn\fmesondefd{
\vev{0\vert A_\mu(0)\vert X(p)}=f_Xp_\mu~,
}
where $A_\mu=\bar q \gamma_\mu\gamma_5 Q$ is the heavy-light axial
current, and the meson has the standard relativistic normalization
\eqn\mesonnorm{
\left<X(p')\vert X(p)\right>=2E\delta^{(3)}({\bf p}-{\bf p'})~.
}
Thus, the states have mass-dimension $-1$. Analogous definitions can
be made for other mesons. For example, for the vector meson $X^\ast$
(the $l ={1/2}$ partner of $X$), has
\eqn\fvectdefd{
\left<0\vert V_\mu(0)\vert
X^\ast(p,\epsilon)\right>=f_{X^\ast}\epsilon_\mu ~.
}
Note that the mass-dimensions of $f_X$ and $f_{X^\ast}$ are $1$ and
$2$, respectively.

Consider the decay constant of the meson state in the HQET.
The effective pseudoscalar decay constant $\tilde f_{X}$ is defined by
\eqn\ftildedefd{
\langle 0\vert \widetilde A_\mu(0)\vert\widetilde X(v)\rangle =\tilde
f_Xv_\mu
}
The state in the HQET, $ |\widetilde X\rangle$, is normalized {\it \`a la}
Bjorken and Drell,\bjdrell\ to $2E/M_X$ rather than to $2E$:
\eqn\IVii{
\langle\widetilde X(v')\vert\widetilde
X(v)\rangle=2v^{0}\delta^{(3)}({\bf  v}-{\bf  v'})~.
}
Actually, defining states in the HQET requires some care, but I will
just assume it all works and merely refer the interested reader to the
literature\dgg.  Obviously, since the normalization of states and the
dynamics are $M_Q$ independent, so is $\tilde f_X$. To relate $\tilde
f_X$ to the physical $f_X$ simply multiply Eq.~\ftildedefd\ by
$\sqrt{M_X}$, to restore the normalization of states of
Eq.~\mesonnorm, and write $v_\mu=p_\mu/M_X$. Thus we arrive at
\eqn\ftoftilde{
f_X=\tilde f_X/ \sqrt{M_X} ,
}
%
\OMIT{
\eqn\ftoftilde{
f_X=\tilde f_X/ \sqrt{M_X} \left({\bar\alpha_s(M_Q)
\over\bar\alpha_s(\mu)}\right)^{a_I},
}
where the last factor comes from the relation between currents in the
full and effective theories.  The `constant' $\tilde f_X$, is in fact
a function of the renormalization point $\mu$; the combination $\tilde
f_X \bar\alpha_s(\mu)^{-a_I}$ is $\mu$-independent to leading-log
order.
We have obtained
$f_X\sim{1/\sqrt{M_X}}$ plus a logarithmic correction. 
}
A useful way of
quoting the result is, for the physical case of $B$ and $D$ mesons,
\eqn\fBoverfD{
{f_B\over f_D}=\sqrt{{M_D\over M_B}}
}
\OMIT{
\eqn\fBoverfD{
{f_B\over f_D}=\sqrt{{M_D\over M_B}}
\left({\bar\alpha_s(M_B)\over\bar\alpha_s(M_D)}\right)^{a_I}
}
}

As a simple application of the spin symmetry, consider the
pseudoscalar decay constant $f_{X^\ast}$. Using the $4\times4$
notation of Section~\cvrepofsts, the matrix element in Eq.~\fvectdefd\ that
defines the pseudoscalar constant is proportional to
\eqn\eIVi{
\hbox{Tr}\,\left(\gamma^\mu\gamma_5\widetilde M(v)\right)=
\hbox{Tr}\,\left(\gamma^\mu\gamma_5\left({1+\vsl\over2}\right)
\gamma_5\right) =-2v^\mu
} 
The matrix element
\eqn\eIVii{
\langle 0\vert \widetilde V^\mu(0)\vert
\widetilde X^\ast(v)\epsilon\rangle=\tilde f_{X^\ast}\epsilon^\mu
}
is proportional to 
\eqn\eIViii{
\hbox{Tr}\,\left(\gamma^\mu\widetilde M^*(v,\epsilon)\right)=
\hbox{Tr}\,\left(\gamma^\mu\left({1+\vsl\over 2}\right)\spur
\epsilon  \right)= 2\epsilon^\mu
}
with the same constant of proportionality. Therefore
\eqn\eIViv{
\tilde f_{X^\ast}=-\tilde f_X
}
The sign is unimportant, since it can be absorbed into a phase
redefinition of either state. It is the magnitude that
matters. Multiplying by $\sqrt{M_{X^\ast}}\approx\sqrt{M_X}$ to
restore to the standard normalization, we have
\eqn\eIVv{f_{X^\ast}=-f_XM_X
}

The predictions Eq.~\fBoverfD\ and Eq.~\eIVv\ have not been tested
experimentally. The difficulty is the small expected branching
fraction for the decays $X\rightarrow\mu\nu$ or
$X^\ast\rightarrow\mu\nu$, for $X=B$ and $D$. Alternatively, the decay
constants $f_X$ and $f_{X^\ast}$ can be measured in Monte Carlo
simulations of lattice QCD. There are indications from such
simulations that the $1/M_Q$ corrections to the relation~\fBoverfD\
are large\chris.

\subsec{Semileptonic decays}
\subseclab\semilepdecsec
The semileptonic decays of a $\bar B$-meson to $D$- or $D^\ast$-mesons
offer the most direct means of extracting the mixing angle $\vert
V_{cb}\vert$.  In order to extract this angle from experiment, theory
must provide the form factors for the $\bar B\rightarrow D$ and $\bar
B\rightarrow D^\ast$ transitions.  Several means of estimating these
form factors can be found in the literature. A popular method consists
of estimating the form factor at one value of the momentum transfer
$q^2=q^2_0$, and then introducing the functional dependence on $q^2$
in some arbitrary, hopefully reasonable, way. In the pre-HQET days it
was customary to estimate the form factor at $q^2_0$ from some model
of strong interactions, like the non-relativistic constituent quark
model.

The HQET gives the form factor at the maximum momentum transfer,
$q^2=q^2_{\rm max}=(M_B-M_D)^2$ ---the point at which the resulting
$D$ or $D^\ast$ does not recoil in the restframe of the decaying
$B$-meson. While the functional dependence on $q^2$ is a
non-perturbative problem, it is already progress to have a prediction
of the form factor at one point. Moreover, the HQET gives relations
between the form factors. One may study these relations experimentally
to test the accuracy of the HQET predictions.

The standard definition of form factors in semileptonic $\bar B$-meson 
decays is
\eqna\fulff
$$\eqalignno{
\left<D(p')\vert V_\mu\vert
\bar B(p)\right>&=f_+(q^2)(p+p')_\mu+f_-(q^2)(p-p')_\mu &\fulff a\cr
\left<D^\ast(p')\epsilon\vert A_\mu\vert
\bar B(p)\right>&=f(q^2)\epsilon^\ast_\mu + a_+(q^2)\epsilon^\ast\ccdot
p(p+p')_\mu+a_-(q^2)\epsilon^\ast\ccdot p(p-p')_\mu\qquad &\fulff b\cr
\left<D^\ast(p')\epsilon\vert V_\mu\vert
\bar B(p)\right>&=ig(q^2)\epsilon_{\mu\nu\lambda\sigma}\epsilon^{\ast\nu}
(p+p')^\lambda(p-p')^\sigma &\fulff c\cr}$$ 
Here, the states have the standard normalization, Eq.~\mesonnorm, and
$q^2\equiv (p-p')^2$.  The contribution to the decay rates from the
form factors $f_-$ and $a_-$ are suppressed by $m^2_\ell /M^2_B$,
where $m_\ell$ is the mass of the charged lepton, and therefore they
are often neglected.

\exercise{Compute the differential decay rates, $d\G/dx\;dy$, where
$x=q^2/M_B^2$ and $y=v\cdot q/M_B$, for 
$\bar B\rightarrow D e \bar\nu$ and $\bar B\rightarrow D^\ast e
\bar\nu$, in terms of these form factors.} 

In the effective theory, we would like to compute the matrix elements
of the effective currents $\tilde V_\mu$ and $\tilde A_\mu$ between
states of the $ l=\half$ multiplet. We can take advantage of the
flavor and spin symmetries to write these matrix elements in terms of
generalized Clebsch-Gordan coefficients and reduced matrix elements,
{\it i.e.,} we use the Wigner-Eckart theorem. We have already
introduced the relevant machinery in Section~\cvrepofsts. The matrix elements
of the operator $\widetilde G=\bar c_{v'}\Gamma b_v$ between $B$ and
$D$ or $D^\ast$ states, are given by ({\it c.f.,\/} Eqs.~\ohyes)
\eqna\eVii
$$\eqalignno{
\vev{D(v')| \widetilde G|\bar B(v)} &=-\xi(v\ccdot v'){\rm Tr}\,
\overline{\widetilde D}(v')\Gamma \widetilde B(v) &\eVii a\cr
\vev{D^*(v')\varepsilon|\widetilde G|\bar B(v)} &=-\xi(v\ccdot v'){\rm Tr}\,
\overline{\widetilde D}\null^*(v',\varepsilon)\Gamma \widetilde B(v). &\eVii 
b\cr}
$$

Before expanding Eqs.~\eVii{ }, we note that the flavor symmetry implies that
the $B$-current form factor between $\bar B$-meson states is given by the 
same reduced matrix element:
\eqn\eViii{
\vev{\bar B(v')|\bar b_{v'}\Gamma b_v|\bar B(v)} =-\xi(v\ccdot v'){\rm Tr}\,
\overline{\widetilde B}(v')\Gamma \widetilde B(v)
}
Using $\Gamma=\gamma^0$, and recalling that $B$-number is conserved,
one finds that $\xi$ is fixed at $v'=v$. With the normalization of
states appropriate to the effective theory, Eq.~\IVii, and expanding
Eq.~\eViii\ at $v=v'$, one has
\eqn\eVv{
\xi(1)=1.
}

The reduced matrix element $\xi$ is the universal function that describes
all of the matrix elements of operators $\widetilde G$ between $ l =\half$
states. It is known as the Isgur-Wise function after the discoverers of the
relations~\eVii\null\ and~\eViii. It is quite remarkable that the Isgur-Wise
function describes both timelike form-factors (as in $\bar B\to De\nu$) 
as well as
spacelike form-factors (as in $\bar B\to \bar B$). The point, of course, is
that in both cases it describes transitions between infinitely heavy
sources at fixed ``velocity-transfer" $(v-v')^2$.

Expanding Eq.~\eVii\null\ for $\Gamma=\gamma^\mu$ or 
$\gamma^\mu\gamma_5$, we
have  
\eqna\eVvi
$$\eqalignno{
\vev{D(v')\vert \widetilde V_\mu\vert \bar B(v)}&=
\xi(v\ccdot v')(v_\mu+v'_\mu) &\eVvi a\cr
\vev{D^\ast(v')\epsilon\vert\tilde A_\mu\vert\ \bar B(v)}&=
-\xi(v\ccdot
v')[\epsilon^\ast_\mu(1+v\ccdot v')-v'_\mu\epsilon^\ast\ccdot v]&\eVvi b\cr
\vev{D^\ast (v')\epsilon\vert\widetilde V_\mu\vert\ \bar B(v)} &=
-\xi(v\ccdot v')[-
i\epsilon_{\mu\nu\lambda\sigma}\epsilon^{\ast\nu}v^\lambda
v^\sigma] &\eVvi c\cr}$$

It remains to express the physical form factors in terms of the Isgur-Wise
functions. \OMIT{We must introduce the
coefficient functions $\widehat C_\Gamma$ of Eq.~\exv,
which in the leading-log approximation are given in \replacevec\ and 
\replaceaxi.  Also, w}We must
multiply by $\sqrt{M_DM_B}$ to restore to the standard normalization of 
states,
and express Eqs.~\eVvi\null\ in terms of momenta using $v=p/M_B$ and 
$v'=p'/M_D$.
For example, one has,
\eqn\eVvii{
\vev{D(p')\vert V_\nu\vert B(p)}
=\xi(v \ccdot v')\sqrt{M_BM_D}\left({p_\nu\over M_B}+{p'_\nu\over 
M_D}\right)
}
\OMIT{\eqn\eVvii{
\vev{D(p')\vert V_\nu\vert B(p)}
=\left({\bar\alpha_s(m_b)\over\bar\alpha_s(m_c)}\right)^{a_I}
\left({\bar\alpha'_s(m_c)\over \bar\alpha'_s(\mu)}\right)^{a_L}
\xi(v \ccdot v')\sqrt{M_BM_D}\left({p_\nu\over M_B}+{p'_\nu\over 
M_D}\right)
}
}
It follows that
\eqn\eVviii{
f_\pm(q^2) =\xi
(v \ccdot v')\left({M_D\pm M_B\over2\sqrt{M_BM_D}}\right)
}
\OMIT{\eqn\eVviii{
f_\pm(q^2) =\left({\bar\alpha_s(m_b)\over\bar\alpha_s(m_c)}
\right)^{a_I}\left({\bar\alpha_s(m_c)\over\bar\alpha_s(\mu)}\right)^{a_L}\xi
(v
\ccdot v')\left({M_D\pm M_B\over2\sqrt{M_BM_D}}\right)
}
}
Similarly, $f$, $a_\pm$ and $g$ can all be written in terms of
$\xi(v\ccdot v')$. Moreover, at $v\ccdot v'=1$, one has $q^2=(M_B
v-M_D v)^2=(M_B-M_D)^2\equiv q^2_{\rm max}$ so the normalization
Eq.~\eVv\ gives
\eqn\eVix{
f_\pm (q^2_{\rm max})=\left({M_D\pm M_B\over 2\sqrt{M_BM_D}}\right)
}
\OMIT{\eqn\eVix{
f_\pm (q^2_{\rm max})=\left({\bar\alpha_s(m_b)\over\bar\alpha_s(m_c)}
\right)^{a_I}\left({M_D\pm M_B\over 2\sqrt{M_BM_D}}\right)
}
}
This remarkable result gives the form factors, in the heavy quark
limit, without uncertainties from hadronic matrix elements. Short
distance corrections will be discussed below; see Sections~\oneloopsec\
and~\xtrnlcs.

\OMIT{We have used $a_L(v\ccdot v')=0$ at $v\ccdot v'=1$. This is as it
should be, for the physical quantity $f_\pm$ is $\mu$-independent.
It should be emphasized that there is no $\mu$-dependence of $f_\pm$ in
Eq.~\eVviii: the explicit dependence through $(\bar\alpha_s(\mu))^{a_L}$ 
is canceled by the implicit dependence on $\mu$ of the Isgur-Wise
function, $\xi(\vvv)=\xi(\vvv,\mu)$.}

\exercise{
The same methods can be used to obtain relations among, and
normalizations of, the form factors relevant to semileptonic decays of
heavy baryons. The case of transitions between $l=0$ states is
simplest.  The case of transitions involving higher $l$ states can be
found elsewhere.\refs{\baryonrefs,\falkspin}
There are three  form factors, $F_i$, for the matrix element of the
vector current between $\Lambda_b$ and $\Lambda_c$ states, and three 
more,
$G_i$, for the matrix element of the axial current:
$$
\eqalign{
\vev{\Lambda_c(v',s')|\bar c\gamma_\mu b|\Lambda_b(v,s)} &=
        \bar u^{(s')}(v')[\gamma_\mu F_1+v'_\mu F_2+v_\mu F_3]
        u^{(s)}(v)~, 
\cr
\vev{\Lambda_c(v',s')|\bar c\gamma_\mu\gamma_5 b|\Lambda_b(v,s)} &=
        \bar u^{(s')}(v')[\gamma_\mu G_1+v'_\mu G_2+v_\mu G_3]\gamma_5
        u^{(s)}(v)~. 
\cr}
$$
Prove that  all six are given in terms of one universal `Isgur--Wise'
function\baryonrefs. Show that the matrix element of the
current is given by  
\eqn\baryonvev{ 
  \vev{\Lambda_c(v',s')|\bar c_{v'}\Gamma
  b_v|\Lambda_b(v,s)} = \zeta(\vvv)\bar u^{(s')}(v')\Gamma u^{(s)}(v)~,
}
and that $\zeta$ is fixed at one point: $\zeta(1)=1$. Thus
\baryonvev\ show that
\eqn\lambdaffszeroth{
F_1=G_1\qquad\hbox{and}\qquad F_2=F_3=G_2=G_3=0,
}
and 
\OMIT{$G_1(1)=\left({\bar\alpha_s(m_b) / \bar\alpha_s(m_c)}\right)^{a_I}$.}
$G_1(1)=1$.
}

\subsec{Beyond Tree Level}
\subseclab\oneloopsec
In the previous sections we have seen how the HQET can be derived from
QCD and used to obtain useful information about physical processes.
The derivation of the HQET in Section~\efflagsec\ involved only tree
level Feynman diagrams. Clearly one must extend this beyond tree level
if the HQET is to be at all useful. After all, we want to use it to
describe mesons made of a confined  heavy quark and light
antiquark.  It is not difficult to extend the HQET to arbitrary order
in perturbation theory\refs{\grinstein,\grinsteinannrevs}. We will
content ourselves with an understanding of how this works at one loop,
although going beyond is not much more complicated.

The alert reader may complain, justifiably, that an all orders proof
is not enough. There are {\it bona fide} non-perturbative effects that
one cannot obtain even in all orders of perturbation theory. I do not
know of a non-perturbative proof of   the validity of the HQET. One
must not forget, though, that many important results of quantum field
theory are proved perturbatively, \eg renormalizability of the
S-matrix in Yang-Mills theories and the operator product expansion.

The generalization of the effective lagrangian of Eq.~\eIIi\ beyond
tree level consists of adding to it counterterms,
\eqn\eIIiloop{
\leff^{(v)} = \bar Q_{v} i v\ccdot D  Q_v + \CL_{\rm light}+
	\CL_{\rm c.t.}~.
}

At tree level the HQET gave us an expression for the Green functions
of QCD as an expansion of powers in the residual momentum $k=p-M_Qv$.
For the two point function we had, Eq.~\eIIv,
\eqn\eIIvagain{
G (p;q) = \widetilde G_v (k;q) + \CO \left({\Lambda / M_Q}\right)
\qquad\qquad\hbox{(tree level)}~.
}

Beyond tree level the corrected version is still close in form to this,
\eqn\eIIvi{
G(p;q;\mu) = C({M_Q / \mu}, g_s) \widetilde G_{v}(k;q;\mu) 
+ \CO\left({\Lambda /  M_Q}\right)\qquad\hbox{(beyond tree level)}~.} 
The Green functions $G$ and $\widetilde G_{v}$ are renormalized, so
they depend on a renormalization point $\mu$. The function 
$C$ is independent of momenta or light quark masses: it is independent
of the dynamics of the light degrees of freedom. It is there because
the left hand side has some terms which grow logarithmicaly with the
heavy mass, $\ln( M_Q/\mu)$. The beauty of Eq.~\eIIvi\ is that {\it
all of the logarithmic dependence on the heavy mass factors out.}
Better yet, since $C$ is dimensionless, it is a function of
the ratio $M_Q/\mu$ only, and not of $M_Q$ and $\mu$
separately\foot{Actually, additional $\mu$ dependence is implicit in
the definition of the renormalized coupling constant $g_s$.  This
reflects itself in the explicit form of $C$.}. To find the
dependence on $M_Q$ it suffices to find the dependence on $\mu$.  This
in turn is dictated by the renormalization group equation.

It is appropriate to think of the HQET as a factorization theorem,
stating that, in the large $M_Q$ limit, the QCD Green functions
factorize into a universal function of $M_Q$, 
$C({M_Q / \mu}, g_s)$, which depends on the short distance
physics only, times a function that contains all of the information
about long distance physics and is independent of $M_Q$, and can be
computed as a Green function of the HQET lagrangian.

\exercise{
There is a factorization theorem in the physics of deep inelastic
scattering, expressing the cross section as a product of parton
distributions times parton cross sections. Draw an analogy between
those quantities in that factorization theorem and the ones in the
HQET.
}

Of course, the theorem holds for any Green functions, and not just for
the two point function. 

\nfig\fulltoefffig{Relation between Green functions in full and
effective field theories.}
To see how this works at one loop, we start by considering Green
functions for a heavy quark with $n$ gluons, with $n>1$. These are
convergent by power counting, and since there are no nested
divergences at one loop, they are convergent. It suffices to consider
one-particle irreducible (1PI) functions. In Fig.~\xfig\fulltoefffig\ 
 the left side
is calculated in the full theory and the right side in the HQET. The
double line stands for the heavy propagator in the HQET.

\INSERTFIG{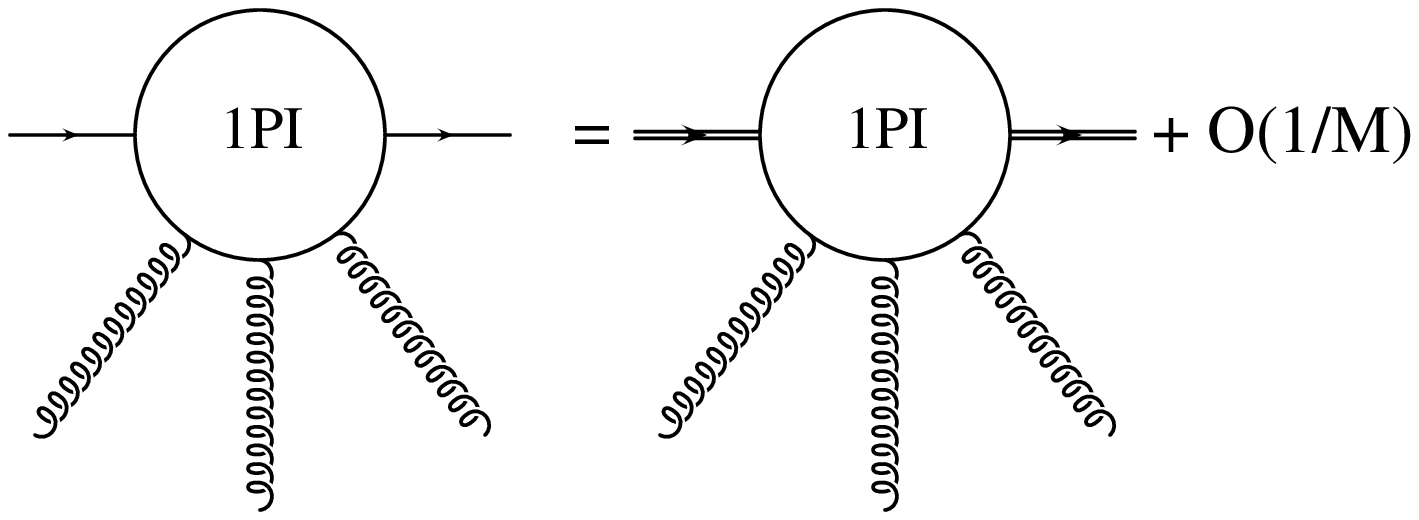}\fulltoefffig{Relation between Green functions in full and
effective field theories.}
\OMIT{
\PASTEFIG{1.7}\fulltoefffig{Relation between Green functions in full and
effective field theories.}
}
\nfig\fourptfnctfig{Relation between Green functions in full and
effective field theories in simple contribution to four point
function, at one loop.} We can prove the validity of the equation
represented in Fig.~\xfig\fulltoefffig, diagram by diagram (there are
several diagrams that contribute to each side of the equation).
Consider, for definiteness, the diagrammatic equation in
Fig.~\xfig\fourptfnctfig.

\INSERTFIG{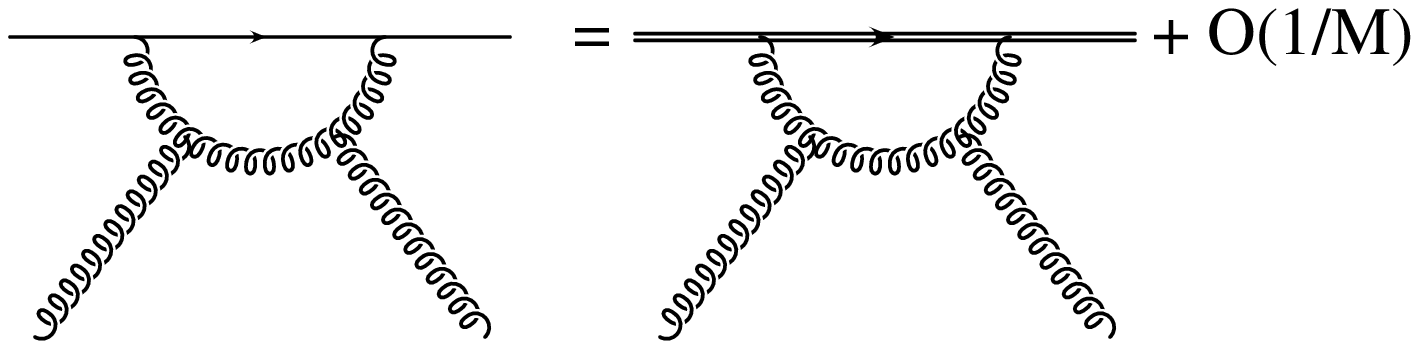}\fourptfnctfig{Relation between Green functions in full and
effective field theories in simple contribution to four point
function, at one loop.}
\OMIT{ 
\PASTEFIG{1.7}\fourptfnctfig{Relation between Green functions in full and
effective field theories in simple contribution to four point
function, at one loop.} 
}

The equation would trivially hold if we could make the propagator
replacement
$${i \over \spur p + \lsl - M_Q} \rightarrow 
\left({1 + \vsl \over 2}\right) {i \over v\cdot (k+l)}$$
even inside the loop integral. Here $p=M_Qv+k$, and $l$ is the loop
momentum.  In other words, in the right hand side of
Fig.~\xfig\fourptfnctfig, we take the limit $M_Q \to\infty$ and then
integrate, while on the left side we first integrate and then take the
limit. Everyone knows that, if both integrals converge, then they
agree. And that is the case for Fig.~\xfig\fourptfnctfig, and, indeed,
it is also the case for any 1--loop integral with a heavy quark and
$n\ge2$ external gluons. We have established Fig.~\xfig\fulltoefffig\
for $n \ge 2$.

\nfig\threeptfnctfig{Relation between infinite Green functions in full and
effective field theories at one loop: three point function.} We are
left with the 2--point ($n= 0$) and 3--point ($n=1$) functions. These
are different from the $n \ge 2$ functions in two ways. First, they
receive contributions at tree level. And second, they are divergent at
1--loop.  Choose some method of regularization. Dimensional
regularization is particularly useful as it preserves gauge invariance
(or, more precisely, BRST invariance). The comparison between full and
effective theories is simplest if the same gauge and regularization
choices are made. For concreteness, consider
Fig.~\xfig\threeptfnctfig.

\INSERTFIG{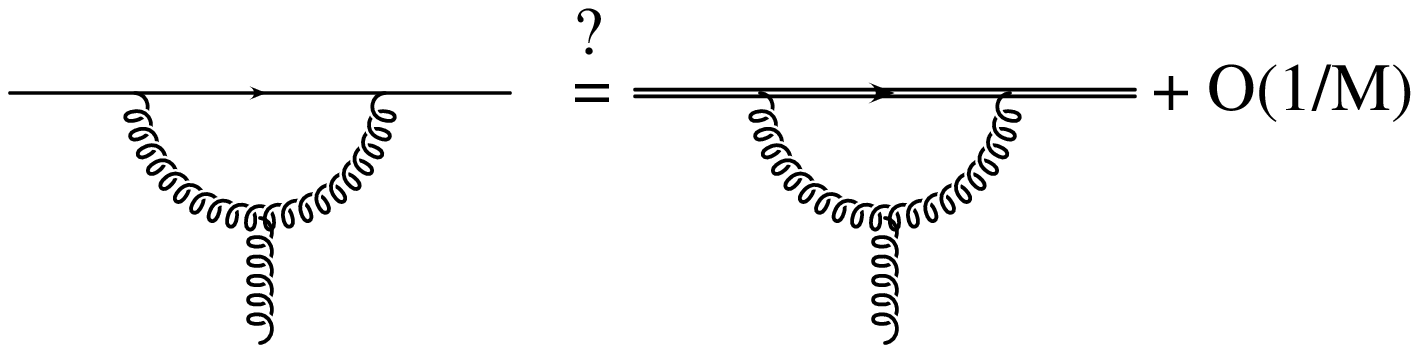}\threeptfnctfig{Relation between infinite Green
functions in full and effective field theories at one loop: three
point function.}
\OMIT{
\PASTEFIG{1.7}\threeptfnctfig{Relation between infinite Green
functions in full and effective field theories at one loop: three
point function.}
}

Since both sides are finite, we can argue as before. But we run into
trouble when we try to remove the regulator. One must renormalize the
Green functions by adding counter terms, but there is no guarantee
that the counterterms satisfy the same relation as the regulated Green
functions of Fig.~\xfig\threeptfnctfig. To elucidate the relation
between counterterms, take a derivative from both sides of
Fig.~\xfig\threeptfnctfig\ with respect to either the residual
momentum $k_{\mu}$ or the gluon external momentum of $q_{\mu}$. This
makes the diagrams finite and the regulator can be removed. Thus, at
1--loop, the relations
\eqn\eIIvii{{\partial \over \partial k_{\mu}} G^{(2,1)} = {\partial \over
\partial k_{\mu}} \tilde G^{(2,1)}_v + \CO\left({\Lambda / M_Q}\right)}
and
\eqn\eIIviii{{\partial \over \partial q_{\mu}} G^{(2,1)} = {\partial \over
\partial q_{\mu}} \tilde G_v^{(2,1)} + \CO ({\Lambda / M_Q})}
hold. The counterterms, or at least the difference between them, are
$k_{\mu}$ and $q_{\mu}$ independent. It is a simple algebraic exercise to
show, then, that the difference between counterterms is of the form 
\eqn\eIIviiii{a G^{(2,1)0} + b \tilde G_v^{(2,1)0}}
where the superscript `0' stands for tree level, and $a$ and $b$ are
infinite constants, {\it i.e.}, independent of $k_{\mu}$ and
$q_{\mu}$.  Thus, one can subtract the 1--loop Green functions by
standard counterterms, and establish the equality of
Fig.~\xfig\threeptfnctfig.

A similar argument can be constructed for the 2--point function. One must
take two derivatives with respect to $k_{\mu}$, but that is as
it should, since the counterterms are linear in momentum.

We have therefore established that, to 1--loop, the renormalized Green
functions in the full and effective theories agree. The alert reader must
be puzzled as to the fate of the function $C({{M_Q / \mu},g_s})$ of 
Eq.~\eIIvi. What has happened is that the constant $b$ in the counterterm in 
Eq.~\eIIviiii\ is, in general, $M_Q$ dependent. Indeed, if we take derivatives
with respect to $M_Q$, as in \eIIvii\ or \eIIviii, the degree of divergence
is not changed, and one cannot argue that $a$ or $b$ are $M_Q$ independent. The
relation between renormalized Green functions that we have derived contains
hidden $M_Q$--dependence in the renormalization prescription for the Green
functions in the HQET.

Given two different renormalization schemes, the corresponding renormalized
Green functions $\tilde G$ and $\tilde G'$ are related by a finite
renormalization 
$$\tilde G = z(\mu,g_s) \tilde G' $$
Choosing $\tilde G$ to be the mass--independent subtracted Green function,
and $\tilde G'$ the one in our peculiar subtraction scheme, we have that
the relation between full and effective theories becomes
$$G^{(2,1)} (p,q;\mu) = C\left({M_Q / \mu}, g_s\right) 
\tilde G_v^{(2,1)}(k,q;\mu) +
\CO\left({\Lambda  / M_Q}\right)$$
as advertised in Section~\efflagsec. Here, $C$ is nothing but
this finite renormalization $z(\mu,g_s)$. That we can use the same
function $C$ for all Green functions can be established by
using the same wave functions renormalization prescription for gluons
in the full and effective theories.  Otherwise, an additional factor
of $z_A^{n/2}$ would have to be included in the relation between
$G^{(2,n)}$ and $\tilde G^{(2,n)}$. This completes the argument.

It is worth mentioning that the discussion above {\it assumes} the
renormalizability, preserving BRST invariance, of the effective theory.
Although, to my knowledge, this has not been established, there is no
obvious reason to doubt that the standard techniques apply in this case.

\subsec{External Currents}
\subseclab\xtrnlcs
We will often be interested in computing Green functions with an insertion
of a current. Consider, the current
\eqn\fullcurrent{
J_\Gamma = \bar q\, \Gamma Q}
in the full theory, where $\Gamma$ is some Dirac matrix, and $q$ a light
quark. In the effective theory, this is replaced according to
\eqn\ex{J_\Gamma (x)\to e^{-i M_Q v\ccdot x}\tilde J_\Gamma(x) ~,}
where
\eqn\exi{\tilde J_\Gamma = \bar q \,\Gamma Q_v~,}
and it is understood that in $\tilde J_\Gamma$ the heavy quark  is that of the
HQET, satisfying, in particular, $\vsl Q_v=Q_v$. The exponential factor in
Eq.~\ex\ reminds us to take the large momentum out through the current,
allowing us to keep the external momentum of light quarks and gluons small.
The relation between full and effective theories takes the form of an
approximate equation between Green functions ---and eventually 
amplitudes--- of
 insertions of these currents:
\eqn\exii{G_{\!J_\Gamma}(p,p';q;\mu)= 
C({M_Q / \mu},g_s)^{1/2}  
C_{\Gamma}({M_Q / \mu},g_s) 
\widetilde G_{\!v,\tilde J_\Gamma}(k,k';q;\mu) 
+ \CO({\Lambda / M_Q}) ~,}
where $p$ and $p'$ are the momenta of the heavy quark and the external
current, $k$ and $k'$ the corresponding residual momenta, $p=M_Qv+k$,
$p'=M_Qv+k'$, and $q$ stands for the momenta of the light degrees of
freedom.  The factor $C^{1/2}\,C_{\Gamma}$
accounts for the logarithmic mass dependence, as explained earlier. We
see that an additional factor, namely, $C_{\Gamma}$, is
needed in this case to account for the different scaling behavior of
the currents in the full and effective theories.  It is convenient to
think of the replacement of currents, not as given by Eq.~\ex, but
rather by
\eqn\exiii{J_\Gamma(x) \to e^{-iM_Qv\ccdot x} C_\Gamma({M_Q / 
\mu}, g_s) \tilde J_\Gamma(x)~.
}

In fact, Eq.~\exii, and therefore the replacement in Eq.~\exiii, are
not quite correct. To reproduce the matrix elements of the current
$J_\Gamma$ of Eq.~\fullcurrent, it is necessary to sum over matrix
elements of several different `currents' in the effective theory. The
operator $\tilde J_\Gamma$ of Eq.~\exi\ is just one of them. In
addition, one may have to introduce such operators as $\bar q
\vsl\Gamma Q_v$. The correct replacement is therefore
\eqn\fulltoeffcurr{J_\Gamma(x) \to e^{-iM_Qv\ccdot x} \sum_i
C_{\Gamma}^{(i)}
({M_Q / \mu}, g_s) \widetilde \CO^{(i)}(x) ~.}
Here $\widetilde \CO^{(i)}(x)$  is the collection of the operators of dimension
3 with appropriate quantum numbers.  The first
operator in the sum, call it $\widetilde \CO^{(0)}$,  is there even at tree
level, and corresponds to the operator $\tilde J_\Gamma$ of Eq.~\exi.

Another case of interest is that of the insertion of a current of two heavy
quarks
\eqn\exiv{J_\Gamma = \bar Q' \Gamma Q ~.}
The replacement now is 
\eqn\exv{J_\Gamma(x) \rightarrow e^{-i M_Q v\ccdot x + i M_{Q'} v'\ccdot x} 
\sum_i\widehat 
C^{(i)}_{\Gamma}({M_Q \over \mu},{M_{Q'} \over M_Q}, v\ccdot v',g_s)
\widehat \CO^{(i)}_\Gamma(x)~.
}
Again, $\widehat \CO^{(i)}(x)$ stands for the complete list of
operators of dimension 3 in the effective theory with the right
quantum numbers. Also, the operator $\widehat\CO^{(0)}=\bar Q'_{v'}
\Gamma Q_v$ appears in the sum at tree level.

This deserves some explanation. The Green functions now include two
heavy quarks. The functions $\widehat C$ connecting these full and
effective Green functions will now, in general, depend on both $M_Q$
and $M_{Q'}$.  Moreover, we can not argue that $\widehat C$ are
independent of the velocities $v$ and $v'$.  In fact, this was true of
the simpler case considered in Section~\oneloopsec; but there,
$C$ could only depend on $v_\mu$ through $v^2=1$. In the
case at hand there is an additional invariant on which $\widehat C$
can depend, namely~$v\ccdot v'$. 

The explicit functional dependence on $M_Q$ in the functions
$C_\Gamma$ and $\widehat C_\Gamma$ can be obtained from a
study of their dependence on the renormalization point $\mu$.
For clarity of presentation we neglect operator mixing for
now. When necessary, this can be incorporated without much difficulty. 
Taking a derivative $d/d\mu$ on both sides of 
Eqs.~\eIIvi\ and~\exiii, we find
\eqn\exvii{\mu {d \over d\mu} C_\Gamma = 
(\gamma_{_\Gamma} - 
\tilde \gamma_{_\Gamma}) C_\Gamma}
where $\gamma_{_\Gamma}$ and $\tilde \gamma_{_\Gamma}$ are the 
anomalous
dimensions of the currents $J_\Gamma$ and $\tilde J_\Gamma$ in the full
and effective theories, respectively. Of particular interest are the cases
$\Gamma= \gamma^\mu$ and $\Gamma = \gamma^\mu \gamma_5$. These 
correspond,
in the full theory, to conserved and partially conserved currents, and
therefore the corresponding anomalous dimensions vanish, giving
\eqn\exviii{\mu {d C_{\Gamma } \over d \mu} = - \tilde 
\gamma_{_\Gamma}\,
\tilde  C_{\Gamma}\qquad \,\,\,\,\,(\Gamma = \gamma^\mu, 
\, \gamma^\mu \gamma_5) ~.}

Before we solve this equation, we recall that
\eqn\exiv{ \mu {d \over d\mu} = \mu {\partial \over \partial\mu} +
\beta (g_s) {\partial \over \partial g_s} ~.}
Here $\beta$ is the QCD $\beta$--function, with perturbative expansion
\eqn\exx{{\beta(g) \over g}= -{b_0} {g^2 \over 16\pi^2} + b_1 \left({g^2 
\over
16\pi^2}\right)^2 + \ldots\quad,}
and
\eqn\exxi{b_0 = 11- {2 \over 3} n_f~,}
where $n_f$ is the number of quarks in the theory. For our purposes,
$n_f$ should {\it not} include the heavy quark. This is explained in
the famous paper by Appelquist and Carrazone\decoupling; it simply
reflects the fact that the logarithmic scaling of $g_s$ is not
affected by heavy quark loops, since these are suppressed by powers of
$M_Q$. Now, the solution to~\exviii\ is standard:
\eqn\exxii{C_{\Gamma}(\mu,g_s) = \exp \left(
- \int^{g_s}_{\bar g_s(\mu_0)}
\!\!dg'\,{\tilde  \gamma_{_\Gamma}(g') \over \beta(g')} \right)\, 
C_\Gamma(\mu_0,\bar g_s(\mu_0))}
where $\bar g_s$ is the running coupling constant defined by
\eqn\exxiiidupl{\mu' {d \bar g_s(\mu') \over d\mu'} = \beta(\bar
g_s(\mu'))\qquad ,\qquad
\bar g_s(\mu) = g_s~.}
Choosing $\mu_0 = M_Q$, and restoring the dependence on $M_Q$, we have 
then
\eqn\exxiii{
C_\Gamma(M_Q/\mu,g_s)=\exp\left(-\int^{\bar g_s(\mu)}_{\bar 
g_s(M_Q)}
\!\! dg'\,{\tilde\gamma_{_\Gamma}(g')\over\beta(g')} \right) \, 
C_\Gamma(1,\bar g_s(M_Q))~.
} 

Therefore, the problem of determining $C\tjg({M_Q/\mu},g_s)$  
breaks down into two parts. One is
the determination of the anomalous dimensions $\tilde\gamma_{_\Gamma}$.
The other is the calculation of $C_\Gamma(1,\bar g_s(M_Q))$. Both 
can
be done perturbatively, and $C_\Gamma(M_Q/\mu, g_s)$ can thus 
be
computed, provided $\mu$ and $M_Q$ are large enough so that $\bar 
g_s(\mu)$ and
$\bar g_s(M_Q)$ are small. One finds, for example, that in leading order 
$C_\Gamma$ is $\Gamma$ independent and there is no mixing:
\eqn\exxvi{
C\tjg({M_Q / \mu}, g_s) = \left({\bar \alpha_s(M_Q) \over \bar
\alpha_s(\mu)}\right)^{\! a_I}~,
}
where $\bar \alpha_s \equiv \bar g_s^2/4 \pi$, 
and\refs{\volotwo{--}\politzerwiseone}\ 
$a_I \equiv -{c_1/2b_0} = -6 / (33-2n_f)$.

\exercise{Obtain an expression for $C_\G$ analogous to that
of Eq.~\exxiii\ but without assuming that the anomalous dimension
$\g_\G$ of Eq.~\exvii\ vanishes.}

We now turn to the computation of the coefficient $\widehat
C_{\Gamma}$ for the current of two heavy quarks in Eq.~\exv. A new
difficulty arises. Because $\widehat C_{\Gamma}$ depends on three
dimensionful quantities, namely the masses $M_Q$ and $M_{Q'}$, and the
renormalization point $\mu$, its functional dependence is not
determined from the renormalization group equation (even if we neglect
the implicit dependence of $g_s$ on $\mu$). Two different
approximations have been developed to deal with this problem:

I) Treat the ratio $M_{Q'}/M_Q$ as a dimensionless parameter, and
study the dependence of $\widehat C_\Gamma$ on $M_{Q'}/\mu$ through
the renormalization group\falkgrinsteina. This is just like what was
done for the heavy-light case, so we can transcribe the result: %
\eqn\exxviv{
\IFBIG{\widehat C_\Gamma \!\left({M_{Q'} \over \mu},{M_{Q'} \over M_Q} , 
v\ccdot v',g_s\!\right) \!\approx\! \exp\left(
\!-\! \int_{\bar g_s(M_{Q'})}^{\bar g_s(\mu)}
\kern-1.5ex dg'\,{\hat \gamma_{_\Gamma} (g') \over \beta (g')} 
\right)\!
\widehat C_\Gamma\!\left(1,{M_{Q'} \over M_Q},v\ccdot v',\bar 
g_s(M_{Q'})\right)
.}{\widehat C_\Gamma \!\left({M_{Q'} \over \mu},{M_{Q'} \over M_Q} , 
v\ccdot v',g_s\!\right) \!\approx\! \exp\left(
\!-\! \int_{\bar g_s(M_{Q'})}^{\bar g_s(\mu)}
\kern-2.5ex dg'\,{\hat \gamma_{_\Gamma} (g') \over \beta (g')} 
\right)\!
\widehat C_\Gamma\!\left(1,{M_{Q'} \over M_Q},v\ccdot v',\bar 
g_s(M_{Q'})\right)
}}
Again
\eqn\exxvv{\widehat C_{\Gamma}\!\left(1,{M_{Q'} 
\over M_Q},v\ccdot v',\bar g_s(M_{Q'})\right) = 1 + \CO
(\bar \alpha_s(M_{Q'}))~.}
But, now, the correction of order $\bar \alpha_s(M_{Q'})$ is a function of
$M_{Q'}/M_Q$. This method has the advantage that the complete
functional dependence on $M_{Q'}/M_Q$ is retained, order by order in
$\bar\alpha_s(M_{Q'})$. Nevertheless, it fails to re--sum the leading--logs 
between
the scales $M_{Q'}$ and $M_Q$, {\it i.e.}, it does not include the effects of
running of the QCD coupling constant between $M_Q$ and $M_{Q'}$. Therefore, 
this
method is useful when $M_{Q'}/M_Q \sim1$, or, equivalently, when $(\bar
\alpha_s(M_{Q'}) - \bar \alpha_s(M_Q))/\bar \alpha_s(M_Q) \ll 1$.

II) Treat the ratio $M_{Q'}/M_Q$ as small. Expand first in a HQET
treating $Q$ as heavy and $Q'$ as light. The corrections are not just
of order $\Lambda/M_Q$ but also $M_{Q'}/M_Q$, but this is assumed to
be small (even if much larger than $\Lambda/M_Q)$. Then expand from
this HQET, in powers of $\Lambda/M_{Q'}$, by constructing a new HQET
where both $Q$ and $Q'$ are heavy\fggw.  The calculation of $\widehat
C_\Gamma$ then proceeds in two steps. The first gives a factor just
like that of the heavy-light current, in~\exxiii\ 
\eqn\factorone{\exp\left( -\int^{\bar g_s(\mu)}_{\bar g_s(M_Q)} \! dg' 
{\tilde \gamma_{_\Gamma} (g')
\over \beta (g')}\right) C\tjg(1,\bar g_s(M_Q)) ~.
}
The second factor is as in method I, above, but neglecting
$M_{Q'}/M_Q$.  Moreover, the current $\tilde J_\Gamma$ is not
conserved, so the anomalous dimension to be used is not
$-\hat\gamma_{_\Gamma}$ but $\tilde\gamma_{_\Gamma} -\hat
\gamma_{_\Gamma}$.  Finally, we must make explicit the fact that in
the first and second steps the appropriate $\beta$-functions differ in
the number of active quarks. We therefore label the one in the second
step $\beta'$ and the corresponding running coupling constant $\bar
g_s'$. The second factor is
\eqn\factortwo{\exp\left(\int^{\bar g_s (\mu)}_{\bar g_s(M_{Q'})} dg' 
{\tilde
\gamma_{_\Gamma}(g')
\over \beta (g')} - \int_{\bar g_s'(M_{Q'})}^{\bar g'_s (\mu)} dg'' {\hat
\gamma_{_\Gamma}(g'') \over \beta'(g'')}\right) \widehat C_\Gamma 
(1,0,v\ccdot v',\bar
g_s(M_{Q'})) ~.}
Combining factors gives
\eqn\exxvviii{\eqalign{
\widehat C_{\Gamma}\left({M_{Q'} \over \mu},\,{M_{Q'} \over M_Q},\, v 
\ccdot
v' , g_s\right) 
\approx \exp &\left(-\int^{\bar g_s (M_{Q'})}_{\bar g_s(M_{Q})}
\kern-2.5ex dg'\, {\tilde\gamma_{_\Gamma} (g') \over \beta (g')} 
- \int^{\bar g'_s (\mu)}_{\bar g'_s (M_{Q'})} 
\kern-2.5ex dg''\, {\hat\gamma_{_\Gamma} (g'') \over \beta'(g'')}\right)\cr
&\null\qquad\times C\tjg(1,\bar g_s(M_Q)) \widehat C_\Gamma
(1,0,v\ccdot v',\bar g_s(M_{Q'})) ~.\cr}
}

The advantage of method II over method I is that it does include the
effects of running between $M_Q$ and $M_{Q'}$. The disadvantage is that it
neglects powers of $M_{Q'}/M_Q$. (Actually, the result can be improved by
reincorporating the $M_{Q'}/M_Q$ dependence, as a power series expansion in
this ratio).

For example, in method II eqn.~\exv\ becomes, in leading order,\fggw
 \eqn\replacevec{
\bar c\gamma^\mu b\to\left({\bar\alpha_s(m_b)\over\bar\alpha_s(m_c)}
\right)^{a_I}\left({\bar\alpha'_s(m_c)\over\bar\alpha'_s(\mu)}\right)^{a_L}
\bar c_{v'}[(1+\kappa)\gamma^\mu+(\lambda_b -\lambda_c(v\ccdot v'))\vsl
\gamma^\mu] b_v
}
 for $\Gamma=\gamma^\mu$, and 
\eqn\replaceaxi{
\bar c\gamma^\mu\gamma_5 
b\to\left({\bar\alpha_s(m_b)\over\bar\alpha_s
(m_c)}\right)^{a_I}\left({\bar\alpha'_s(m_c)\over\bar\alpha'_s(\mu)}\right)^
{a_L}
\bar c_{v'}[(1+\kappa)\gamma^\mu\gamma_5-(\lambda_b 
+\lambda_c(v\ccdot v'))\vsl
\gamma^\mu\gamma_5] b_v}
for $\Gamma=\gamma^\mu\gamma_5$, 
where
\eqn\rdefined{\eqalign{
\lambda_b={\alpha_s(m_b)\over 3\pi}\,,\quad & \quad
\lambda_c(\vvv)={2\alpha_s(m_c)\over 3\pi}\,r(\vvv)~, \cr
a_L(\vvv) &= {8\over33-2n_f}[\vvv r(\vvv) -1]~,\cr
r(x) &\equiv { 1 \over \sqrt{x^2 -1}} \ln \left(x + \sqrt{x^2-1}\right)~,\cr}
}
and $\kappa$ is of order $\alpha_s$ but a subleading log.

\exercise{Why is the term involving $\kappa$ in Eqs.~\replacevec\
and~\replaceaxi\  a subleading log, while those involving
$\lambda_b$ and $\lambda_c$ are leading logs?}

\subsec{Form factors in order $\alpha_s$}
The predicted relations between form factors, and normalizations at
$q^2_{\rm max}$, are only approximate. Indeed, several approximations
were made in obtaining those results. Corrections that arise from
subleading order in the $1/M$ expansion will be considered in
Chapter~\oneoverm. Here we will discuss corrections of
order~$\alpha_s$.

As observed in Section~\xtrnlcs, the vector and axial--vector currents
of the full theory, $\bar c\Gamma b$, match onto a linear combination
of `currents', {\it i.e.,} dimension~3 operators, in the effective
theory. At one loop, the correspondence between vector and axial
currents in the full and effective theories is given by
Eqs.~\replacevec\ and~\replaceaxi.  The constant $\lambda_b$ and the
function $\lambda_c$ arise only from $1$-loop matching, and are scheme
independent. The constant $\kappa$ receives contributions both from
matching at $1$-loop, and from $2$-loops anomalous dimensions. Leaving
out the latter would give a meaningless, scheme dependent, result.
Although $\kappa$ has been computed, it is interesting to note that
predictions can be made solely form the $1$-loop matching computation.

Indeed, comparing Eqs.~\eVvi\null\ with Eqs.~\fulff\null, we see that
at zeroth order in $\bar\alpha_s(m_b)$ or $\bar\alpha_s(m_c)$ we have
\eqn\eVxiv{
		a_++a_-=0.
}
Plugging Eq.~\replaceaxi\ into Eq.~\eVii\null\ we see that, to order
$\bar\alpha_s(m_c)$ and $\bar\alpha_s(m_b)$ there is a computable
correction to this combination of form factors, namely
\eqn\eVxv{
{a_++a_-\over a_+}=-4{m_c\over m_b}\left[{\bar\alpha_s(m_b)\over 3\pi} +
{2\bar\alpha_s(m_c)\over3\pi}r(v\ccdot v')\right]
}

The constant $\kappa$, although difficult to compute, does not change
the relations between form factors since it simply rescales the
leading order predictions in Eq.~\eVvi\null\ by the common factor of
$(1+\kappa)$. It does, however, affect the predicted normalization of
form factors at $q^2_{\rm max}$.  Since at $v'=v$ the effective vector
current is again $\bar c_{v'}\gamma_\mu b_v$, but rescaled by
$(1+\kappa+\lambda_b-\lambda_c(1))$, the correction to Eq.~\eVix\ is
\eqn\eVxvi{
f_\pm(q^2_{\rm max})=(1+\kappa+\lambda_b-\lambda_c(1))\left(\alpha
_s(m_b) \over \alpha_s(m_c)\right)^{a_I}\left({M_D\pm M_B\over
2\sqrt{M_BM_D}}\right).
}
We emphasize that retaining the constant $\kappa$ in Eq.~\eVxvi\ is
inconsistent, because not all next to leading logs are included.
For a calculation of the sub-leading logs, including the constant $\kappa$,
see Ref.~\anatoli.

\newsec{$1/M_Q$}
\seclab\oneoverm
\subsec{The Correcting Lagrangian}
One of the main virtues of the HQET is that, in contrast to {\it
models} of the strongly bound hadrons, it lets us study systematically
the corrections arising from the approximations we have made. To be
sure, we've made several approximations already, even within the
zeroth order expansion in $\Lambda/M_Q$. For example, we have computed
the logarithmic dependence on $M_Q$, {\it i.e.,} the functions
$C^{(i)}_\Gamma$ and $\widehat C^{(i)}_\Gamma$ of Eqs.~\fulltoeffcurr\
and \exv, using perturbation theory. In this Section we turn to the
corrections of order $\Lambda /M_Q$.

The HQET lagrangian was derived, in Section~\efflagsec,
by putting the heavy quark
almost on-shell and expanding in powers of the residual momentum, 
$k_\mu$, or
light quark or gluon momentum, $q_\mu$, over $M_Q$, which we generally 
wrote
as $\Lambda/M_Q$. Let us again derive the effective lagrangian, keeping 
track, this time, of the terms of order $\Lambda/M_Q$.

We will rederive $\leff^{(v)}$, including $1/M_Q$ corrections, working 
directly in configuration
space\fgl. The heavy quark equation of motion is
\eqn\eVIi{
(i\Dsl-M_Q)Q=0
}
We can put the quark almost on shell by introducing the redefinition
\eqn\eVIii{
Q=e^{-iM_Qv\ccdot x}\tilde Q_v
}
In terms of $\tilde Q_v$, the equation of motion is
\eqn\eVIiii{
[i\Dsl+M_Q(\vsl -1)]\widetilde Q_v=0
}
If we separate the $(1+\vsl)$ and $(1-\vsl)$ components of $\widetilde Q_v$, 
we 
see that, as expected, the latter is very heavy and decouples in the infinite
mass  limit. To project  out the components, 
\eqn\eVIiv{
\widetilde Q_v=\widetilde Q^{(+)}_v+\widetilde Q^{(-)}_v
}
where
\eqn\eVIv{
\widetilde Q^{(\pm)}_v = \projpm\widetilde Q_v~,
}
we multiply  Eq.~\eVIiii\ by $({1\pm\vsl\over2})$. Thus  we have the 
equations 
\eqn\eVIvi{
iv\ccdot D \widetilde Q^{(+)}_v=-\projp i\Dsl\qt{-}
}
and
\eqn\eVIvii{
iv\ccdot D\widetilde Q^{(-)}_v+2M_Q\widetilde Q^{(-)}_v= \projm
i\Dsl\widetilde Q^{(+)}_v
}
These equations can be solved self-consistently by assuming that $\widetilde
Q^{(+)}_v$ is
order $(M_Q)^0$ while $\widetilde Q^{(-)}_v$ is order $M^{-1}_Q$.
A recursive solution follows. From
Eq.~\eVIvii\
\eqn\eVIviii{
\widetilde Q^{(-)}_v={1\over2M_Q}\projm i\Dsl\widetilde
Q^{(+)}_v-i{v\ccdot D\over 2M_Q}\widetilde Q^{(-)}_v
}
Substituting into Eq.~\eVIvi\ and dropping terms of order $1/M^2_Q$, we
have 
\eqn\eVIix{
iv\ccdot D\widetilde Q^{(+)}_v=-\left({1+\vsl\over 2}\right)i\Dsl
{1\over2M_Q}\left({1-\vsl\over2}\right)i\Dsl\tilde 
Q^{(+)}_v
}
The right hand side involves
\eqn\eVIx{
\IFBIG{\projp\Dsl\projm  \Dsl\projp
=\projp\left[ D^2- (v\ccdot D)^2 +\half g_s\sigma^{\mu\nu}G_{\mu\nu}
\right]\projp}%
{\projpsm\Dsl\!\projmsm \Dsl\!\projpsm
=\projpsm\!\left[ D^2- (v\ccdot D)^2 +\half g_s\sigma^{\mu\nu}G_{\mu\nu}
\right]\!\projpsm}
}
where $\sigma^{\mu\nu}={i\over 2}[\gamma^\mu,\gamma^\nu]$ and
$G_{\mu\nu}={1\over ig_s}[D_\mu,D_\nu]$ is the QCD field strength tensor.
 This equation of motion is
obtained from the lagrangian
\eqn\eVIxi{
\leff^{(v)}=\bar Q_viv\ccdot DQ_v+{1\over 2M_Q}\bar Q_v \left [D^2-(v\ccdot
D)^2+{g_s\over 2}\sigma^{\mu\nu} G_{\mu\nu}\right]Q_v
}
Here I have reverted to the notation $Q_v$ for $\widetilde Q^{(+)}_v$.
How to include higher order terms in the $1/M_Q$ expansion
into $\leff^{(v)}$ should be clear.

\exercise{Find the effective lagrangian describing a heavy scalar field
to first order in~$1/M$.}

The $1/M_Q$ term in $\leff^{(v)}$ is treated as small. If it is not, it
does not make sense to talk about a HQET in the first place.
It is therefore appropriate to use perturbation theory to compute its
effects. In this perturbative expansion, the corrections of order $1/M_Q$
to Green functions, and therefore to physical observables,
are computed by making a single insertion of the perturbation
\eqn\eVIxii{
\Delta \CL={1\over 2M_Q}\bar Q_v\left[ D^2-(v\ccdot D)^2+{g_s\over
2}\sigma^{\mu\nu} G_{\mu\nu}\right] Q_v~.
}

The symmetries of the HQET, discussed at length in Sections~\intuitive\
and~\symmtrssec, are
broken by $\Delta\CL$. Under the $SU(N_f)$--flavor symmetry, $\Delta \CL$
transforms as a combination of the Adjoint and Singlet representations,
while only the chromo-magnetic moment operator
\eqn\colormagmom{
\bar Q_v\sigma^{\mu\nu}G_{\mu\nu}Q_v
}
breaks the spin-$SU(2)$ symmetry: it transforms as a 3 of spin--$SU(2)$.

A single insertion of $\Delta \CL$ does include all orders in QCD, and it will
often prove difficult to make precise calculations of $1/M_Q$ effects. Since
$\Delta\CL$ is treated as a simple insertion
in Green functions, its treatment in
the HQET is entirely analogous to that of current operators of
Section~\xtrnlcs.
There are coefficient functions that connect the HQET
results with the full theory. It is convenient to include them directly
into the effective lagrangian 
as{\refs{\fgl,\falkgrinsteinb,\eichtenhillb}}
\eqn\coefsinlag{
\Delta\CL={1\over2M_Q}\bar Q_v\left[c_1D^2+c_2(v\ccdot D)^2+ \half c_3
g_s\sigma^{\mu\nu}G_{\mu\nu}\right]Q_v.
}
Here
\eqn\ccoefsa{
c_i=c_i(M_Q/\mu,g_s)}
can be determined through the methods discussed extensively in
Section~\xtrnlcs.
In leading-log, one finds\fgl
\eqn\ccoefsb{
\eqalign{
c_1 &= -1\cr
c_2 &= 3\left({\bar\alpha_s(\mu)\over\bar\alpha_s(M_Q)}
\right)^{\!\!-8/(33-2n_f)}-2\cr
c_3 &= -\left({\bar\alpha_s(\mu)\over\bar\alpha_s(M_Q)}
\right)^{\!\!-9/(33-2n_f)}.\cr}
}

\subsec{ The Corrected Currents}
Just as the lagrangian is corrected in order $1/M_Q$, any other operator is
too. In particular, the current operators studied in Section~\xtrnlcs,
are modified
in this order. At tree level, these corrections are given by the change of
variables of last Section: 
\eqn\subhlcur{
J_\Gamma = \bar q\Gamma Q\to\bar q\Gamma e^{-iM_Qv\ccdot
x}\left[Q_v+{1\over2M_Q}\projm i\Dsl Q_v\right].
}

Beyond tree level, this sum of two terms has to be replaced by a more
general sum over operators of the right dimensions and quantum numbers.
The replacement is
\eqn\fullsubhlcur{
J_\Gamma\to e^{-iM_Qv\ccdot x}\left(\sum_i \widetilde 
C^{(i)}_{\Gamma}\bar
q\Gamma_i  Q_v + {1\over2M_Q}\sum_j \widetilde 
D^{(j)}_{\Gamma}\CO_j\right)
}
where $\CO_j$ are operators of dimension 4 that include, for example, the
operators
\eqn\opexamples{
\bar q\Gamma i \Dsl Q_v \qquad,\qquad
\bar q\Gamma i (v\ccdot D)  Q_v \qquad,\qquad
\bar q \vsl\Gamma i \Dsl Q_v ~.
}
A complete set of operators, and the corresponding coefficients, $\widetilde
D^{(i)}_\Gamma$, for the cases $\Gamma=\gamma^\mu$
and $\Gamma=\gamma^\mu\gamma_5$, can be found in
Refs.~{\refs{\falkgrinsteinb,\goldenhill}}\ in the
leading-log approximation. 

The case of two heavy currents is similar. A straightforward
calculation gives
\eqn\subhhcur{
\eqalign{
J_\Gamma=\bar Q'\Gamma Q \to &
e^{-iM_Qv\ccdot x + iM_{Q'}v'\ccdot x } \Big[ \bar Q'_{v'}\Gamma Q_v \cr
&\left. +{1\over2M_Q}\bar Q'_{v'}\Gamma\projm i\Dsl Q_v +
{1\over2M_{Q'}}\bar Q'_{v'}i\overleftarrow\Dsl\projm\Gamma  
Q_v\right]\cr}
}
Again, beyond tree level we must replace this expression by a more general
sum over operators of dimension four,
\eqn\fullsubhhcur{
\eqalign{
J_\Gamma\to
e^{-iM_Qv\ccdot x + iM_{Q'}v'\ccdot x } \left[ \sum_i \widehat
C^{(i)}_{\Gamma} \bar Q'_{v'}\Gamma_i Q_v           \right.
&\left. +{1\over2M_Q}\sum_j \widehat D^{(j)}_{\Gamma}\CO_j \right.\cr
& \left.+{1\over2M_{Q'}}\sum_j \widehat
D^{\prime(j)}_{\Gamma}\CO_j\right]\cr}
}

It is worth pointing out that, in the computation of the coefficient
functions $\widetilde D^{(j)}_{\Gamma}$, $\widehat D^{(j)}_{\Gamma}$ and
$\widehat D^{\prime(j)}_{\Gamma}$, there is a contribution from the
term of order $(1/M_Q)^0$.  
In computing the coefficient functions to order
$1/M_Q$ one must not forget graphs with one insertion of the zeroth order
term
 in the current and one insertion of the first order term in the HQET
lagrangian.

\subsec{ Corrections of order $m_c/m_b$}
In the case of semileptonic decays of a beauty hadron to charmed hadron, we
introduced earlier an approximation method (``Method II'' in
Section~\oneloopsec) in
which $m_c/m_b$ was treated as a small parameter. Now, 
$m_c/m_b\sim1/3$
and you may justifiably worry that this is not a
good expansion parameter. We will see in this Section that the corrections
are actually of the order of $\alpha_s/\pi(m_c/m_b)$ and therefore small.
Moreover, they are explicitly calculable.

The strategy is\falkgrinsteinb\ to look at those corrections of order
$1/m_b$ which may be
accompanied by a factor of $m_c$. In the first step of the approximation
scheme we construct a HQET for the $b$-quark, treating the $c$-quark as
light. We must, of course, keep terms of order $1/m_b$ in this first step. The
second step is to go over to a HQET in which the $m_c$-quark is also heavy.
For now, we care only about terms in this HQET that have positive powers of
$m_c$.

In the first step, the hadronic current $\bar c\Gamma b$, with $\Gamma =
\gamma^\mu$ or $\Gamma=\gamma^\mu\gamma_5$, is
replaced according to   Eq.~\fullsubhlcur. The question is, which
terms in Eq.~\fullsubhlcur\ can give factors of $m_c$ when we replace the
$c$-quark by a HQET   quark, $c_{v'}$. Recall that, once we complete the 
second
step, all of the $m_c$ dependence is explicit.  The answer is that any 
operators in
Eq.~\fullsubhlcur\ which have a derivative acting on the $c$-quark will give a
factor of $m_c$. From Eq.~\eVIii\ we see that a derivative $i\partial_\mu$ 
acting
on the charm quark becomes, in the effective theory, the operation
$m_cv'_\mu+i\partial_\mu$. So the
prescription is simple: take $J_\Gamma$ in Eq.~\fullsubhlcur\  and replace
\eqn\replacement{
i\partial_\mu\to m_cv'_\mu
}
in those terms where $i\partial_\mu$ is  acting on the charm quark.

For example, if the operator
\eqn\exampleope{
{1\over m_b} \bar c \,i\!\overleftarrow\Dsl\Gamma b_v
}
is generated at some order in the loop expansion, it gives an operator
\eqn\exampleopegives{
-{m_c\over m_b}\bar c_{v'} \vsl'\Gamma b_v =-
{m_c\over m_b}\bar c_{v'} \Gamma b_v}
after step two is completed.

It is really interesting to note that the resulting correction does not
introduce any new unknown form factors. For example, the matrix element of
 \exampleopegives\ between a $\bar B$ and a $D$ is given by Eq.~\eVii{}\ %
only with an additional factor of $-m_c/m_b$ in front.

The calculation described here has been performed in the leading-log
approximation in Ref.~\falkgrinsteinb. The correction to the vector current
is  \eqn\deltavec{
\Delta V_\mu= {m_c\over m_b} \bar c_{v'}(a_1  \gamma_\mu
+ a_2 v_\mu + a_3 v'_\mu )b_v
}
where the coefficients $a_i=a_i(\mu)$, written in terms of
\eqn\aovera{
z={\bar\alpha_s(m_c)\over\bar\alpha_s(m_b)}
}
are
\eqn\bcoefs{
\eqalign{
a_1 &= {5\over9}(\vvv-1)-{1\over18}z^{-{6\over25}}+
{2\vvv+12\over27}z^{-{3\over25}}-{34\vvv-9\over54}z^{6\over25}
- {8\over25}\vvv z^{6\over25}\ln z\cr
a_2 &= {5\over9}(1-2\vvv)-{13\over9}z^{-{6\over25}}-
{44\vvv-6\over27}z^{-{3\over25}}-{14\vvv-18\over27}z^{6\over25}\cr
a_3 &= {15\over9}-{2\over3}z^{-{3\over25}}-z^{6\over25}\cr}
}
In particular, this gives a contribution to the form factor, at $v=v'$, of
\eqn\bsum{
{m_c\over m_b}(a_1+a_2+a_3)|_{\vvv=1}\simeq .07
}
This is not negligible! It is reassuring that this type of corrections can be
extracted explicitly. On the other hand, it should be remembered that both
corrections of order $(m_c/m_b)^2$ and of subleading-log order can still be
considerable and should be, but have not been, computed.

\subsec{ Corrections of order $\bar\Lambda/m_c$ and 
$\bar\Lambda/m_b$.}
\subseclab\lukesthm
Corrections to the form factors for semileptonic decays of $B$'s and
$\Lambda_b$'s that arise from the terms of order $1/m_c$ in
the effective lagrangian Eq.~\eVIxi\ and the currents Eqs.~\fullsubhlcur\ and
\fullsubhhcur\  are, in principle, as large or larger than those considered in
the previous Section. It is a welcome
surprise that the corrections to the combination of form factors that
contribute to the  semileptonic decay vanish at the endpoint
$\vvv=1$. Thus, the predicted normalization of form factors persists, 
although,
as we will see, not so the relations between form factors.

The decay\ggw\
$\Lambda_b\to\Lambda_c e\nu$ is simpler to analyze than the
decays\luke\ $\bar B\to De\nu$ and
$\bar
B\to D^*e\nu$. Moreover, it turns out that for the baryonic decay some 
relations
between form factors survive at this order. For these reasons, we will present
here the baryonic case. We will briefly return to the decay of the meson at 
the
end of this section, where we will describe the result.

There are two types of corrections to consider\ggw,
coming from either the modified lagrangian of from the modified current. We
start by considering the former. 
The $c_1$ and $c_2$ terms in the effective lagrangian \coefsinlag\ 
transform  trivially under the spin symmetry, contributing to the form
factors in the same proportion as the leading term in Eq.~\baryonvev. This
effectively renormalizes the function $\zeta$ but does not affect relations
between form factors.

Moreover, the normalization at the symmetry point $\vvv=1$ is not affected.
This is a straightforward application of the Ademollo-Gatto theorem. If
$j_\mu$ is a symmetry generating current of a hamiltonian $H_0$, then
corrections to the matrix element of the current, at zero momentum, from
a symmetry-breaking   perturbation to the hamiltonian, $\epsilon H_1$,
are of order $\epsilon^2$.
In the case at hand the Ademollo-Gatto theorem implies that
corrections to the normalization of $\zeta$ at the symmetry point are of order
$(1/m_c)^2$.

The chromomagnetic moment operator in the lagrangian \coefsinlag\  does 
not give
a contribution at all. The spin symmetries imply
\eqn\magneticvev{
\eqalign{
\vev{\Lambda_c(v',s')|{\rm T} &\int d^4\!x\,
(\bar c_{v'}\sigma^{\mu\nu}G_{\mu\nu} c_{v'})(x)
(\bar c_{v'}\Gamma b_v)(0)|\Lambda_b(v,s)}\cr
&= \zeta_{\mu\nu}(v,v')\bar u^{(s')}(v')\sigma^{\mu\nu}\projp\Gamma 
u^{(s)}(v)
.\cr}
}
The function $\zeta_{\mu\nu}$ must be an antisymmetric tensor and must 
therefore
be proportional to $v'_\mu v_\nu-v'_\nu v_\mu$. But
\eqn\magneticzero{
\pprojp\sigma^{\mu\nu}\pprojp v'_\mu=0.}
This, we see, is an enormous simplification. There is no analogous simple 
reason
for the matrix element of the chromomagnetic moment operator to vanish in 
the
case of a  meson  transition. The  chromomagnetic
matrix element gives, in that
case, uncalculable corrections to the relations between form factors.

We turn next to the contribution from the modification to the current. We
need the matrix element of the local operators of order $1/m_c$ in
Eq.~\fullsubhhcur. These operators are all constructed of one derivative
acting on either heavy quark in the quark bilinear.
Consider the matrix element
\eqn\baryonvevb{
\vev{\Lambda_c(v',s')|\bar c_{v'}i\overleftarrow D_\mu \Gamma b_v
|\Lambda_b(v,s)}
= \bar u^{(s')}(v')\Gamma u^{(s)}(v)[Av_\mu + B v'_\mu],}
where the form of the right hand side follows again from the spin
symmetries. The form factors $A$ and $B$ are not independent. Rather,
they are given in terms of $\zeta$. To see this, note that, contracting with
$v'_\mu$ and using the equations of motion,
\eqn\BandA{
B=-\vvv A.}
Also, if the mass of the $l=1/2$ state in the effective theory is
$\bar \Lambda$, then
\eqn\baryonvevc{
\vev{\Lambda_c(v',s')|i\partial_\mu(
\bar c_{v'}\Gamma b_v) |\Lambda_b(v,s)}
=\bar\Lambda(v_\mu-v'_\mu)\zeta(\vvv) \bar u^{(s')}(v')\Gamma u^{(s)}(v).
}
Contracting with $v_\mu$, using the equations of motion and Eq.~\BandA\
we have
\eqn\Aandzeta{
A(1-(\vvv)^2)=\bar\Lambda(1-\vvv)\zeta}
Therefore, the matrix element of       interest is
\eqn\eVIviii{
\langle\Lambda_c(v',s')\vert \bar c_{v'}i \overleftarrow
\Dsl \Gamma b_v\vert\Lambda_b(r,s)\rangle = \bar \Lambda \zeta(v\ccdot
v'){v_\nu-(v\ccdot v')v'_\nu\over 1+v\ccdot v'}\bar 
u^{(s')}(v')\gamma^\nu\Gamma
u^{(s)}(v)
}
 where $\Gamma =\gamma^{\mu}$ or $\gamma^{\mu}\gamma_5.$ 
Putting it all together, using the leading log expression for
the coefficients $\widehat D^{(j)}_{\Gamma}$ in
the current of Eq.~\fullsubhhcur, one finds
\eqn\baryonffs{
\eqalign{
F_1&=G_1\left[1+{\bar \Lambda
\over m_c}\left({1\over 1+v\ccdot v'}\right)\right]\cr
F_2&=G_2=-G_1{\bar \Lambda \over m_c}\left({1 \over 1+v \ccdot
v'}\right)\cr
  F_3&=G_3=0\cr}
  }
Moreover,
\eqn\baryonffnorm{
G_1(1)=\left({\bar\alpha _s(m_b) \over \bar\alpha _s(m_c)}\right)^{a_I}
}
as before. Up to an unknown constant, $\bar \Lambda$, there are still
five relations among six form factors. We can estimate $\bar \Lambda$ by
writing $\bar \Lambda =M_{\Lambda_c}-m_c=(M_{\Lambda_c}-M_D)+(M_D-
m_c)$. If
the `constituent' quark mass in the $D$ meson is $\simeq 300MeV$, then 
$\bar
\Lambda \simeq 700MeV$.  With this, we can estimate the next order
corrections to be of the order of $\left({\bar \Lambda /
2m_c}\right)^2\sim 5\%$. There are, of course, additional computable
corrections, of order ${\bar\Lambda/2m_b}$ and
${\alpha_s(m_c)/\pi}\left({\bar\Lambda/2m_c}\right)$.\chog

The result of ${1 / m_c}$ corrections to the mesonic transitions is
quite different. There both the matrix elements of the correction to the
current and of the time order product with the chromo-magnetic moment
operator lead to new form factors. The result is that there are
incalculable corrections, of order $\bar\Lambda/2m_c$, to all the leading 
order
relations between form factors. Even if $\bar \Lambda$ is smaller in this 
case,
presumably $\bar\Lambda\sim300$MeV, these corrections may be large, say
10\%--20\%. Remarkably, at the symmetry point, $v'\ccdot v=1$, there are no
corrections of order $\bar\Lambda/2m_c$ to the leading order predictions.
Thus, one may still extract the mixing angle $\vert V_{cb}\vert$ with high
precision from measurements at the
 end of the
spectrum of the semileptonic decay rates for $B\rightarrow De\nu$ and
$B\rightarrow D^{\ast}e\nu$.

\newsec{Inclusive Semileptonic $B$-Meson Decays.}

Well before the development of HQETs, it was commonly held that the
inclusive rate of decay of a heavy meson should be well approximated by
the decay rate of the heavy quark. Intuitively, the heavy quark just
sits at rest (in the rest frame of the meson) surrounded by `brown muck'
for which it acts as a static source of color. When the heavy quark
decays, the complications of strong interactions come in, dictating how
the rate is divided up between exclusive channels, but the sum total must
be the quark's rate of decay.

With the advent of HQET and HQ symmetries one may prove the validity of
this assertion. Moreover, and more importantly, this is done by setting
up an expansion for the decay rate in inverse powers of the heavy mass.
With a systematic expansion one may investigate the accuracy of the
result for the not-really-so-heavy $D$ and $B$ mesons.

\subsec{Kinematics.}
The decay rate is
\eqn\inclusiverate{
d\G={\kappa\over2M}\sum_X|\vev{X|j_\mu|\bar B}\ell^\mu|^2 d\Phi_X
}
where $j_\mu$ is the hadronic $\Delta B=-1$ current,
\eqn\phspdefd{
d\Phi_X\propto\delta^{(4)}(P-p_X-p_e-p_{\bar\nu})
        \;\Pi_i{d^3\!p_i\over2E_i}
}
is the phase space for final states and
\eqn\ellmu{
\ell_\mu =  \bar v(p_\nu)\g_\mu(1-\g_5) u(p_e)
}
is the lepton current.
The factor $\kappa$ includes the coupling constants $G_F^2|V_{qb}|^2$.

All of the interesting dynamics is in the hadronic tensor
\eqn\hmunudefd{
h_{\mu\nu} = \sum_X (2\pi)^3\delta^{(4)}(P-p_x-q)
                \vev{\bar B|j^\dagger_\nu(0)|X}
                \vev{X|j_\mu(0)|\bar B}
}
where the sum symbol includes the integral over phase space for the
final state $X$.

We will investigate the properties of the hadronic tensor $h_{\mu\nu}$
indirectly, by studying the related tensor
\eqn\tmunudefd{
T_{\mu\nu}= i \int d^4\!x\, e^{-iq\cdot x}\vev{\bar B(p)|
                T(j^\dagger_\nu(x)j_\mu(0))|\bar B}
}
Both tensors can be written in terms of Lorentz invariant form factors,
thus:
\eqna\handtffs
$$\eqalignno{
h_{\mu\nu}&=g_{\mu\nu}h_1+p_\mu p_\nu h_2 +\cdots &\handtffs a\cr
T_{\mu\nu}&=g_{\mu\nu}T_1+p_\mu p_\nu T_2 +\cdots &\handtffs b\cr}
$$
\exercise{I have intentionally left the list of form factors incomplete.
The ellipsis in Eqs.~\handtffs\null\ stand for a finite number of terms.
Ennumerate them. There is one term that  can be eliminated by the
requirement of time reversal invariance of the strong interactions.
Which one is it?}

The form factors $T_i$ are more amenable to theoretical investigations
because $T_{\mu\nu}$ is a Green function. The $h_i$ form factors can be
obtained from these by cutting:
\eqn\hfromt{
h_i={1\over\pi}\im T_i\qquad\hbox{for}\qquad p\cdot q<m_B^2
}

\exercise{To see this,
\item{(i)}Separate the integral in Eq.~\tmunudefd\ into $x^0>0$ and
$x^0<0$ parts. Insert a complete set of states between the two currents,
and make the dependence on $x$ explicit.
\item{(ii)}Carry out the integrations, using, eg,
$$\int d^4\!x\;\theta(x^0)e^{i(P-p_X-q)\cdot x}
=(2\pi)^3\delta(\vec P-\vec p_X-\vec q){i\over P^0-p^0_X-q^0+i\epsilon}
$$
and use the standard identity
$${1\over z\pm i\epsilon} = \hbox{PP}{1\over z} \mp i \pi\delta(z)~,
$$
where PP stands for the principal part, to separate ``imaginary'' from
``real'' parts (these are in quotation marks because by them I just mean
the parts that come from the PP terms and from the $\delta$ term,
respectively).
\item{(iii)}In the ``imaginary'' part you will recognize one term is
just $h_{\mu\nu}$. Write out explicitly the other term in the
``imaginary'' part and show that it vanishes if $P\cdot q< m_B$. (Hint:
What is the lowest mass of the intermediate state X?).
\item{(iv)}Complete the proof by relating form factors and taking, where
appropriate, the real or imaginary parts.
\item{ }{ }
\vskip-8pt
}

\nfig\fiveone{The location of the physical and resonant regions in
the complex  z-plane. The variable z is the energy of the final hadronic
system in the semileptonic $B$ decay, in the $B$ restframe.}
\IFBIG{\null}%
{\INSERTFIG{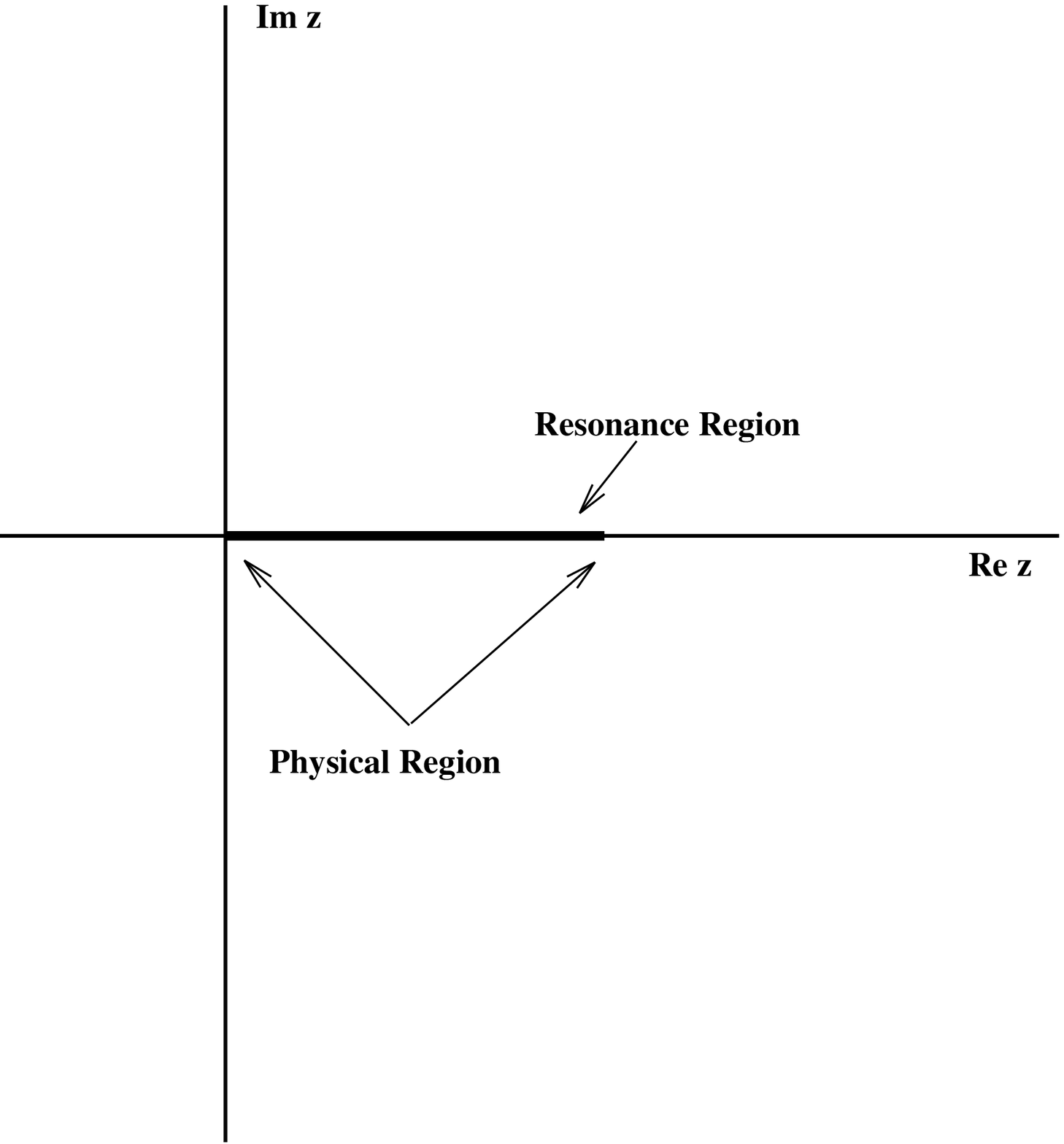}\fiveone{The location of the physical and resonant
regions in the complex z-plane. The variable z is the energy of the
final hadronic system in the semileptonic $B$ decay, in the $B$
restframe.}}

\subsec{The Analytic Structure of The Hadronic Green Function}
The hadronic form factors $T_i$ and $h_i$ are Lorentz invariant
functions of the momenta $P$ and $q$. There are only two invariant
variables, namely $\hat Q\equiv\sqrt {q^2/m_B^2}$ and $z\equiv P\cdot
q/m_B^2$. Here and throughout a hat on a dimensionful quantity means
taking out the dimensions by apropriate powers of $m_B$. We
will study the behaviour of $T_{\mu\nu}$ for fixed $\hat Q^2$ as a function
of $z$. This variable is of interest because $1-z$ measures how far one is
from the resonant region. To see this, consider any final state $X$
contributing to the inclusive decay, so $q=P-p_X$. Then $z=1-P\cdot
p_X/m_B^2$, or, in  the rest frame of the decaying meson, $1-z=E_X/m_B$.
Only states with small energy contribute to the region of $z\approx1$,
and only single resonances contribute when $z$ is large enough.

\exercise{Show that the physical region extends from $z=0$ (neglect the
lepton mass) to $$z=\half(1+\hat Q^2-\hat m_\pi^2)\equiv z_0$$
}

Since $T_{\mu\nu}$ is a Green function we can study its behaviour as a
function of complex momenta. In particular we can hold $\hat Q$ fixed
and study $T_i(z)$ as functions of complex $z$. What do we know about
the complex $z$ plane? You established above that the segment of the
real line $(0,z_0)$ corresponds to the physical region for the
inclusive semileptonic $\bar B$ decay. And we know that the right end
of the segment corresponds to the resonance region.
See Fig.~\xfig\fiveone.

\nfig\fivetwo{The location of the  cuts in the complex z-plane.}

\IFBIG{\INSERTFIG{5_1.eps}\fiveone{The location of the physical and resonant
regions in the complex z-plane. The variable z is the energy of the
final hadronic system in the semileptonic $B$ decay, in the $B$
restframe.}}{\INSERTFIG{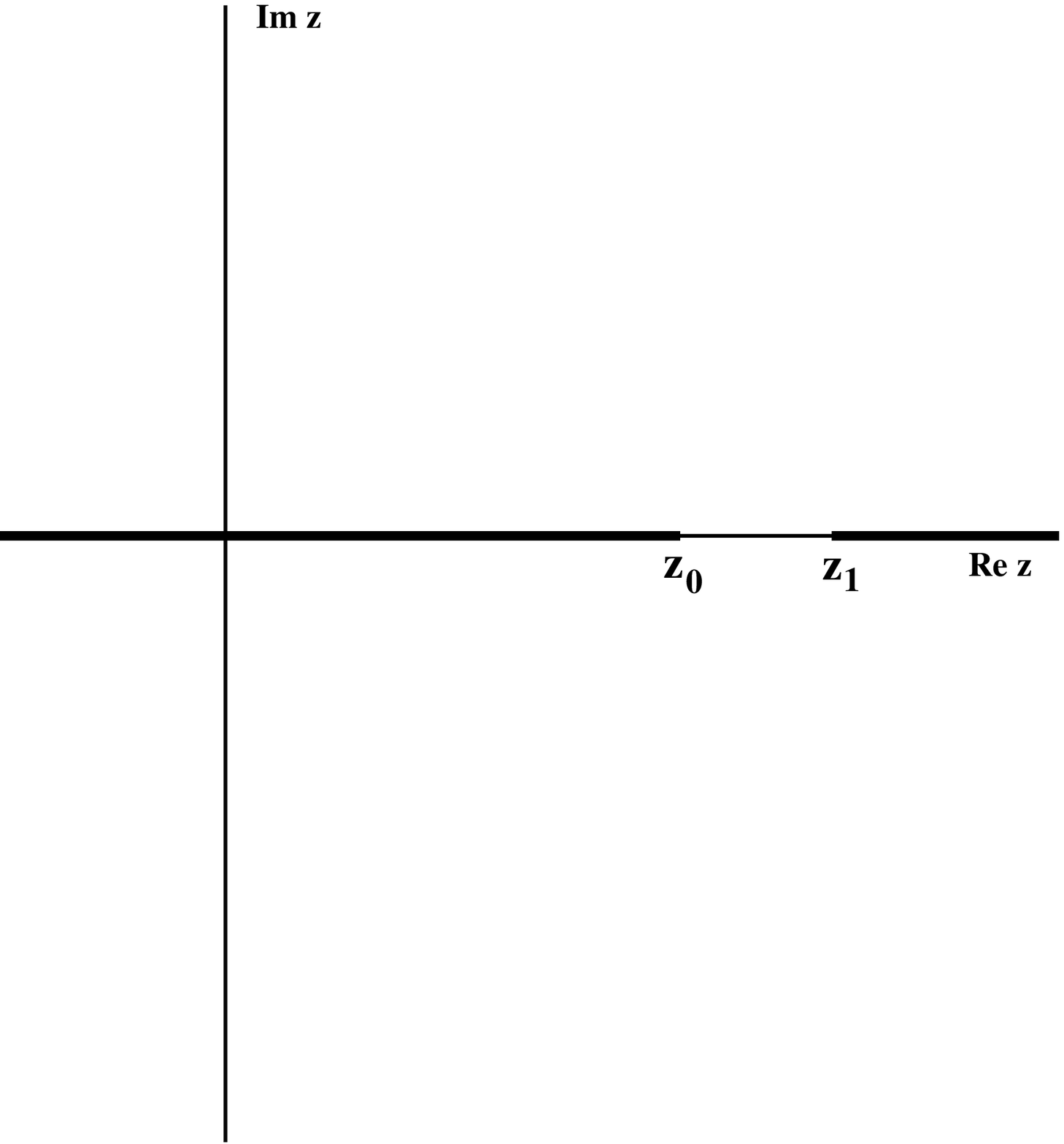}\fivetwo{The location of the  cuts
in the complex z-plane.}}%

Associated with the physical process there is a cut in the $z$ plane.
The cut extends from $z_0=\half(1+\hat Q^2-\hat m_\pi^2)$ to $-\infty$
on the real axis. There is, in addition, a cut that extends from
$z_1=\half(3-\hat Q^2)$ out to $+\infty$ on thereal axis. See
Fig.~\xfig\fivetwo.

\exercise{What are the physical processes that correspond to the cuts in
the z plane?}

If the presence and location of these cuts does not seem obvious, the
reader may use Landau conditions\foot{See, for example,
Ref.~\bjdrell}\ to determine the analytic properties of a Green
function in perturbation theory. She will then find the same cuts,
save for the fact that the masses will turn out to be those of quarks
rather than of the physical bound states.
\IFBIG{\INSERTFIG{5_2.eps}\fivetwo{The location of the  cuts
in the complex z-plane.}}{\null}

As will become evident later, a perturbative computation will be
trustworthy provided $z$ is nowhere near the resonance region.
We should stay away from the resonance region by at least a
typical hadronic scale, say the rho mass, $m_\rho$, or the chiral
symmetry breaking scale, $\Lambda_\chi$. This is why the problem at hand
is non-trivial: to compute the form factors $h_i$ from the imaginary
part of a Green function one needs to deal with the resonant region
where perturbation theory is not valid.

The situation is much better if we are willing to settle for slightly
less, namely for averages over the variable $z$ of the decay rate (at
fixed $\hat Q$).   Integrating both sides of Eq.~\hfromt\ one has
\eqn\inthtof{ \eqalign{
\int_0^{z_{0}} dz\; h_i(z)&= \int_0^{z_{0}}dz\; {1\over\pi} \im T_i(z)\cr
                          &={1\over\pi}\int_C dz\; T_i(z)\cr}
}
\nfig\fivethree{Contour of integration, C, in the dispersion
relation \inthtof.}
where the contour C runs above the cut from 0 to $z_{0}$, around the
branch point and back to 0 under the cut; see
Fig.~\xfig\fivethree.

\INSERTFIG{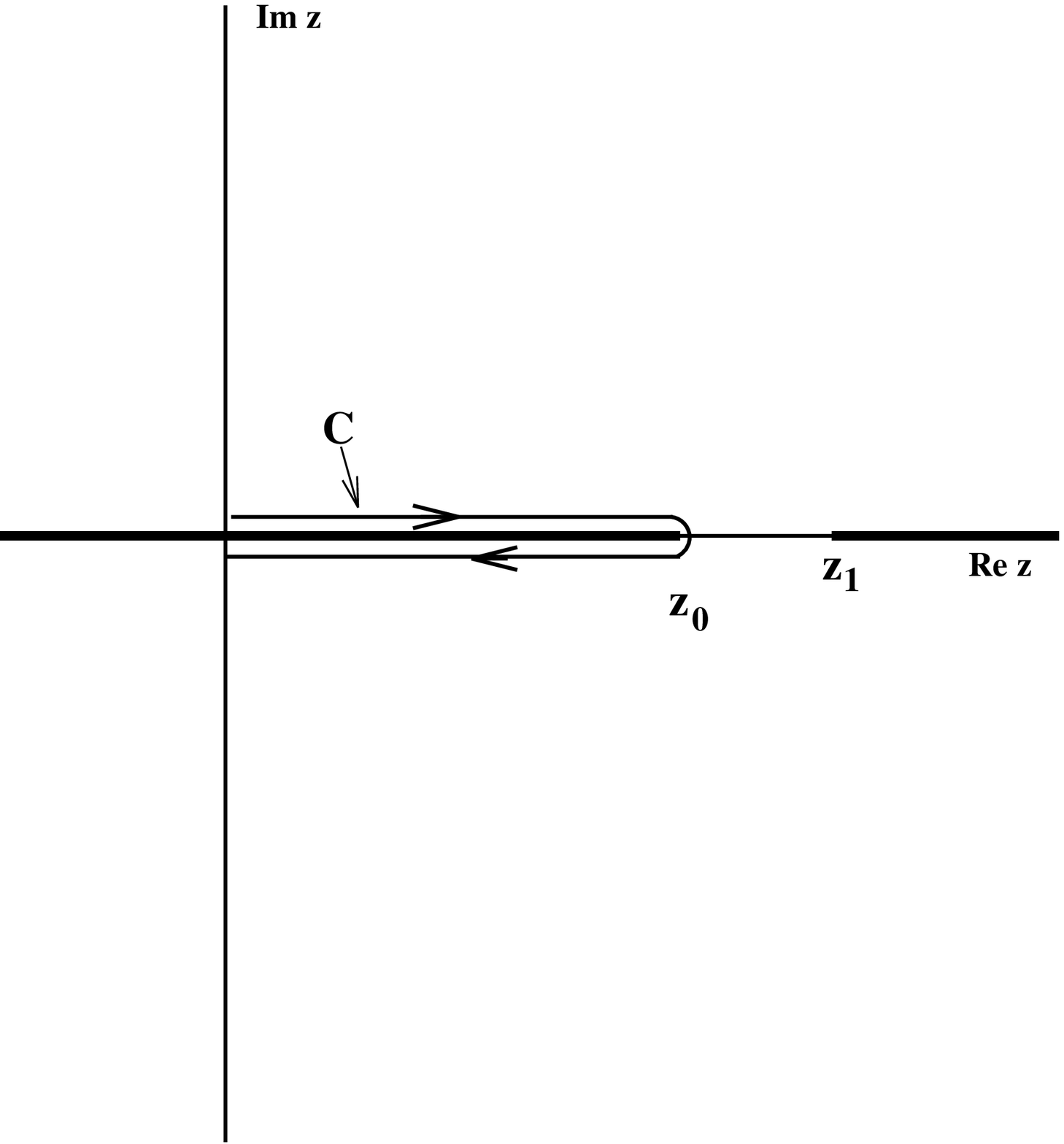}\fivethree{Contour of integration, C, in the dispersion
relation \inthtof.}

\nfig\fivefour{Deformed contour of integration, C', in the dispersion
relation \inthtof. Note that it stays away from the physical region
everywhere except near z=0, provided $Z_0$ and $z_1$ are well
separated.} We can trade the contour C for a new one C' that stays
away from the resonance region. C' is restricted to start just above
the origin of the complex z plane and to finish just under it. This is
justified because there are no singularities other than the cuts.  See
Fig~\xfig\fivefour.  Provided $z_0$ and $z_1$ are well separated there
is no problem getting the contour C' to lie well away from the cut,
exept in the vicinity of $z=0$. But in particular one can have the cut
away from the resonance region in its entirety, and one may apply
perturbation theory to the computation of the Green function.

\exercise{Under what circumstances can the cuts meet? In other words,
determine the kinematic regime for which $z_1 < z_0$. What does this
imply for inclusive $b\to c e\bar\nu$ and $b\to u e\bar\nu$?
}

The only thing missing is a means of computing the Green function
perturbatively. The problem is that we need to compute a matrix element
between meson states, and this usually involves complicated dynamics. As
we will see in the next section it is possible to use the HQET and the
HQ symmetries to compute the matrix element. Before turning to that
problem, we conclude with a simple observation that extends the above
result. The average in Eq.~\inthtof\ uses one of many possible measures,
namely simply $dz$. One could just as well consider
\eqn\avggnrlzd{
\int_0^{z_0} dz \;f(z) h_i(z)
}
Clearly we want the function $f(z)$ to be regular on the region of
integration. More generally we would like its singularities in the
complex z plane to be as far away from the contour C' as possible. This
suggest having them at infinity, ie, using an entire function.
Particularly interesting is the set of functions $f_n(z)=(z_0-z)^n$,
which measure the relative importance of the non-resonant and resonant
regions.

\INSERTFIG{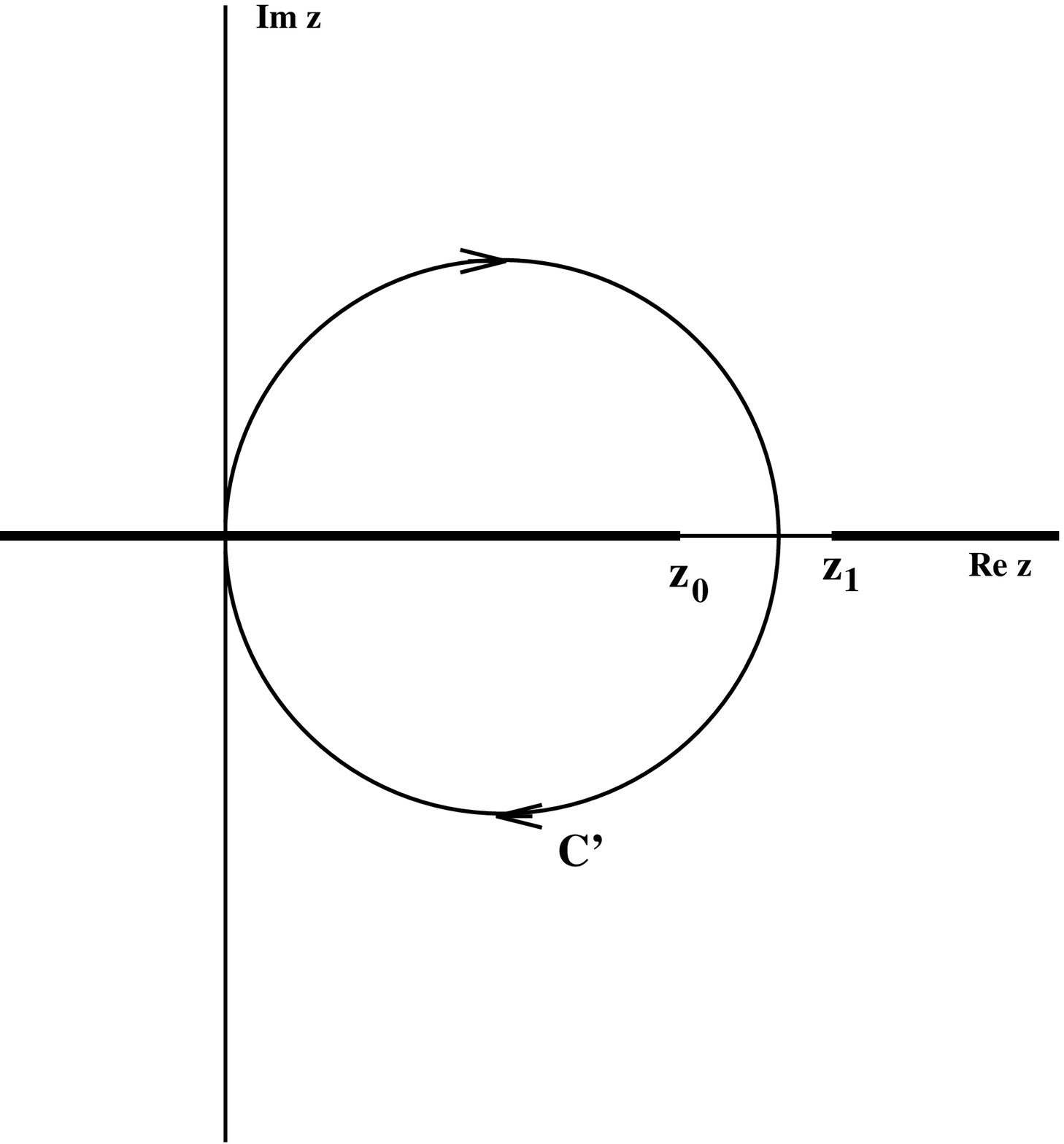}\fivefour{Deformed contour of integration, C', in
the dispersion relation \inthtof. Note that it stays away from the
physical region everywhere except near z=0, provided $Z_0$ and $z_1$
are well separated.}

\subsec{An HQET based OPE}
The Operator Product Expansion (OPE) has been used with great success to
separate perturbative from non-perturbative effects in deep inelastic
nucleon scattering. The coefficient functions may be computed
perturbatively, while the non-perturbative information is encoded in the
matrix elements of the operators.

The common OPE cannot be applied to $T_{\mu\nu}$. The problem is that
it is an expansion in inverse powers of $Q$. The matrix elements of
operators of heavy mesons can in fact grow with powers of the heavy
mass $m_B$. Since in the physical region $Q<m_B$, the ratio~$m_B/Q$ is
not a good expansion parameter.

But there is a way one can produce a useful OPE for $T_{\mu\nu}$. The
idea is simple: expand in inverse powers of the largest scale around,
namely $m_B$. Clearly we should be using the techniques of the HQET
to produce such an expansion. Let $P=m_Bv$. Consider any matrix
element (physical or unphysical) of the time ordered product
$T(j^\dagger_\nu(x)j_\mu(0))$. The idea is to take the momentum of the
$b$ quark to be $p=m_bv+k$ and to expand in inverse powers of $m_b$.

\nfig\fivefive{Tree level Green function for the product of currents.}
Consider, in particular, the matrix element between $b$ quarks of
momentum~$p$. The tree level Feynman graph is in Fig.~\xfig\fivefive.
We insist that the momentum in the intermediate light quark, $q=u$ or
$c$, be $m_bv+\cdots$ and expand in inverse powers of the mass.
Arbitrary quantities, like $Q^2$, may scale with $m_b$ (so we hold,
eg, $\hat Q^2$ fixed as $m_b\to\infty$).

\INSERTFIG{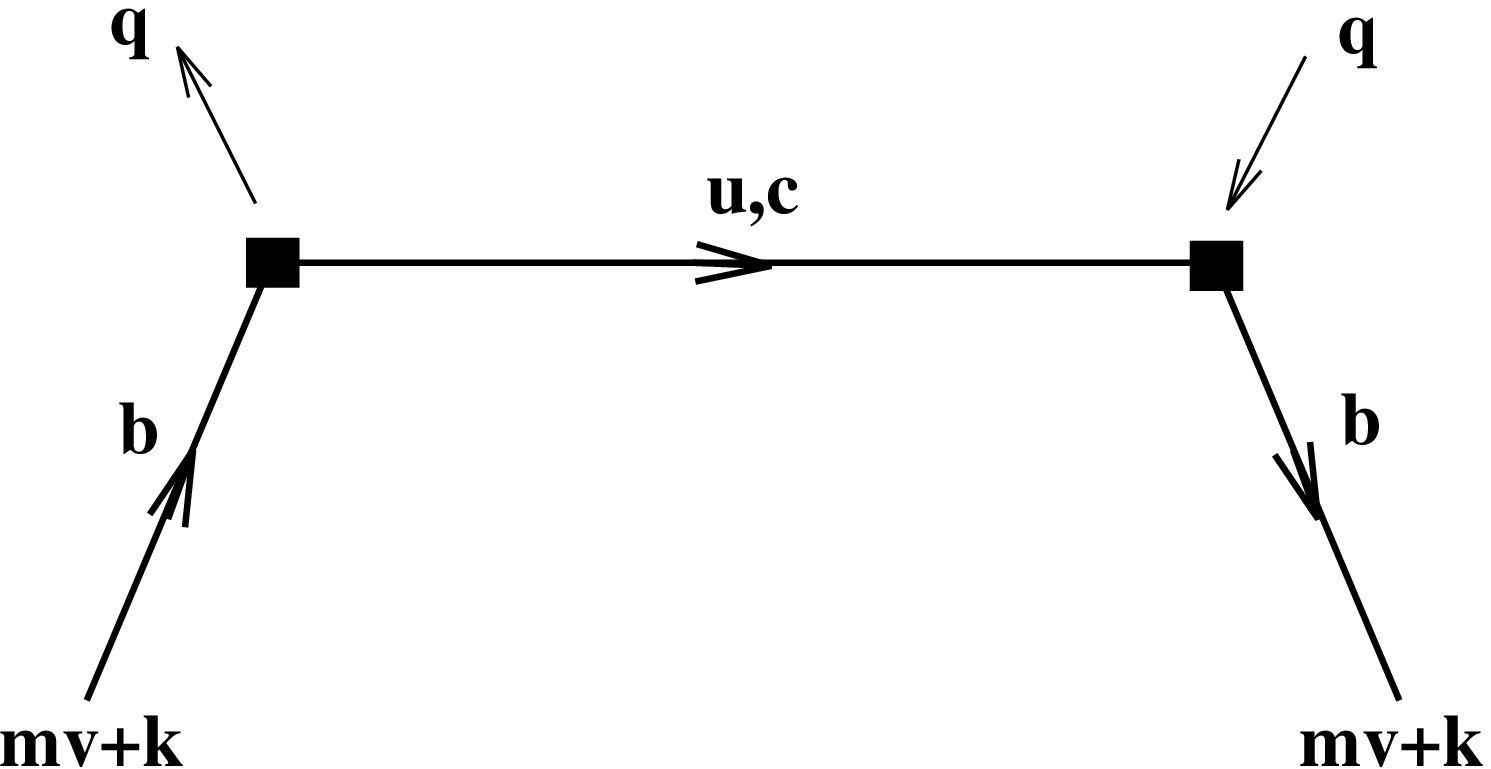}\fivefive{Tree level Green function for the product
of currents.}

To see how the expansion works, let us compute the Feynman diagram of
Fig.~\xfig\fivefive. To be concise we consider the current $j_\Gamma=\bar
q\Gamma b$, where $\Gamma$ stands for any Dirac gamma matrix. Then we
have the tree level Green function and its expansion:
\eqn\opetree{\eqalign{
\Gamma &{i\over m_b \vsl +\spur k -\spur q - m_q}\Gamma =
{1\over m_b}\Gamma{i\over \vsl-\spur{\hat q}-\hat m_q +
\spur k/m_b}\Gamma\cr
&={1\over m_b}\Gamma{i\over \vsl-\spur{\hat q}-\hat m_q}\left[
1-{ \spur k\over m_b}{1\over \vsl-\spur{\hat q}-\hat m_q}+\cdots
\right] \Gamma \cr}
}
This matrix element can then be used to determine the coefficients of
operators in  an OPE-like relation between operators:
\eqn\hqetope{\eqalign{
\int d^4\!x\; &e^{-iq\cdot x} T(\bar b\Gamma q(x)\; \bar q \Gamma b(0)) \cr
&=
{1\over m_b}\left[\bar b_v\Gamma{i\over \vsl-\spur{\hat q}-\hat
m_q}\Gamma b_v -
{1\over m_b}\bar b_v\Gamma{i\over \vsl-\spur{\hat q}-\hat
m_q}i\Dsl{1\over \vsl-\spur{\hat q}-\hat m_q}\Gamma
b_v+\cdots\right]\cr}
}
where the ellipsis stand for an expansion in powers of $1/m_b$ which has
operators of ever increasing dimensions. Going beyond tree level will
modify the specific coefficient functions but the form of the expansion
will remain practically the same (some new operators may be introduced
beyond tree level).

\nfig\fivesix{Representation of the combined OPE-HQET at tree level.}

\INSERTFIG{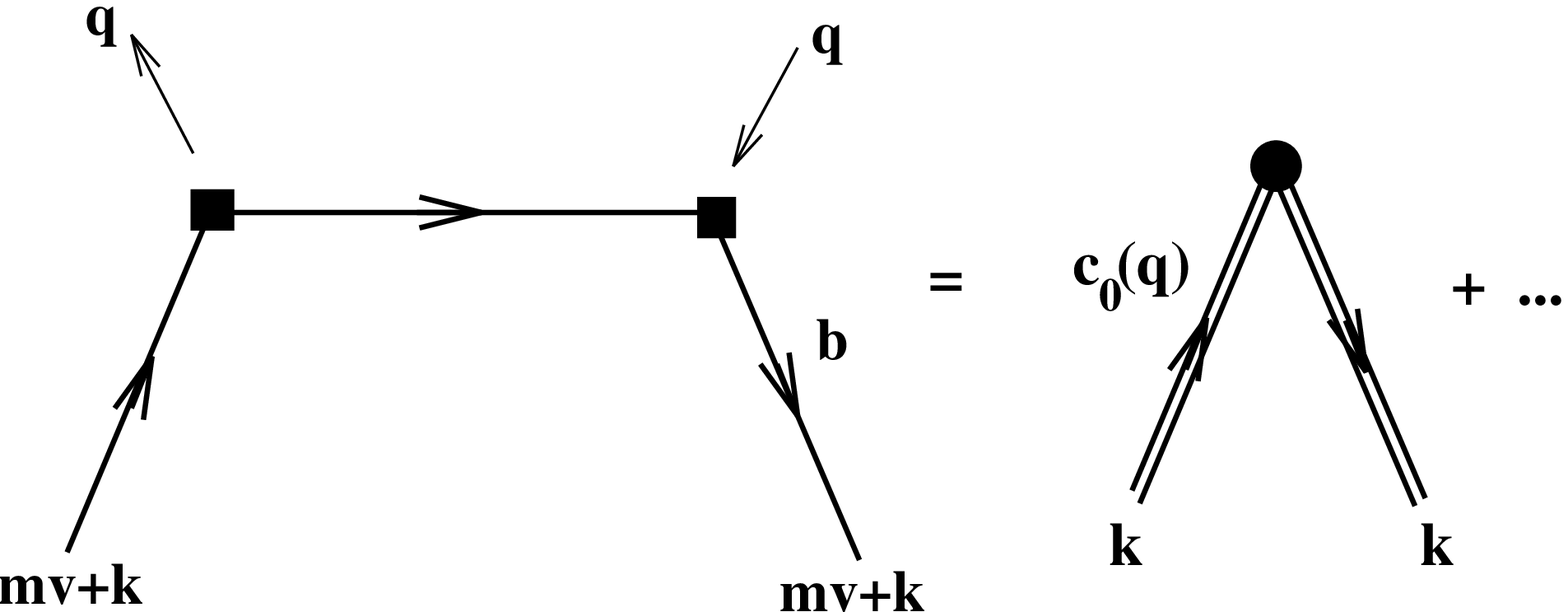}\fivesix{Representation of the combined OPE-HQET at
tree level.}

Note that the right hand side of Eq.~\hqetope\ is written in terms of
the quark operators of the HQET, while the left hand side involves the
original quark fields. This is what makes this OPE different from the
common one (used, say, in deep inelastic scattering). And there is no
way out of it: since we are expanding the full theory time ordered
product in powers of $1/m_b$, we must use HQET fields for the result.
A useful pictorial representation of this is presented in Fig.~\xfig\fivesix,
which shows the HQET-OPE at tree level. The diagrams on the right hand
side of the equation have double lines, reminding us they represent
quarks in the HQET.

The coefficients in the OPE of Eq.~\hqetope\ have poles in $z-z_0$ of
increasingly higher order. Symbolically we may write for the expansion
of the operator product
\eqn\opepoles{
{1\over m_b}\sum_n{1\over(2m_b)^n}c_n(\hat Q^2,z) \CO^{(n)}
}
with coefficient functions that behave as 
\eqn\csgolike{
c_n\sim{1\over(z-z_0)^{n+1}}
}
Beyond tree level there will be cuts in additions to these poles.
Although the nature of the singularities change, the discussion that
follows remains essentially the same. 

The first two terms in the OPE expansion of $T_{\mu\nu}$ can
be computed! The first one involves the matrix element between $\bar
B$-mesons of a dimension-3 operator. But this is fixed by spin/flavor
symmetries:
\eqn\oneopdone{
\vev{\overline B|\bar b_v\G(\vsl+\spur{\hat q}+\hat m_q)\G b_v|\overline B} =
-\xi(1)
\tr\overline{\widetilde B}(v)\G(\vsl+\spur{\hat q}+\hat m_q)\G \widetilde B(v)
}
with $\xi(v\cdot v')$ the Isgur-Wise function, $\xi(1)=1$.  The second
one involves a dimension-4 operator. The matrix element of this vanishes:
\eqn\twoopdone{
\vev{\overline B|\bar b_v\G D_\mu b_v|\overline B} =0
}
There are therefore no corrections of order $1/m_b$. More on this
below.

\exercise{Prove Eq.~\twoopdone. {\sl Hint:\/ }Consult section
\lukesthm.}

We see that one can compute the rate for inclusive semileptonic $B$
decay. With the tree level coefficient functions just computed, 
the matrix elements in Eqs.~\oneopdone\ and~\twoopdone, and neglecting
higher order terms in $1/m_b$, one obtains for the decay rate the same
result as if computing the decay rate of a free b quark!

\exercise{Prove this assertion. In fact, we can  do even better because we
need not integrate over~$Q^2$. Prove that the moments
$$\int_0^{z_0}dz\;(z_0-z)^n 
{d\G(B\to Xe\nu)\over {\rm d}z\;{\rm d}Q^2} $$
are the same as for the free quark decay. {\bf This is an important
exercise.} You will have to put together all the pieces of the
argument discussed above, and then more.}

The deviations from the free decay rate are parameterized by the
matrix elements of the operators of dimension 5 and higher. In this
sense we have managed to separate long from short distance effects.

Note that for $z$ on the real axis, the expansion breaks down as $z\to
z_0$. More precisely, in this limit the coefficients $c_n$ of
Eq.~\opepoles\ grow arbitrarily large. This is in accord with the
statement we had made previously that the perturbative calculation of
the Green function breaks down in the resonant region, $z\sim z_0$. 

A physically interesting choice of the measure $f$ in Eq.~\avggnrlzd\
is $f_1=z-z_0$, because it measures the deviation from the free quark
result. The quantity $m_b^2 f_1(v\cdot q/m_b)$ is the invariant
mass-squared of the hadronic final state minus $m_q^2$. If the final
state quark produced by the $b$ decay is on its mass shell, then
$f_1=0$.

\exercise{Prove that with $f_1=z-z_0$ the contribution from the parton
model to the average in Eq.~\avggnrlzd\ vanishes. Discuss corrections
to this result arising from the nonperturbative terms and from the
perturbative corrections to the coefficients in the expansion of
Eq.~\opepoles.}

\exercise{$|V_{ub}|$ is determined from a measurement of inclusive
charmless semileptonic $B$ decay. Since $V_{ub}|\ll |V_{cb}|$, the
rate for $B\to Xe\bar\nu$ is dominated by the decay into charm. Short
of reconstructing the hadronic state $X$, one is forced to consider
the spectrum of electron energies, $E_e$. Find the maximum electron
energies, $E_e^{c,{\rm max}}$ and $E_e^{u,{\rm max}}$, in the decay
into charm and charmless final states, respectively.  What does the
region $E_e^{c,{\rm max}} < E_e <E_e^{u,{\rm max}}$, correspond to
in terms of our variables $z$ and $Q$?}

\newsec{Chiral Lagrangian with Heavy Mesons}

\subsec{Generalities}

Chiral symmetry and soft pion theorems have been used in particle
physics for several decades now with great success. The most efficient
way of extracting information from chiral symmetry is by writing a
phenomenological lagrangian for pions that incorporates both the
explicitly realized vector symmetry and the non-linearly realized
spontaneously broken axial symmetry.\georgibook\ Theorems that
simultaneously use heavy quark symmetries and chiral symmetries are
most expediently written by means of a phenomenological lagrangian for
pions and heavy mesons that incorporates these
symmetries.\refs{\chiral,\yanetal}

In the limit $m_b\to\infty$, the $\ol B$ and the $\ol B^*$ mesons are
degenerate, and to implement the heavy quark symmetries it is
convenient to assemble them into a ``superfield'' $H_a(v)$:
\eqn\eqsuperf{
H_a(v)={1+\vslash\over2}\left[
    \ol B_a^{*\mu}\g_\mu-\ol B_a\g^5\right]\,.}
Here $v^\mu$ is the fixed four-velocity of the heavy meson, and $a$ is
a flavor $SU(3)$ index corresponding to the light antiquark. Because
we have absorbed mass factors $\sqrt{2m_B}$ into the fields, they have
dimension 3/2; to recover the correct relativistic normalization, we
will multiply amplitudes by $\sqrt{2m_B}$ for each external $\ol B$ or
$\ol B^*$ meson.

The chiral lagrangian contains both heavy meson superfields and
pseudogoldstone bosons, coupled together in an $SU(3)_L\times SU(3)_R$
invariant way.  The matrix of pseudogoldstone bosons appears in the
usual exponentiated form $\xi=\exp(\imi{\cal M}/f)$, where
\eqn\calmeq{
    {\cal M}=\pmatrix{{\textstyle{1\over\sqrt{2}}}\pi^0+{\textstyle
    {1\over\sqrt{6}}}\eta&\pi^+&K^+\cr\pi^-
    &{\textstyle{-{1\over\sqrt{2}}}}\pi_0+
    {\textstyle{1\over\sqrt{6}}}\eta&K^0\cr K^-&\overline K^0&-
    {\textstyle\sqrt{2\over 3}}\,\eta}\,,}
and $f$ is the meson decay constant.  The bosons couple to the heavy
fields through the covariant derivative and axial vector field,
$$
\eqalign{ D_{ab}^\mu&=\delta_{ab}\del^\mu+V_{ab}^\mu
=\delta_{ab}\del^\mu+\half\left(\xi^\dagger
\del^\mu\xi+\xi\del^\mu\xi^\dagger\right)_{ab}\,, \cr
A_{ab}^\mu&={\imi\over 2}\left(\xi^\dagger\del^\mu\xi
-\xi\del^\mu\xi^\dagger\right)_{ab} =-{1\over f}\partial_\mu{\cal
M}_{ab}+{\cal O}({\cal M}^3)\,.}$$
Lower case roman indices correspond to flavor $SU(3)$. Under chiral
$SU(3)_L\times SU(3)_R$, the pseudogoldstone bosons and heavy meson
fields transform as 
$$
\eqalign{\xi&\to L\xi U^\dagger=U\xi R^\dagger\cr
A^\mu&\to UA^\mu U^\dagger\cr
H&\to HU^\dagger\cr
(D^\mu H)&\to (D^\mu H)U^\dagger\cr}
$$
where the matrix $U$ is a nonlinear function of the
pseudogoldstone boson matrix ${\cal M}$, implicitly defined by the
transformation law for $\xi$. Then $\Sigma=\xi^2$transforms as follows: 
$$\Sigma\to L\Sigma R^\dagger~.$$

The effective lagrangian is an expansion in derivatives and in inverse
powers of the heavy quark mass.  The kinetic energy terms take the
form 
$$
{ {\cal L}_{\rm kin}={1\over8}
f^2\,\partial^\mu\Sigma_{ab}\,\partial_\mu \Sigma^{\dagger}_{ba}
-\Tr\left[\overline H_a(v)i v\cdot D_{ba} H_b(v)\right]\,.} $$
Here the trace is in the space of $4\times 4$ Dirac matrices that
define the ``superfields'' $H_a(v)$ in Eq.~\eqsuperf .  The leading
interaction term is of dimension four,
\eqn\eqgterm{
	{\cal L}_{\rm int}= 
    g\,\Tr\left[\ol H_a(v)H_b(v)\,\Aslash_{ba}\g_5\right]\,,
} 
where $g$ is an unknown parameter, of order one in the constituent
quark model. The analogous term in the charm system is responsible for
the decay $D^*\to D\pi$. Expanding the term in the lagrangian
in~\eqgterm\  to linear order in the Goldstone Boson fields,
${\cal M}$, we find the explicit forms for the $D^*D{\cal M}$ and
$D^*D^*{\cal M}$ couplings
\eqn\eqgexpand{
\left[\left({-2g\over f}\right)D^{*\nu}\del_\mu{\cal M} D^\dagger +
{\rm h.c.}\right] + \left({2gi\over f}\right)\emnlk
D^{*\mu}\del^\nu{\cal M} D^{*\lambda} v^\kappa\,.
}
Using this one can compute the partial width
$$\eqalign{
\Gamma(D^{*+}\to D^0\pi^+) &= {g^2\over6\pi f^2}|\vec p_\pi|^3 \cr
\Gamma(D^{*+}\to D^+\pi^0) &= {g^2\over12\pi f^2}|\vec p_\pi|^3
} $$
The ACCMOR collaboration has reported an upper limit of 131~KeV on the
$D^*$ width.\accmor\ The branching fractions for $D^{*+}\to
D^0\pi^+$ and $D^{*+}\to D^+\pi^0$ are $(68.1\pm1.0\pm1.3)$\% and
$(30.8\pm0.4\pm0.8)$\%, respectively, as measured by the CLEO
collaboration.\cleobfs\ Using $f=130$~MeV, one obtains the
limit $g^2<0.5$. Even if the $D^*$ decay width is too small to
measure, radiative $D^*$ decays provide an indirect means for
determining the coupling $g$, and provide a lower bound
$g^2\gs0.1$.\jimcho

Since charmed and beauty baryons are long lived, one can write down
phenomenological lagrangians for their interactions with pions. These
are as well justified and should be as good an approximation as the
lagrangian for heavy mesons discussed above. The treatment is rather
similar, and we refer the interested reader to the
literature.\yanetalbaryons

\nfig\polediagfig{Feynman  diagrams for $B\to D\pi e\nu$}
\INSERTFIG{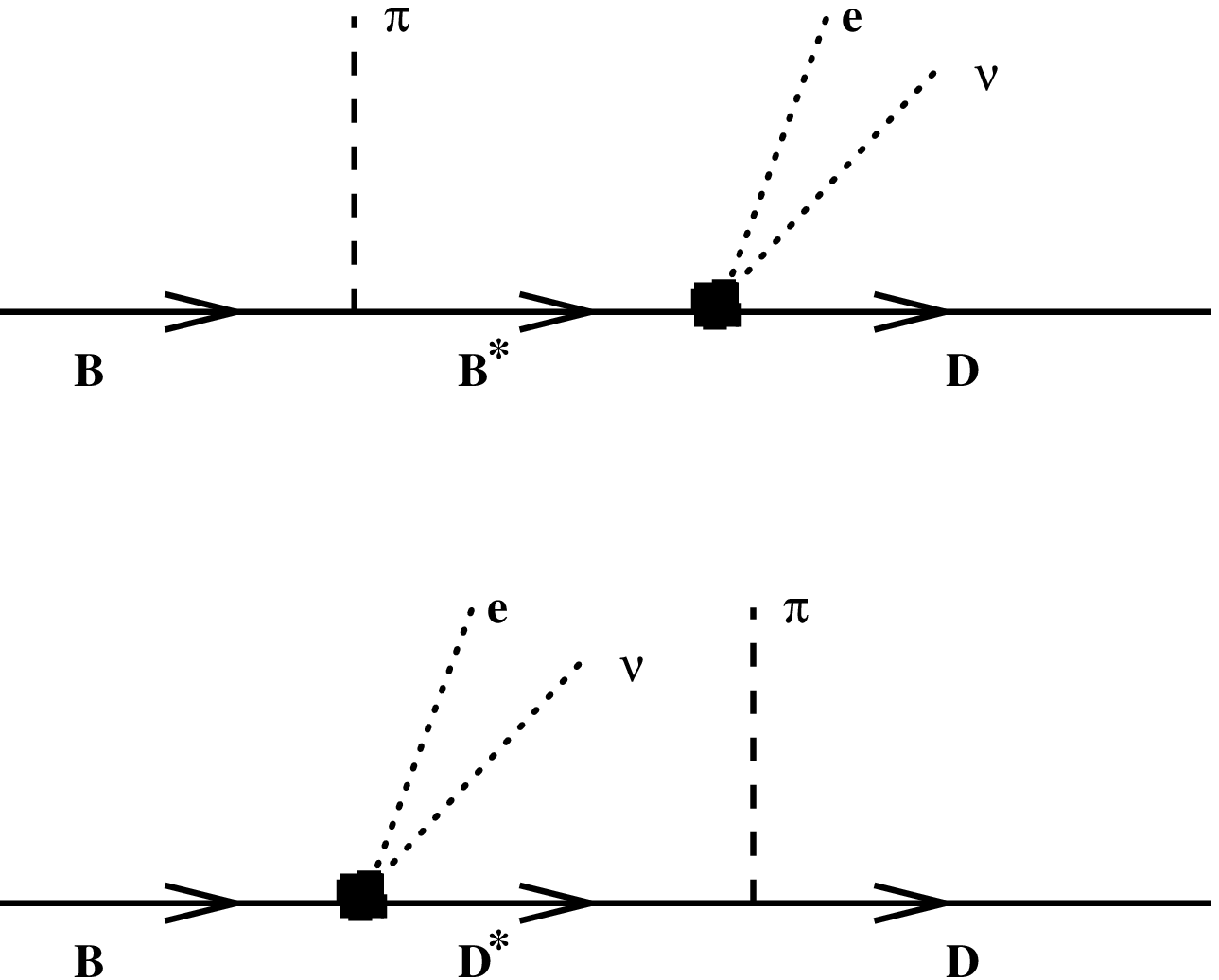}\polediagfig{Feynman  diagrams for $B\to
De\nu$}

\subsec{$B\to De\nu$ And $B\to D^*\pi E \nu$}
As a first example of an application consider a soft pion theorem
that relates the amplitudes for $B\to D^*e \nu$ and $B\to D^*\pi e
\nu$, for small pion momentum.\yanetal\ The heavy quark current is
represented in the phenomenological lagrangian approach
by
\eqn\eqeffcurr{
J^{\bar c b}_\mu = \bar h^{(c)}_{v'} \g_\mu(1-\g_5)h^{(b)}_v \to 
-\xi(\vv)\Tr\overline H_a^{(c)}(v')\g_\mu(1-\g_5) H_a^{(b)}(v)
+\cdots 
} 
where the ellipsis denote terms with derivatives, factors of light
quark masses $m_q$, or factors of $1/M_Q$, and $\xi(\vv)$ is the
Isgur-Wise function.\foot{The symbol $\xi$ is traditionally used
both for the Isgur-Wise function and for the exponential of the meson
fields. To distinguish between them, whenever context may not be
sufficient, we denote the former as a function of velocities,
$\xi(\vv)$.}\ The leading term in Eq.~\eqeffcurr\ is independent of
the pion field. Therefore, it is pole diagrams that dominate the
amplitude for semileptonic $B\to D\pi$ and $B\to D^*\pi$ transitions;
see Fig.~\xfig\polediagfig. These pole diagrams are calculable in this
approach, and are determined by the Isgur-Wise function and the
coupling $g$.

A straightforward calculation gives\yanetal $$\eqalign{
\vev{D(v')\pi^a(q)|J^{\bar c b}_\mu|B(v)} &=
iu(B)^*\half\tau^au(D)\sqrt{M_BM_D}{g \over f} \xi(\vv) \cr
&\times \left\{{1\over v\ccdot q}[i\emnlk q^\nu
v^{\prime\lambda}v^\kappa + q\cdot(v+v')v_\mu - (1+\vv)q_\mu
                                                \right. \cr
&{}\left. -{1\over v'\ccdot q}[i\emnlk q^\nu
v^{\prime\lambda}v^\kappa + q\cdot(v+v')v'_\mu - (1+\vv)q_\mu \right\}
} $$
where $u(M)$ stands for the isospin wavefunction of meson $M$. A
similar but lengthier expression is found for $B\to D^*\pi e
\nu$. Even if the coupling $g$ is close to its upper limit,
this expression makes a small contribution to the inclusive semileptonic
rate. 
\OMIT{It does not seem to account for some of the anomalously
large ``$D^{**}$'' contributions observed by CLEO.\morecleo}

\subsec{Violations To Chiral Symmetry}
Phenomenological lagrangians are particularly well suited to explore
deviations from symmetry predictions.  We begin by introducing
symmetry breaking terms into the phenomenological lagrangian. The
light quark mass matrix $m_q={\rm diag}(m_u, m_d, m_s)$ parametrizes
the violations to flavor $SU(3)_V$. To linear order in $m_q$ and
lowest order in the derivative expansion, the correction to the
phenomenological lagrangian is
\eqnn\eqlcoup%
$$\eqalignno{
\Delta{\cal L} &=
\lambda_0\ \left[ m_q \Sigma + m_q \Sigma^\dagger \right]^a{}_a
			\cr
&+\lambda_1 \Tr \overline H^{(Q)a} H_b^{(Q)} \left[ \xi m_q \xi + \xi^\dagger
m_q \xi^\dagger \right]^b{}_a		\cr
&+ \lambda_1^\prime \Tr \overline H^{(Q)a} H_a^{(Q)}\ 
\left[ m_q \Sigma + m_q \Sigma^\dagger \right]^b{}_b &\eqlcoup}$$
The coefficients $\lambda_0$, $\lambda_1$ and $\lambda'_1$ are
fixed by non-perturbative strong interaction effects, but may be
determined phenomenologically. We postpone consideration of mass
relations obtained from this lagrangian until we have introduced
heavy quark spin symmetry breaking terms into the lagrangian too.

\exercise{The operator $\left[ \xi m_q \xi - \xi^\dagger
m_q \xi^\dagger \right]$ is parity odd. One could add a term
$\Tr \overline H^{(Q)a} H_b^{(Q)}\g_5 \left[ \xi m_q \xi - \xi^\dagger
m_q \xi^\dagger \right]^b{}_a$. Why did I leave it out?}

The decay constants for the $D$ and $D_s$ mesons, defined by
\eqn\fDdefd{
\vev{ 0 | \bar d \gamma_{\mu} \gamma_5 c| D^+(p)}
= i f_D p_{\mu}
}
and
\eqn\fDSdefd{
\vev{ 0 | \bar s \gamma_{\mu} \gamma_5 c | D_s(p)}
= i f_{D_s} p_{\mu}~,
}
determine the rate for the purely leptonic decays $D^+ \rightarrow
\mu^+ \nu_{\mu}$ and $D_s \rightarrow \mu^+ \nu_{\mu}$. These are
likely to be measured in the future.\cleofds\ In the chiral
limit, where the up, down and strange quark masses go to zero, flavor
$SU(3)_V$ is an exact symmetry and so $f_{D_S} / f_{D^+} = 1$.  However
$m_s \neq 0$, so this ratio will deviate from unity.  

In the chiral lagrangian approach the current operators in
Eqs.~\fDdefd\ and~\fDSdefd\ are represented by 
\eqn\chiralqQcurr{
 J_{a(M)}^\lambda =  \frac{i\alpha}2 
          \Tr[\g^\lambda\g_5 H_b(v)\xi^\dagger_{ba}] 
}
\nfig\fdsubsfig{Feynman Diagrams in the calculation of $f_{D_s} / f_D$.}
This is the lowest order only in an expansion in derivatives. Chiral
symmetry violation in the relation between~$f_{D_S}$ and~$f_{D^+} $
arises directly from symmetry breaking terms that can be added to the
current in Eq.~\chiralqQcurr\ and indirectly in loop diagrams through
the symmetry breaking in the masses of the virtual particles.  The
latter involves, at one loop, the Feynman diagrams in
Fig.~\xfig\fdsubsfig, where a dashed line stands for a light
pseudoscalar propagator. Neglecting the up and down quark masses in
comparison with the strange quark mass, the loop graphs
give\refs{\fiveus,\goity}
\eqn\eqratiodecays{
f_{D_S} / f_{D^+} = 1 - {5\over 6}\left(1+3g^2\right) {M_K^2 \over
{16\pi^2 f^2}}\ln\left(M_K^2/ \mu^2 \right) + \cdots
}
Here $\mu$ is the renormalization point. The contribution from $\eta$
loops has been written in terms of $M_K$ using the Gell-Mann--Okubo
formula $M_\eta^2=4M_K^2/3$, and the contribution from pion loops,
proportional to $M_\pi^2 \ln M_\pi^2$, has been neglected.  The
ellipsis denote terms proportional to the strange quark mass (recall
$M_K^2 \sim m_s$) without a log.  Theoretically, as $m_s\to0$ the log
term, which has non-analytic dependence on $m_s$, dominates. The
neglected terms, as well as the contribution from the symmetry
breaking terms in the current are analytic in $m_s$.

\exercise{Display the symmetry breaking additions to the current in
Eq.~\eqratiodecays\ to lowest order in $m_q$. This will introduce new
unknown parameters. Compute the modification to $f_{D_S} / f_{D^+}-1$
from these terms.}

The dependence of the new parameters in the current on the subtraction
point $\mu$ cancels that of the logarithm. Without additional
information on these new parameters we have no real predictive power.
But one can venture a guess by arguing, as above, that the log term
dominates. What value should we assign the renormalization point
$\mu$? If $\mu$ is of order the chiral symmetry breaking scale then
the new parameters have no implicit large logarithms.  
Numerically, using $\mu=1$~GeV, the result is that
\eqn\fsubdsubs{
f_{D_s} / f_{D^+} = 1 + 0.064\ (1+3g^2),
}
or $f_{D_s} / f_{D^+} = 1.16$ for $g^2=0.5$. 

\INSERTFIG{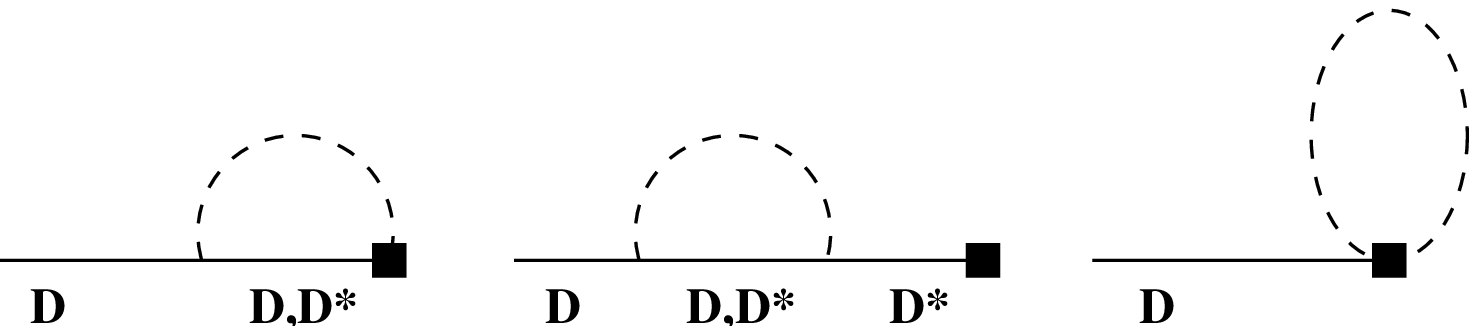}\fdsubsfig{Feynman Diagrams in the
calculation of $f_{D_s} / f_D$.}

The same formula also holds for $f_{B_s}/f_B$. In fact, to leading
order in $1/M_Q$ the ratio is independent of the the flavor of the
heavy quark. Consequently, 
\eqn\eqratioratio{
{f_{B_s}/f_B \over f_{D_s} / f_D} =1 
}
to leading order in $1/M_Q$ and all orders in the light quark masses.
Now, Eq.~\eqratioratio\ also holds as a result of chiral
symmetry, for any $m_c$ and $m_b$. That is $f_{B_s}/f_B$ and
$f_{D_s}/f_D$ are separately unity in the limit in which the light
quark masses are equal. This means that  deviations from unity in
Eq.~\eqratioratio\ must be small, $O(m_s)\times
O(1/m_c-1/m_b)$.\ratioofbs\ This ratio of ratios is observed to
be very close to unity in a variety of calculations.\soni\ This
may be very useful, since it suggests obtaining the ratio
$f_{B_s}/f_B$ of interest in the analysis of $B-\overline B$ mixing (see
below) from the ratio $f_{D_s}/f_D$, measurable from leptonic $D$ and
$D_s$ decays.

\exercise{%
The hadronic matrix elements needed for the analysis of $B-\overline B$
mixing are
$$\eqalign{
\vev{\overline B(v)| \bar b \gamma^\mu (1-\gamma_5) d
\ \bar b \gamma^\mu (1-\gamma_5) d | B(v)} &= {8\over 3} f_B^2 B_B ~,\cr
\vev{\overline B_s(v)| \bar b \gamma^\mu (1-\gamma_5) s
\ \bar b \gamma^\mu (1-\gamma_5) s | B_s(v)}&=
 {8\over 3} f_{B_s}^2 B_{B_s} ~,
}$$
where the right hand side of these equations define the parameters
$B_{B_s}$ and $B_{B}$. In the $SU(3)_V$ symmetry limit
$B_{B_s}/B_{B}=1$.  For non-zero strange quark mass, the ratio is no
longer unity.  Show that the chiral log is
$${
{B_{B_s} \over B_{B} } = 1 - {2\over 3}\left(1 - 3g^2\right)
{M_K^2 \over {16\pi^2 f^2}}\ln\left(M_K^2/ \mu^2 \right)+\cdots
}$$
Again, $M^2_{\eta}= 4 M^2_K/3$ has been used. Using $\mu=1$~GeV,
$f=f_K$, and $g^2=0.5$, the correction is $B_{B_s} / B_{B}\approx
0.95$. {\sl Hint:\/ } The hard part is to 
represent the four-fermion operators in the effective theory. If
stuck, consult Ref.~\fiveus}

Violations to chiral symmetry in $B\to D$ semileptonic decays have
also been studied. One obtains that a different Isgur-Wise function
must be used for each flavor of light spectator quark\goity
\eqn\xisoverxiud{
\frac{{\xi_{s}(\vv)}}{{\xi_{u,d}(\vv)}}=
1+ {5\over 3}g^2\Omega(\vv) {M_K^2 \over {16\pi^2
 f^2}}\ln\left(M_K^2/ \mu^2 \right) + \lambda'(\mu,\vv) M_K^2+\cdots
}
where $\lambda'(\mu,\vv)$ is the analytic counter-term, and
$$
\IFBIG{\Omega(x) =
 -1+\frac{2+x}{2\, \sqrt{x^2-1}} \;\;
 \ln \left( \frac{x+1+\sqrt{x^2-1}}{x+1-\sqrt{x^2-1}}
\right)
+ \frac{x}{4\,\sqrt{x^{2}-1}}\;\;
\ln\left( \frac{x-\sqrt{x^2-1}}{x+\sqrt{x^2-1}} \right)
}{\eqalign{\Omega(x) =
 -1+\frac{2+x}{2\, \sqrt{x^2-1}} \;\;
 &\ln \left( \frac{x+1+\sqrt{x^2-1}}{x+1-\sqrt{x^2-1}} \right)\cr
&+ \frac{x}{4\,\sqrt{x^{2}-1}}\;\;
\ln\left( \frac{x-\sqrt{x^2-1}}{x+\sqrt{x^2-1}} \right)\cr}
}$$
or, expanding about $x=1$,
$${
\Omega(x)= 
-\frac{1}{3}(x-1)+\frac{2}{15}(x-1)^{2}-
\frac{2}{35}(x-1)^{3}+\cdots
}$$
Using $g^2=0.5$ and $\mu=1$~GeV, and neglecting the counterterm one
obtains
$${
\frac{{\xi_{s}(\vv)}}{{\xi_{u,d}(\vv)}}=
1- 0.21\, \Omega(\vv) +\cdots}
$$
or a 5\% correction at $\vv=2$.

\exercise{Display the Feynman graphs that contribute to the ratio in
Eq.~\xisoverxiud.} 

\subsec{Violations To Heavy Quark Symmetry}
In a similar spirit one can consider the corrections in chiral
perturbation theory to predictions that follow from heavy quark spin
and flavor symmetries. These are effects that enter at order $1/M_Q$,
so the first step towards this end is to supplement the
phenomenological lagrangian with such terms. In particular, there are
$SU(3)_V$ preserving terms of order $1/M_Q$ that violate spin symmetry
in the lagrangian, such as\refs{\fiveus,\boydbg}
\eqn\eqspinsymviol{
\Delta{\cal L}_{\rm int} =
 {\lambda_2 \over M_Q} \Tr \overline H^{(Q)a} \sigma^{\mu \nu} H_a^{(Q)}
\sigma_{\mu \nu}
+\frac{g_2}{M_Q} \Tr\left[\overline H_a(v) \spur A_{ba}
            \gamma_5 H_b(v)\right]~.}
In addition there are contributions to the lagrangian in order $1/M_Q$
that violate flavor but not spin symmetries. These can be
characterized as introducing $M_Q$ dependence in the couplings $g$,
$\lambda_1$ and $\lambda'_1$ of Eqs.~\eqgterm\ 
and~\eqlcoup.  At the same order as these corrections, there is
a term that violates both spin and $SU(3)_V$ symmetries
\eqn\eqlthree{
\Delta{\cal L}_{\rm int} ={\lambda_3\over M_Q} 
\Tr\left[ \overline H^{(Q)a}\sigma^{\mu \nu} H_b^{(Q)}\sigma_{\mu\nu}\right] 
\left[ \xi m_q \xi + \xi^\dagger
m_q \xi^\dagger \right]^b{}_a	}

\exercise{Make a complete analysis of terms that violate $SU(3)_V$ and
heavy-quark symmetries simultaneously, to leading order in both $m_q$
and $1/M_Q$. Note that our list is shorter since we are going to
concentrate shortly on the spectrum only.}

Spin symmetry violation is responsible for ``hyperfine'' splittings in
spin multiplets. To leading order these mass splittings are computed
in terms of the spin symmetry violating coupling of
Eq.~\eqspinsymviol
\eqn\eqhyperfsplit{
\Delta_B\equiv M_{B^*}-M_B = -{8\lambda_2\over m_b}}
That the mass splittings scale like $1/M_Q$ seems to be well verified
in nature:
$${
{M_{D^*}-M_D \over M_{B^*}-M_B} \approx {M_B\over M_D}}
$$

\bigskip
\vbox{
\centerline{Measured Mass Splittings}
\vskip2ex
\centerline{\vbox{
\offinterlineskip
\hrule
\halign{&\vrule#&\strut\quad\hfil#\hfil\quad\cr
height2pt&\omit&&\omit&\cr
&$X-Y$\hfil&&$M_{X}-M_Y$(MeV)&\cr
height2pt&\omit&&\omit&\cr
\noalign{\hrule}
height2pt&\omit&&\omit&\cr
&$D_s - D^+$&&$ 99.5 \pm 0.6$\PDG&\cr
&$D^+ - D^0$&&$4.80 \pm 0.10 \pm 0.06  $\cleomass &\cr
&$D^{*+} - D^{*0}$&&$3.32 \pm 0.08 \pm 0.05  $\cleomass&\cr
&$D^{*0} - D^{0}$&&$142.12 \pm 0.05 \pm 0.05 $\cleomass &\cr
&$D^{*+} - D^{+}$ &&$140.64 \pm 0.08 \pm 0.06$\cleomass &\cr
&$D_s^{*} - D_s$&&$ 141.5 \pm 1.9$\PDG &\cr
height2pt&\omit&&\omit&\cr
\noalign{\hrule}
height2pt&\omit&&\omit&\cr
&$ B_s - B  $&&$ 82.5 \pm 2.5  $\cleomass\ or $121 \pm 9$\cusb&\cr
&$  B^0 - B^+ $&&$ 0.01 \pm 0.08  $\PDG &\cr
&$ B^{*} - B $&&$46.2 \pm 0.3 \pm 0.8$\cleotwomass\ or%
                                 $  45.4 \pm 1.0$\cusb &\cr 
&$  B_s^{*} - B_s $&&$ 47.0 \pm 2.6 $\cusb &\cr
height2pt&\omit&&\omit&\cr
\noalign{\hrule}
height2pt&\omit&&\omit&\cr
&$( D^{*0} - D^0)$&&{} &\cr
&$-( D^{*+} - D^+ )$&&$1.48 \pm 0.09 \pm 0.05 $\cleomass  &\cr
height2pt&\omit&&\omit&\cr}
\hrule
}}
\vskip2ex}

Armed with the machinery of chiral lagrangians that include both spin
and chiral symmetry violating terms, one can compare hyperfine
splitting for different flavored mesons. There is a wealth of
experimental information to draw from; see Table~3. Breaking of flavor
$SU(3)_V$ and heavy quark flavor symmetries by electromagnetic effects
is not negligible. It is readily incorporated into the lagrangian in
terms of the charge matrices $Q_Q={\rm diag}(2/3,-1/3)$ and $Q_q={\rm
diag}(2/3,-1/3,-1/3)$,\rosnerwise\ which must come in
bilinearly. For example, terms involving $Q_q^2$ correspond to
replacing $m_q\to Q_q$ in Eqs.~\eqlcoup\  and~\eqlthree.
The electromagnetic effects of the light quarks can be neglected if
one considers only mesons with $d$ and $s$ light quarks. The
electromagnetic shifts in the hyperfine splittings $\Delta_{X_q}$ and
$\Delta_{X_q}$ ($X=D,B$, $q=d,s$) differ on account of different $b$ and $c$
charges, but they cancel in the difference of splittings
$$
 \Delta_{X_s}-\Delta_{X_d}= 
(M_{X^*_s}-M_{X_s}) - (M_{X^*_d}-M_{X_d})
$$
The only term in the phenomenological lagrangian that enters this
difference is Eq.~\eqlthree. This immediately leads to
\IFBIG{\eqn\eqmassrel{
(M_{B^*_s}-M_{B_s}) - (M_{B^*_d}-M_{B_d}) = 
{m_c \over m_b}
\left({\bar\alpha_s(m_c)\over\bar\alpha_s(m_b)} \right)^{\!\!{9\over25}}
\left[ (M_{D^*_s}-M_{D_s}) - (M_{D^*_d}-M_{D_d})\right]
}}%
{\eqn\eqmassrel{
(M_{B^*_s}\!-\!M_{B_s})\!-\!(M_{B^*_d}\!-\!M_{B_d})\! =\! {m_c \over m_b}\!\!
\left({\bar\alpha_s(m_c)\over\bar\alpha_s(m_b)} \right)^{\!\!{9\over25}}
\kern-1.3ex \left[ (M_{D^*_s}\!-\!M_{D_s})\!-\!(M_{D^*_d}\!-\!M_{D_d})\right]
}}
We have included here the short distance QCD effect.\fgl

One expects  Eq.~\eqmassrel\ holds  to much
better accuracy than the separate relations for each hyperfine splitting  in
Eq.~\eqhyperfsplit. Recall that $SU(3)_V$ breaking by light
quark masses and electromagnetic interactions have been accounted for
in leading order. Moreover, the result is trivially generalized by
replacing the quark mass matrix in Eqs.~\eqlcoup\ 
and~\eqlthree, by an arbitrary function of the light quark mass
matrix.  It is seen from Table~3 that this relation works well. The
left side is $1.2\pm2.7$~MeV while the right side is $3.0\pm6.3$~MeV.

Since both sides of Eq.~\eqmassrel\ are consistent with zero and
both are proportional to the interaction term in Eq.~\eqlthree,
it must be that the coupling $\lambda_3$ is very
small.\rosnerwise\ From the difference of hyperfine splittings in
the charm sector
$${
-{8\lambda_3\over m_c}(m_s-m_d) = 0.9\pm1.9~{\rm MeV}}$$
while
$${
M_{D_s}-M_{D_d}=4\lambda_1 (m_s-m_d)- {12\lambda_3\over
m_c}(m_s-m_d) = 99.5\pm0.6~{\rm MeV}}$$
leading to $|\lambda_3/\lambda_1|$ less than $\sim20$~MeV. This is
smaller than expected by about an order of magnitude. With such a
small coefficient it is clear that the next-to-leading terms and the
loop corrections may play an important role. In particular they may
invalidate the simple $1/M_Q$ scaling of
Eq.~\eqmassrel.\ransath\ There is no obvious breakdown of chiral
perturbation theory, even though the leading coupling ($\lambda_3$) is
anomalously small.\jenk

At one loop, the expressions for the mass shifts involve large
$O(m_s\ln m_s)$ and $O(m_s^{3/2})$ (non-analytic)
terms.\refs{\goity,\jenk} The coupling $\lambda_3$ is not anomalously
small at one loop. Instead, the smallness of the difference of
hyperfine splittings in Eq.~\eqmassrel\ is the result of a
precise cancellation between one loop and tree level graphs.
Explicitly,\jenk
$${
\left( M_{X_s}-M_{X_s^*} \right)
-\left( M_{X_d}-M_{X_d^*} \right)  =
 \frac 5 3 g^2 \left(\frac{8\lambda_2}{M_Q}\right)
{M_K^2 \over {16 \pi^2 f^2}}
\ln \left( M_K^2/ \mu^2 \right)
-{8\lambda_3\over M_Q}m_s
}$$
With $g^2=0.5$ and $\mu=1$~GeV, the chiral log is 30~MeV, so the
$\lambda_3$ counterterm must cancel this to a precision of better than
10\%.

The $1/M_Q$ corrections to the masses $M_X$ and $M_{X^*}$ drop out of
the combination $M_{X}+3M_{X^*}$. The combination
$(M_{X_s}+3M_{X_s^*})-(M_{X_d}+3M_{X_d^*})$ is a measure of $SU(3)_V$
breaking by a non-vanishing $m_s$ (or $m_s-m_d$ if the $d$ quark mass
is not neglected). It can be computed in the phenomenological
lagrangian. To one loop\jenk
\eqnn\eqspinindepmass
$$\eqalign{
\frac 1 4 \left( M_{X_s}+3M_{X_s^*} \right)
-\frac 1 4 \left( M_{X_d}+3M_{X_d^*} \right) = 4\lambda_1 m_s
- g^2 \left( 1 + \frac 8 {3 \sqrt 3} \frac 1 2
\right) {{M_K^3} \over {16 \pi f^2}}\cr
- 4\lambda_1 m_s \left( \frac{25}{18} + \frac 9 2 g^2 \right)
{M_K^2 \over {16 \pi^2 f^2}} \ln \left( M_K^2/ \mu^2
\right)
}\eqno\eqspinindepmass $$
The pseudoscalar splittings $(M_{D_s}-M_{D_d})$ and
$(M_{B_s}-M_{B_d})$ have been measured; see Table~3. Also,
$\frac14(M_{X_s}+3M_{X_s^*}) -\frac14(M_{X_d}+3M_{X_d^*})=
\frac34[(M_{X^*_s}-M_{X_s})-(M_{X^*_d}-M_{X_d})] + (M_{X_s}-M_{X_d})$,
and the term in square brackets is less than a few MeV, as we saw
above. The combination $(M_{X_s}+3M_{X_s^*})-(M_{X_d}+3M_{X_d^*})$ in
Eq.~\eqspinindepmass\ is first order in $m_s$ but has no
corrections at order $1/M_Q$. Thus, one expects a similar numerical
result for $B$ and $D$ systems. Experimentally,
$(M_{B_s}-M_{B_d})/(M_{D_s}-M_{D_d})$ is consistent with unity; see
Table 3. The formula in Eq.~\eqspinindepmass\ has a significant
contribution from the $M_K^3$ term which is independent of the
splitting parameter $\lambda_1$.  The $M_K^3$ term gives a negative
contribution to the splitting of $\sim-250$~MeV for $g^2=0.5$.  The
chiral logarithmic correction effectively corrects the tree level
value of the parameter $\lambda_1$; for $\mu = 1$~GeV and $g^2=0.5$,
the term $4\lambda_1 m_s$ gets a correction $\approx 0.9$ times its tree
level value.  Thus, the one-loop value of $4\lambda_1m_s$ can be
significantly greater than the value determined at tree-level of
approximately $100$~MeV.

Chiral perturbation theory can be used to estimate the leading
corrections to the form factors for semileptonic $B\to D$ or $D^*$
decays which are generated at low momentum, below the chiral symmetry
breaking scale. Of particular interest are corrections to the
predicted normalization of form factors at zero recoil, $\vv=1$.
According to Luke's theorem (see Section~\lukesthm), long distance
corrections enter first at order $1/M_Q^2$. Deviations from the
predicted normalization of form factors that arise from terms of order
$1/M_Q^2$ in either the lagrangian or the current are dictated by
non-perturbative physics. But there are computable corrections that
arise from the terms of order $1/M_Q$ in the lagrangian.  These must
enter at one-loop, since Luke's theorem prevents them at tree level,
and result from the spin and flavor symmetry breaking in the hyperfine
splittings $\Delta_D$ and $\Delta_B$. Retaining only the dependence on
the larger $\Delta_D$, the correction to the matrix elements at zero
recoil are\ranwise
\eqna\eqranwise
$$\eqalignno{ 
\vev{D(v) | J^{\bar c b}_\mu|B(v)} &= 2v_\mu\left(1- 
{3g^2\over2}\left({\Dc\over4\pi f}\right)^2\left[ F(\Dc/M_\pi)
+\ln(\mu^2/M_\pi^2)\right] + C(\mu)/m_c^2\right)&\cr
& &\eqranwise a\cr
\vev{D^*(v,\epsilon) | J^{\bar c b}_\mu|B(v)} &=
2\epsilon^*_\mu\left(1-  {g^2\over2}\left({\Dc\over4\pi
f}\right)^2\left[ F(-\Dc/M_\pi) +\ln(\mu^2/M_\pi^2)\right] +
C'(\mu)/m_c^2\right)& \cr & &\eqranwise b \cr}  $$
where $C$ and $C'$ stand for tree level counter-terms and 
\eqn\eqranwisef{
F(x)\equiv\int_0^\infty dz {z^4\over (z^2+1)^{3/2}}
\left(\frac{1}{[(z^2+1)^{1/2}+x]^2}-
\frac{1}{z^2+1}\right)}
As before, no large logarithms will appear in the functions $C$
and $C'$ if one takes $\mu\approx4\pi f\sim1$~GeV. With this
choice, formally, their contributions are dwarfed by the term that is
enhanced by a logarithm of the pion mass. Numerically, with $g^2=0.5$
the logarithmically enhanced term is $-2.1$\% and $-0.7$\% for $D$
and $D^*$, respectively.

The function $F$ accounts for effects of order $(1/m_c)^{2+n}$,
$n=1,2,\ldots$\ \ It is enhanced by powers of $1/M_\pi$ over terms that
have been neglected. Consequently it is expected to be a good
estimate of higher order $1/m_c$ corrections. With
$\Dc/M_\pi\approx1$, one needs $F(1)=14/3-2\pi$ and
$F(-1)=14/3+2\pi$ for a numerical estimate; with $\mu$ and $g^2$ as
above, this term is 0.9\% and $-2.0$\% for $D$ and $D^*$, respectively.

To put it differently, if the function $F$ is expanded in powers of
$1/m_c$, then since the resulting terms have inverse powers of
$1/M_\pi$ they cannot be confused with local counterterms. These
terms play the same role as the non-analytic logs of the previous
sections.

\subsec{Trouble On The Horizon?}

Frequently the non-analytic corrections to relations that follow from
the symmetries are uncomfortably large. A case of much interest is the
relation between the form factors $f_\pm$ and $h$ for $B\to K$
transitions, relevant to the short distance process $b\to se^+e^-$,
$$\IFBIG{\eqalign{
    \langle \ol K(p_K)\,|\,\ol s\gamma^\mu b\,|\,\ol B(p_B)\rangle
    &=f_+\,(p_B+p_K)^\mu + f_-\,(p_B-p_K)^\mu\,,\cr
    \langle \ol K(p_K)\,|\,\ol s\sigma^{\mu\nu} b\,|\,\ol B(p_B)\rangle
    &= {\rm i} h\,[(p_B+p_K)^\mu (p_B-p_K)^\nu -
    (p_B+p_K)^\nu (p_B-p_K)^\mu]\,,\cr}}%
{\eqalign{
    \langle \ol K(p_K)|\ol s\gamma^\mu b|\ol B(p_B)\rangle
    &=f_+(p_B+p_K)^\mu + f_-(p_B-p_K)^\mu,\cr
    \langle \ol K(p_K)|\ol s\sigma^{\mu\nu} b|\ol B(p_B)\rangle
    &= {\rm i} h[(p_B+p_K)^\mu (p_B-p_K)^\nu -
    (p_B+p_K)^\nu (p_B-p_K)^\mu],\cr}}$$
and the form factors for $B\to\pi e\nu$,
$${
   \langle \ol \pi(p_\pi)\,|\,\ol u\gamma^\mu b\,|\,\ol B(p_B)\rangle
    =\hat f_+\,(p_B+p_\pi)^\mu + \hat f_-\,(p_B-p_\pi)^\mu\,.}$$
In the combined large mass and chiral limits  only one of these form
factors is independent:
\eqn\eqkpirel{
m_b h = f_+ = -f_- = \hat f_+ = -\hat f_- }
In this limit, the ratio of rates for $B\to K e^+e^-$ and $B\to\pi
e\nu$ is simply given, in the standard model of electroweak
interactions, by $|V_{ts}/V_{ub}|^2$, times a perturbatively
computable function of the top quark mass. If the relation~\eqkpirel\
held to good accuracy one could thus measure a ratio of fundamental
standard model parameters.\foot{Another application of this relation
has been discussed by I. Dunietz\dunietz. Assuming factorization in
$B\to \psi X$, ratios of CKM elements can be extracted from these two
body hadronic decays.}

The non-analytic, one-loop corrections to the relations in
Eq.~\eqkpirel\ have been computed.\falkben\ The results are
too lengthy to display here. Numerically, the violation to $SU(3)_V$
symmetry is found to be at the 40\% level.\foot{The large
violation of $SU(3)_V$ symmetry affects as well the results of 
Dunietz (see previous footnote).} 

The phenomenological lagrangian that we have been considering
extensively neglects the effects of states with heavy-light quantum
numbers other than the pseudoscalar -- vector-meson multiplet. The
splitting between multiplets is of the order of 400~MeV and is hardly
negligible when one considers $SU(3)_V$ relations involving both $\pi$ and
$K$ mesons. For example, consider the effect of the scalar --
pseudovector-meson multiplet. One can incorporate its effects into the
phenomenological lagrangian. To this end, assemble its components into a
``superfield'', akin to that in Eq.~\eqsuperf\  for the
pseudoscalar -- vector multiplet:\flf
\eqn\eqSsuperf{
S_a(v)={1+\vslash\over2}\left[
     B_{1a}^{\prime\mu}\g_\mu\g^5 - B^*_{0a}\right]\,.
                                        } 
The phenomenological lagrangian has to be supplemented with a kinetic
energy and mass for $S$,
$${
  \Tr\left[\overline S_a(v)(i v\cdot D_{ba}-\Delta\delta_{ba}) 
			S_b(v)\right]\,,}$$
where $\Delta$ is the mass splitting  for the excited $S$ from the
ground state $H$, and with coupling terms
$${
 g'\,\Tr\left[\ol S_a(v)S_b(v)\,\Aslash_{ba}\g^5\right] + 
 (h\,\Tr\left[\ol H_a(v)S_b(v)\,\Aslash_{ba}\g^5\right] + {\rm
h.c.})\,.}$$
In terms of these one can now compute additional corrections to
quantities such as $f_{D_s}/f_D$ in Eq.~\eqratiodecays.
Numerically the corrections are not small,\falk\
$ f_{D_s}/f_D = 1 + 0.13h^2 $ for $M_{D^*_0}=2300$~MeV (or
$ f_{D_s}/f_D = 1 + 0.08 h^2 $ for $M_{D^*_0}=2400$~MeV),
assuming the strange mesons to be 100~MeV heavier. Similarly,
corrections to the Isgur-Wise function can be computed, and are not
negligible.\falk

\newsec{Acknowledgments}
This work is supported in part by a Sloan Foundation Fellowship and by
the Department of Energy  grant DOE-FG03-90ER40546.

\footatend\vfill\supereject\immediate\closeout\rfile\writestoppt
\baselineskip=14pt\centerline{{\bf References}}\bigskip{\frenchspacing%
\parindent=20pt\escapechar=` \input refs.tmp\vfill\eject}\nonfrenchspacing
\ifx\answfig\yesanswfig{  }
\else
\vfill\eject\immediate\closeout\ffile{\parindent40pt
\baselineskip14pt\centerline{{\bf Figure Captions}}\nobreak\medskip
\escapechar=` \input figs.tmp\vfill\eject}
\fi

\bye